\documentclass{article}
\usepackage{ijcai24}

\usepackage{times}
\usepackage{soul}
\usepackage{url}
\usepackage[hidelinks]{hyperref}
\usepackage[utf8]{inputenc}
\usepackage[small]{caption}
\usepackage{graphicx}
\usepackage{amsmath}
\usepackage{amsthm}
\usepackage{booktabs}
\usepackage{algorithm}
\usepackage{algorithmic}
\usepackage[switch]{lineno}

\usepackage{amssymb}
\usepackage{subcaption}
\usepackage{enumitem}
\usepackage{xspace}
\usepackage{csquotes}
\usepackage{latexsym}
\usepackage{dsfont}

\newtheorem{example}{Example}

\newtheorem{proposition}{Proposition}
\newtheorem{conjecture}{Conjecture}

\newtheorem{imr}{}
\newenvironment{cmr}[1]
  {\renewcommand\theimr{#1}\imr}
  {\endimr}

\usepackage{pgfplots}
\usepackage{mathtools}

\newcommand{\real}{\mathbb{R}}
\newcommand{\indicator}{\mathds{1}}
\newcommand{\targetval}{r}
\newcommand{\edgeadj}{R}

\newcommand{\lgsg}{LGSG\xspace}
\newcommand{\lgsgs}{LGSGs\xspace}

\DeclareMathOperator*{\argmax}{arg\,max}
\DeclareMathOperator*{\argmin}{arg\,min}

\newcommand{\citet}[1]{\citeauthor{#1}~\shortcite{#1}}
\newcommand{\supertiny}[1]{{\fontsize{4.5pt}{4.5pt}\selectfont {#1}}}

\definecolor{amber}{rgb}{1.0, 0.75, 0.0}
\definecolor{rust}{rgb}{0.72, 0.25, 0.05}
\definecolor{teal}{rgb}{0.0, 0.5, 0.5}
\definecolor{tealblue}{rgb}{0.21, 0.46, 0.53}
\definecolor{burntorange}{rgb}{0.8, 0.33, 0.0}
\definecolor{navyblue}{rgb}{0.0, 0.0, 0.5}

\newcommand{\gcolor}{rust!40} 
\newcommand{\pcolor}{tealblue!40} 
\newcommand{\ncolor}{burntorange!5} 

\def\fscale{0.41}
\def\fmarksize{2pt}
\def\lpmark{o}
\def\domark{square}
\def\dolmark{triangle}
\def\sgmark{*}
\def\spmark{diamond}
\def\doname{DO}
\def\lpname{LP}
\def\dolname{DO+l}
\def\APlin{$g_{\text{LIN}}$}
\def\APexp{$g_{\text{EXP}}$}
\def\subgamename{SG}
\def\supportname{SP}
\def\graphsizename{LG}
\def\labfont{\huge} 
\def\tickfont{\huge}
\def\titfont{\huge}
\def\legfont{\footnotesize}
\def\legcols{5}
\def\sizescolor{rust}
\def\minnewaskaexpcolor{teal}

\def\computationlabel{Runtime [s]}
\def\sparsitylabel{Sparsity [\% of total paths]}
\def\gamesizelabel{Game size [$\#$ of paths]}

\title{Layered Graph Security Games}

\author{
Jakub \v{C}ern\'{y}$^\dagger$
\and
Chun Kai Ling$^\dagger$
\and
Christian Kroer
\And
Garud Iyengar \\
\affiliations
Department of Industrial Engineering and Operations Research, Columbia University
\emails
\{jakub.cerny, chunkai.ling, christian.kroer\}@columbia.edu,
garud@ieor.columbia.edu \\
}

\begin{document}

\maketitle
\def\thefootnote{$\dagger$}\footnotetext{Equal contribution. }\def\thefootnote{\arabic{footnote}}

\begin{abstract}
Security games model strategic interactions in adversarial real-world applications. Such applications often involve extremely large but highly structured strategy sets (e.g., selecting a distribution over all patrol routes in a given graph). In this paper, we represent each player's strategy space using a \textit{layered graph} whose paths represent an exponentially large strategy space. Our formulation entails not only classic pursuit-evasion games, but also other security games, such as those modeling anti-terrorism and logistical interdiction. We study two-player zero-sum games under two distinct utility models: linear and binary utilities. We show that under linear utilities, Nash equilibrium can be computed in polynomial time, while binary utilities may lead to situations where even computing a best-response is computationally intractable. To this end, we propose a practical algorithm based on incremental strategy generation and mixed integer linear programs. We show through extensive experiments that our algorithm efficiently computes $\epsilon$-equilibrium for many games of interest. We find that target values and graph structure often have a larger influence on running times as compared to the size of the graph per se.
\end{abstract}

\section{Introduction}

Security games model strategic interactions between a defender, typically representing governmental entities, and an attacker engaged in illicit activities. They have served as the foundation for deployed solutions in numerous real-world scenarios, spanning both physical and cyber security domains, such as scheduling air marshals to protect flights \cite{Tsai2009}, or protecting wildlife in natural parks \cite{fang2017paws}. These applications feature complex strategy spaces and reward functions that are application specific. 
In this paper, we introduce \textit{Layered Graph Security Games} (\lgsgs), a class of games where each player selects a path in a layered directed acyclic graph (henceforth \textit{layered graph}) and receive payoffs depending on how ``close'' these two paths were. 
\lgsgs strike a good balance between model expressiveness and computational complexity.
On one hand, many security games and their variants can be easily reframed as \lgsgs, despite not being explicitly defined in such terms. 
These include patrolling games, which have a natural time-component as well as cybersecurity applications, where layered graphs model dependencies in attack chains. 
Yet, being relatively compact structures, layered graphs retain, and in some cases expose much of the ``nice'' combinatorial aspects of the underlying security game, resulting in practically efficient game solvers. This stands in contrast to more heavy-handed formulations such as extensive form games.

Our contributions are summarized as follows.
(i) We introduce Layered Graph Security Games (\lgsg), 
and show how they lead to compact representations of an otherwise exponentially large space (Section~\ref{sec:prelims}).
(ii) We demonstrate how many security problems may be reformulated as \lgsgs, including various pursuit-evasions games and two novel settings relating to anti-terrorism and logistical interdiction (Section~\ref{sec:applications}).
(iii) We study the computational complexity of solving \lgsgs in two regimes: linear and binary utilities (Section~\ref{sec:equilibria}), give polynomial-time algorithms for the former, and prove hardness results for the latter. 
(iv) For binary utilities, we propose a solver based on incremental strategy generation and efficient best-response oracles formulated as mixed integer linear programs. 
(v) Experiments on a range of applications 
using both synthetic and real-world maps from various cities and parks (Section~\ref{sec:experiments}), show that our strategy generation method scales favorably. 
We find that 
equilibria exhibit a tiny support relative to the number of paths, validating our hypothesis that in practical domains, it is structure and not game size that governs computational costs.

\section{Related Work}
This paper is related to several fields spanning across disciplines. 
Since these are huge research areas in and of themselves, we focus on those most related to security games.

\textbf{Pursuit-Evasion games (PEGs)} model scenarios when one group (e.g., robots) locates and captures members of another group, often within a specified timeframe. 
This rich line of work goes by many names (e.g., Cops and Robbers, Differential Games, Games of Pursuit), dealing with a variety of environments, such as those exhibiting perfect or imperfect information, and on discrete and continuous time/space \cite{isaacs1999differential,friedman2013differential,weintraub2020introduction,bopardikar2008discrete,bonato2011game,parsons2006pursuit}. 
The mathematics behind PEGs is deep and attractive. Nonetheless, computational costs become a hindrance in all but the simplest environments. 
Furthermore, models in PEGs can be fairly rigid; indeed, research in PEGs is often centered around the underlying geometry of the environment or proving theoretical bounds on metrics such as task-completion time.

\textbf{Extensive-Form games (EFGs)} are played on a game tree, with each player choosing actions to take at each of their information sets~\cite{shoham2008multiagent}. Extremely expressive and successful in practice, EFG solvers are responsible for superhuman poker bots today \cite{brown2018superhuman,brown2019superhuman}. 
However, as trees, computing exact equilibria in EFGs incurs computational costs that is exponential in time horizon. Conversely, while the number of possible paths in \lgsgs taken is also exponential in horizon, its reward functions are constrained by the layered graph structure, allowing for more efficient computation of best-responses and equilibrium.
Recently, there have been significant work scaling up EFG solvers by incorporating machine learning \cite{lanctot2017unified,perolat2022mastering,moravvcik2017deepstack,wang2019deep,xue2021solving}. While powerful, such approaches rarely yield theoretical guarantees on quality of the equilibria, making them less suited for high-stakes security applications.

\textbf{Network Interdiction games (NIGs)} study the optimal arcs in a network to remove or interdict in order to prevent an evader from traversing the graph. First studied by \citet{wollmer1964removing}, NIGs now come in a variety of objectives, such as increasing the evader's shortest path to an exit, minimizing the maximum flow between two vertices, as well as a range of applications, including cybersecurity, cyberphysical security and supply-chain attacks \cite{washburn1995two,smith2020survey,smith2008algorithms}. Most of the existing literature consider attacker paths, but allow the defender to select arbitrary vertices or edges to interdict. This is unrealistic, particularly in physical patrolling such as anti-poaching, since it amounts to the defender ``teleporting'' to a location rather moving in to intercept the attacker.

\textbf{Security games} are the namesake of this paper, and have enjoyed much attention owing to a number of successful deployments in the real-world \cite{jain2013security,pita2008armor,shieh2012protect,an2017stackelberg}. In its vanilla form, security games feature a defender choosing a distribution over targets to defend and an attacker choosing one of them to attack. This simple setting enjoys polynomial-time solvers, even in the general-sum case \cite{kiekintveld2009computing,conitzer2006computing}. 
Much developments have been made to account for large, but structured strategy spaces such as defender target schedules \cite{korzhyk2010complexity} and repeated interactions \cite{fang2015security}. 
Unfortunately, just like NIGs, most security games focus on the special case where only the defender possesses structured strategies --- the attacker still ``teleports'' to the target. 
This assumption reins in computational costs, but at the price of realism. 

One of the few works which do handle structured strategies for both players is the Escape Interdiction Game (EIG) studied by \citet{zhang2017optimal} who investigate interdiction specifically in transport networks. \lgsgs partially generalizes EIGs by allowing for richer interdiction, reward functions and most importantly goes beyond interdiction to include applications such as anti-terrorism, delayed interdiction and advanced persistent threats (Section~\ref{sec:applications}).
\section{Layered Graph Security Games}\label{sec:prelims}

The fundamental structure that we will work with is the \textit{layered directed graph}. 
Let $\mathcal{V}$ be a finite set of vertices, and  $\mathcal{G}_a = (\mathcal{V}, \mathcal{E}_a),$ $\mathcal{G}_d = (\mathcal{V}, \mathcal{E}_d)$ graphs for the attacker and defender, where the sets $\mathcal{E}_a$ and $\mathcal{E}_d$ contain \textit{directed} edges. 
$\mathcal{V}$ comprises $L > 1$ layers, meaning that $\mathcal V$ can be partitioned into non-empty sets $\mathcal{V}_1, \dots, \mathcal{V}_L$ where
all edges $e_a \in \mathcal{E}_a$ lie in $\mathcal{V}^\ell \times \mathcal{V}^{\ell+1}$ for some $\ell \in [1, L-1]$.
The same holds for all $e_d \in \mathcal{E}_d$. For an edge $e=(v_l,v_{l+1})$, we denote the vertex $v_l$ by $e^-$ and the vertex $v_{l+1}$ by $e^+$. For a player $i\in\{a,d\}$, the incoming edges for a vertex $v$ are denoted as $\mathcal{E}_i^+(v)$ and the outgoing edges as $\mathcal{E}^-_i(v)$.
Note that $\mathcal{G}_a$ and $\mathcal{G}_d$ share the vertex set $\mathcal{V}$, though we allow $\mathcal{G}_a$ and $\mathcal{G}_d$ to be disconnected.

We assume that the first layer is a singleton, i.e., $|\mathcal{V}^1| = 1$; this restriction does not impose any significant modeling restrictions. However, the last layer may comprise multiple vertices. We call these \textit{terminal states} or \textit{targets} $\mathcal{V}^\odot = \mathcal{V}^{L}$. 

\textbf{Player strategies.} We denote by $\mathcal{P}_a$ and $\mathcal{P}_d$ the set of paths for the attacker and defender. Each path for the attacker is of the form $(v_1, v_2, \dots v_{L}) \in \mathcal{V}^1 \times \mathcal{V}^2 \times \dots, \times \mathcal{V}^{L}$ where $(v_\ell, v_{\ell+1}) \in \mathcal{E}_a$. 
Consider some path $p \in \mathcal{P}_a \cup \mathcal{P}_d$. 
It traverses vertices in increasing order of their layer number and terminates at a target $v \in \mathcal{V}^\odot$. We denote by $p(i)$ the $i$-th vertex of $p$ and $p[i]$ its $i$-th edge, i.e., $(v_i, v_{i+1})$. With a slight abuse of notation we write $v \in p$ and $e \in p$ if vertex $v$ or edge $e$ lies in $p$.
In general, $|\mathcal{P}_a|,|\mathcal{P}_d|$ are exponential in $L$; dealing with these large strategy spaces is our key contribution. 

\textbf{Targets and interdiction function.} 
We assume that each target has an associated value given by $\targetval^\odot:\mathcal{V}^\odot\to\real$. Typically, this value will be nonnegative. For some path $p \in \mathcal{P}_a \cup \mathcal{P}_d$, we write $\targetval^\odot(p)$ as a shorthand for $\targetval^\odot(p(L))$. We also introduce an interdiction function $\edgeadj: \mathcal{E}_d \times \mathcal{E}_a \rightarrow \{ 0, 1 \}$, which is equal to $1$ when two edges are in ``close proximity'' such that the defender is able to interdict the attacker. While $\edgeadj$ is usually defined in terms of distance metrics (e.g., the range of a sensor and/or physical distance between two edges), we impose no such restriction in our framework.

\subsection{Player Utilities}
In general, utilities are a function of paths $u_a : \mathcal{P}_d \times \mathcal{P}_a \rightarrow \mathbb{R}$. We assume that the game is \textit{zero-sum}, i.e., $u_a(p_d, p_a) = -u_d(p_d, p_a)$. For simplicity, we will write $u=u_a$. Solving a game with arbitrary $u$ is computationally intractable (in terms of the size of the graph 
$|\mathcal{V}|, |\mathcal{E}_a|,|\mathcal{E}_d|$); indeed, it takes an exponential amount of space to even specify $u$. Nonetheless, for many security games $u$ takes more compact forms involving edges and possibly $\mathcal{V}^{\odot}$ and $\targetval^\odot$. The form of $u$ greatly influences equilibrium structure and its computation.

\begin{itemize}[leftmargin=*] \item \textit{Linear utilities} may be additively decomposed in terms of pairs of edges shared between $p_d$ and $p_a$,
\begin{equation}
    u_\textsc{Lin}(p_d,p_a) = \sum_{e_d\in p_d}\sum_{e_a\in p_a} Q(e_d, e_a),
    \label{eq:linear-utils}
\end{equation}
where $Q: \mathcal{E}_d \times \mathcal{E}_a \to \real$ maps payoffs between edges between $\mathcal{E}_a$ and $\mathcal{E}_d$. 
We are particularly interested in the case where
$Q=-\edgeadj(e_d, e_a),$ i.e.,  the attacker incurs a penalty of $1$ each time it is interdicted.
Linear utility models allow the attacker to be caught repeatedly. For example, a driver fined for speeding may be fined again in the future, with penalties accumulating additively.
\item \textit{Binary utilities} avoid rewarding multiple interdictions,
\begin{align*}
    u_\textsc{Bin}(p_d,p_a) = \targetval^{\odot}(p_a) \cdot \indicator\left[\sum_{e_d\in p_d}\sum_{e_a\in p_a} R(e_d, e_a) = 0\right],
\end{align*}
where $\indicator$ is the indicator function, i.e., the attacker receives a reward of $\targetval^{\odot}(p_a)$ if and only if $p_a$ and $p_d$ do not share \textit{any} edge that are ``close''. Binary utilities are often more suited for security applications. For example, a driver is arrested for drink driving, not released back to the public. 
\end{itemize}

\subsection{Nash Equilibrium}
Our goal is to find a Nash equilibrium (NE), possibly mixed, over player paths. Denote by $\Delta_a$ and $\Delta_d$ the probability simplices over $\mathcal{P}_a$ and $\mathcal{P}_d$ respectively. Then, for some distribution over paths $x_i \in \Delta_i$, $x_i(p_i)$ is the probability that $p_i$ is played by player $i\in\{a,d\}$. 
The NE problem reduces to solving the bilinear saddle point problem
\begin{align}
    &\min_{x_d \in \Delta_d} \max_{x_a \in \Delta_a} \mathbb{E}_{p_d \sim x_d, p_a \sim x_a} \left[ u (p_d, p_a) \right] \label{eq:general-min-max}\\
    = &\min_{x_d \in \Delta_d} \max_{x_a \in \Delta_a} \sum_{p_d \in \mathcal{P}_d} \sum_{p_a \in \mathcal{P}_a} x_a(p_a) \cdot x_d(p_d) \cdot u(p_d, p_a). \nonumber
\end{align}
Since $\Delta_d$ and $\Delta_a$ are convex and compact and the objective is convex-concave, the minimax theorem \cite{v1928theorie} holds. Thus, the game has a unique value.

\begin{figure}[t]
\centering
\begin{subfigure}[t]{0.15\textwidth}
    \centering
    \begin{tikzpicture}[auto,node distance=2.5cm,
                    thick,main node/.style={circle,fill=\gcolor,draw,font=\sffamily\small\bfseries,inner sep=2}, 
                    gray node/.style={circle,fill=\ncolor,draw,font=\sffamily\small\bfseries,inner sep=2},
                    scale=0.25,minimum size=10pt]

  \node[main node] (s) at (0,0) {};
  
  \node[main node] (A) at (2,2) {};
  \node[gray node] (B) at (2,0) {};
  \node[main node] (C) at (2,-2) {};
  
  \node[main node] (D) at (4,2) {};
  \node[main node] (E) at (4,0) {};
  \node[main node] (F) at (4,-2) {};
  
  \node[main node] (G) at (6,2) {};
  \node[gray node] (H) at (6,0) {};
  \node[main node] (I) at (6,-2) {};
  
  \node[main node] (t) at (8,0) {};

  \foreach \source/\dest in {s/A, s/C, A/D, A/E, C/E, C/F, D/G, E/G, E/I, F/I, I/t, G/t}
     \draw [->] (\source) -- (\dest);

\end{tikzpicture}
    \caption{$\mathcal{V}$ and $\mathcal{E}_d \cup \mathcal{E}_a$}
    \label{fig:hexa-eight-full}
\end{subfigure}
\begin{subfigure}[t]{0.15\textwidth}
    \centering
    \begin{tikzpicture}[auto,node distance=2.5cm,
                    thick,main node/.style={circle,fill=\gcolor,draw,font=\sffamily\small\bfseries,inner sep=2}, 
                    gray node/.style={circle,fill=\ncolor,draw,font=\sffamily\small\bfseries,inner sep=2},
                    scale=0.25,minimum size=10pt]

  \node[main node] (s) at (0,0) {};
  
  \node[main node] (A) at (2,2) {};
  \node[gray node] (B) at (2,0) {};
  \node[main node] (C) at (2,-2) {};
  
  \node[main node] (D) at (4,2) {};
  \node[gray node] (E) at (4,0) {};
  \node[main node] (F) at (4,-2) {};
  
  \node[main node] (G) at (6,2) {};
  \node[gray node] (H) at (6,0) {};
  \node[main node] (I) at (6,-2) {};
  
  \node[main node] (t) at (8,0) {};

  \foreach \source/\dest in {s/A, s/C, A/D, C/F, D/G, F/I, I/t, G/t}
    \draw [->] (\source) -- (\dest);

\end{tikzpicture}
    \caption{$\mathcal{G}_d$}
    \label{fig:hexa-eight-def}
\end{subfigure}
\begin{subfigure}[t]{0.15\textwidth}
    \centering
    \begin{tikzpicture}[auto,node distance=2.5cm,
                    thick,main node/.style={circle,fill=\gcolor,draw,font=\sffamily\small\bfseries,inner sep=2}, 
                    gray node/.style={circle,fill=\ncolor,draw,font=\sffamily\small\bfseries,inner sep=2},
                    scale=0.25,minimum size=10pt]

  \node[main node] (s) at (0,0) {};
  
  \node[main node] (A) at (2,2) {};
  \node[gray node] (B) at (2,0) {};
  \node[main node] (C) at (2,-2) {};
  
  \node[gray node] (D) at (4,2) {};
  \node[main node] (E) at (4,0) {};
  \node[gray node] (F) at (4,-2) {};
  
  \node[main node] (G) at (6,2) {};
  \node[gray node] (H) at (6,0) {};
  \node[main node] (I) at (6,-2) {};
  
  \node[main node] (t) at (8,0) {};

  \foreach \source/\dest in {s/A, s/C, A/E, C/E, E/G, E/I, I/t, G/t}
    \draw [->] (\source) -- (\dest);

\end{tikzpicture}
    \caption{$\mathcal{G}_a$}
    \label{fig:hexa-eight-atk}
\end{subfigure}
\caption{Layered graphs of Example~\ref{eg:hexa-eight}. $\mathcal{V}$ has 5 layers with a single source and sink. Disconnected vertices in $\mathcal{G}_a$ and $\mathcal{G}_d$ are in white.}
\label{fig:hexa-eight}
\end{figure}
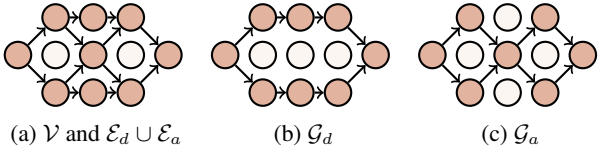

\begin{example}
    Consider the game in Figure~\ref{fig:hexa-eight}, $\targetval^\odot=1$ and $\edgeadj(e_d, e_a) = \indicator\left[ e_d = e_a \right]$, i.e., interdiction occurs when $p_a$ and $p_d$ share an edge.  
    There are 2 non-trivial ``decision points'' for $\mathcal{G}_a$, one in layer $1$ and another in layer $3$. Each has two (independent) decisions, \textbf{U}P or \textbf{D}OWN. This gives $\mathcal{P}_a = \{ UU, UD, DU, DD \}$. Conversely, $\mathcal{G}_d$ has only one nontrivial decision point at the source, and $\mathcal{P}_d = \{U, D\}$. 

    We can derive the following:
    (i) For both $u_\textsc{LIN}$ and $u_\textsc{BIN}$, the defender Nash strategy is $x^*_d(U) = x^*_d(D) = 0.5$.
    (ii) Under $u_\textsc{LIN}$, the attacker Nash strategies are exactly distributions that satisfy $x^*_a(UU) = x^*_a(DD)$. This includes the uniform strategy  $x^*_a(UU)=x^*_a(UD)=x^*_a(DU)=x^*_a(DD) = 0.25$.
    (iii) Under $u_\textsc{BIN}$ the unique attacker Nash strategy is $x^*_a(UU)=x^*_a(DD) = 0.5$.
    \label{eg:hexa-eight}
\end{example}
Example~\ref{eg:hexa-eight} illustrates the equilibrium differences between $u_\textsc{BIN}$ and $u_\textsc{LIN}$. Under $u_\textsc{LIN}$ there exists an equilibrium where $x_a$ and $x_d$ are ``Markovian'', i.e., 
the strategy can be expressed as a distribution over actions at each vertex, independently of the path. Such an equilibrium does not exist for $u_\textsc{BIN}$. We discuss Markovianity in more detail in Section~\ref{sec:equilibria}.

\subsection{Applications}
\label{sec:applications}
\begin{figure}[t]
    \centering
\begin{subfigure}[t]{0.32\linewidth}
    \centering
    \includegraphics[width=\linewidth]{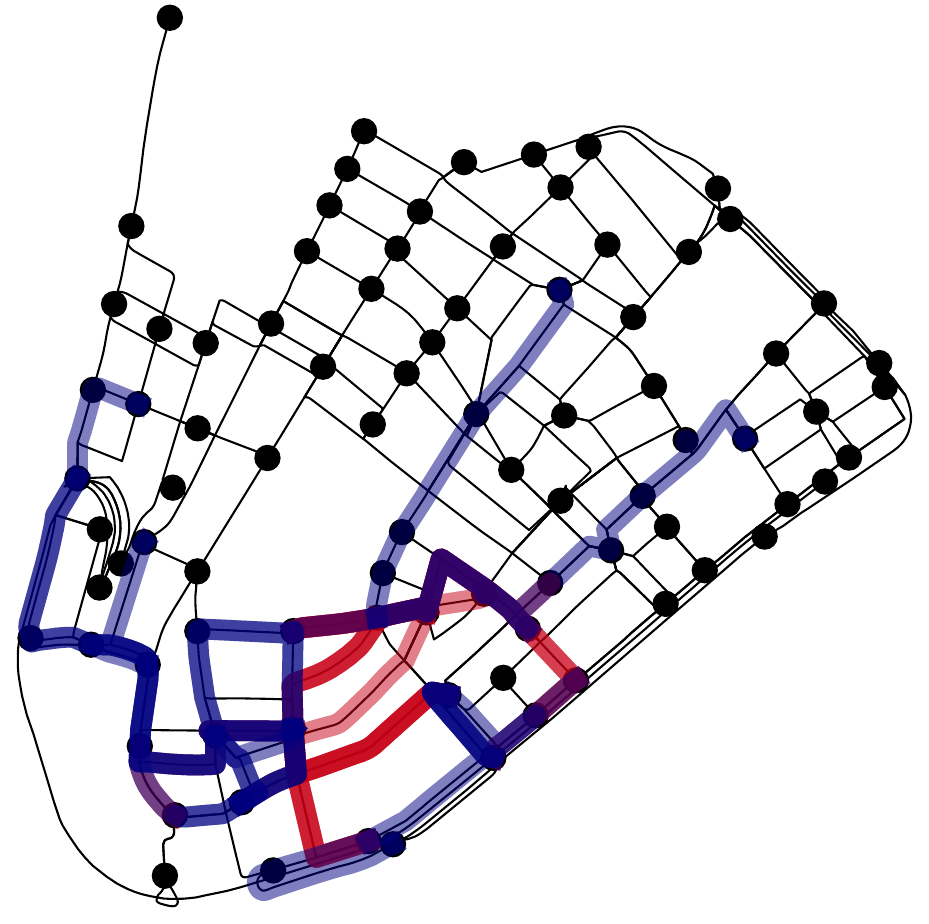}
    \caption{L. Manhattan}
\end{subfigure}
\begin{subfigure}[t]{0.32\linewidth}
    \centering
    \includegraphics[width=\linewidth]{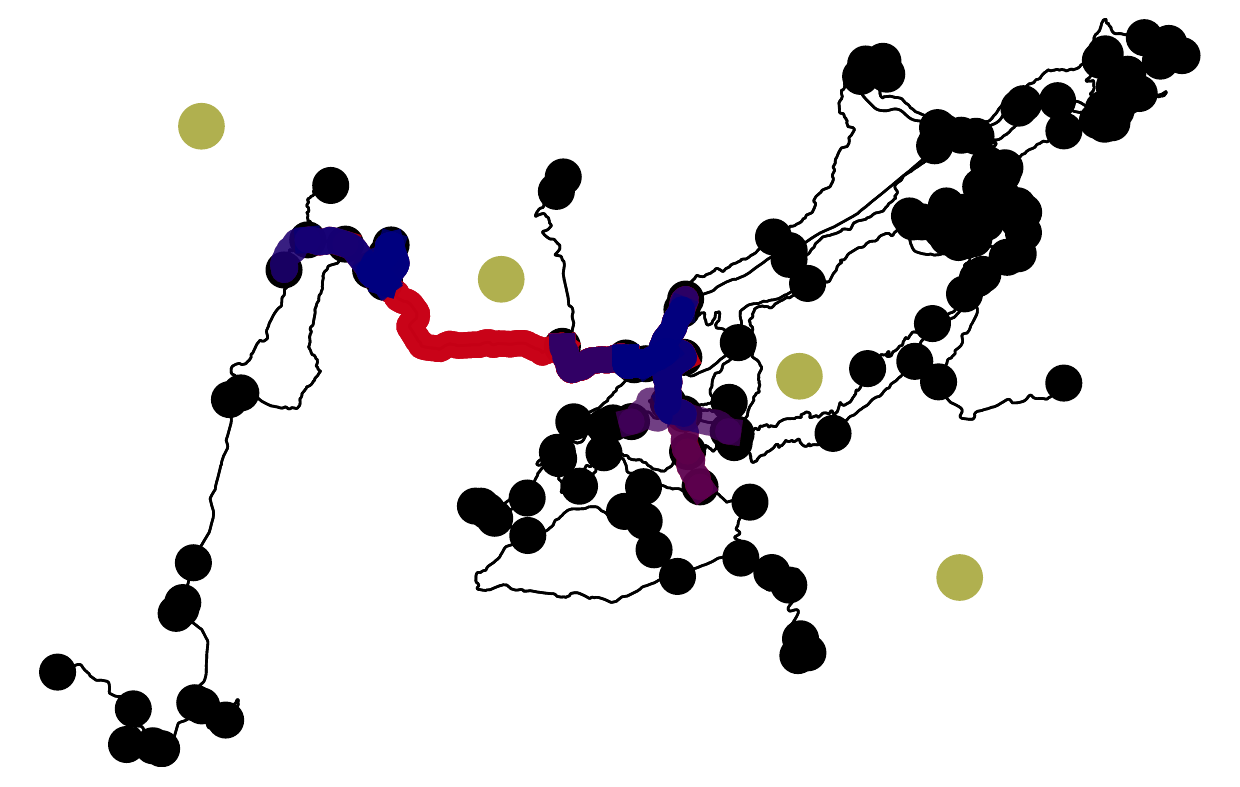}
    \caption{Minnewaska SP}
\end{subfigure}
\begin{subfigure}[t]{0.32\linewidth}
    \centering
    \includegraphics[width=\linewidth]{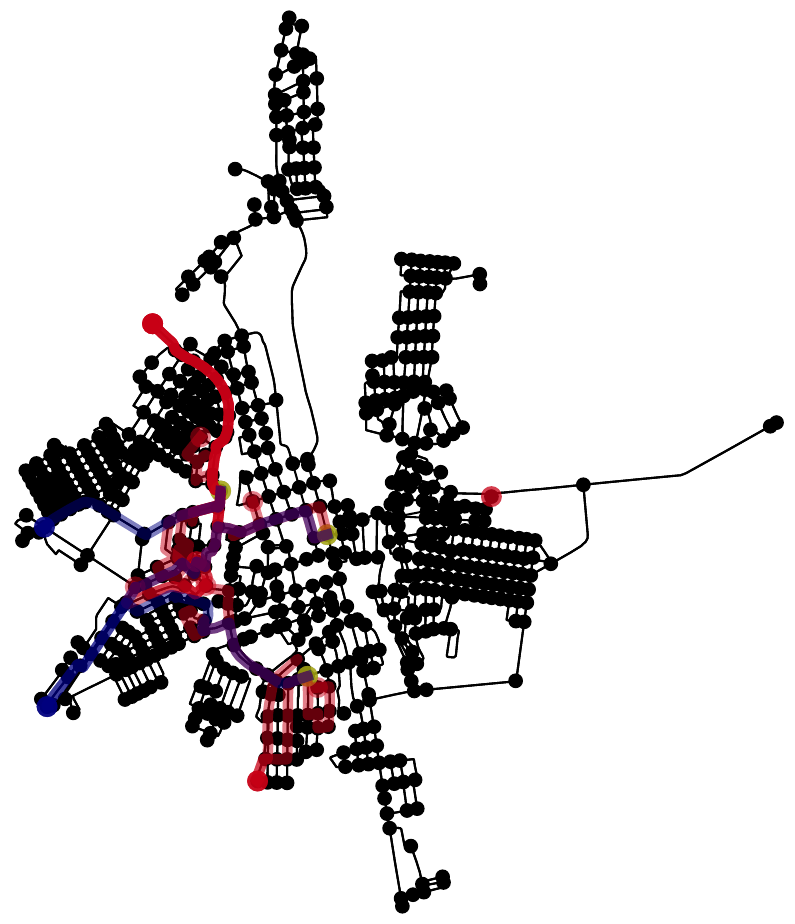}
    \caption{City of Bakhmut}
\end{subfigure}
    \caption{Real-world physical graphs used to generate \lgsgs for our application domains, together with examples of defender's (red) and attacker's (blue) equilibrial paths in (a) PE, (b) AT, and (c) LI.}
    \label{fig:physical_graphs}
\end{figure}

Now we give three examples of layered directed graphs and how the rewards $u_\textsc{LIN}$ and $u_\textsc{BIN}$ arise.
These examples involve \textit{physical graphs} for each player $\mathsf{G}_a = (\mathsf{V}, \mathsf{E}_a)$ and $\mathsf{G}_d=(\mathsf{V}, \mathsf{E}_d)$ (note that we use $\mathsf G$ for  physical graphs and $\mathcal G$ for layered graphs). They may be directed, undirected, or include loops. For simplicity, we assume that $\mathsf{G}_a$ and $\mathsf{G}_d$ share vertices, but not necessarily edges.
The vertices in $\mathsf{G}_a, \mathsf{G}_d$ represent locations in the physical world which the attacker and defender traverse over a finite set of timesteps $T \geq 1$. 
For example, Figure~\ref{fig:physical_graphs} illustrates the road networks used as $\mathsf{G}_a, \mathsf{G}_d$ in the case of Lower Manhattan, Minnewaska State Park, and part of the Ukrainian city of Bakhmut, respectively.

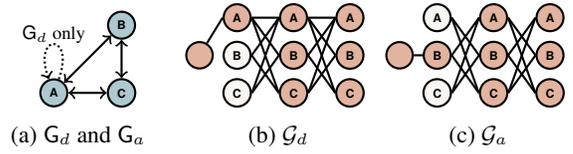
\begin{figure}[t]
    \centering
    \begin{subfigure}[b]{0.3 \linewidth}
    \centering
    \begin{tikzpicture}[auto,node distance=2.5cm,
                    thick,main node/.style={circle,fill=\pcolor,draw,font=\sffamily\small\bfseries,inner sep=2, }, 
                    scale=0.3,minimum size=10pt]

  \node[main node] (A) at (0,0) {\supertiny{A}};
  
  \node[main node] (B) at (3,3) {\supertiny{B}};
  
  \node[main node] (C) at (3,0) {\supertiny{C}};
  
  


\foreach \source/\dest in {A/B, B/C, C/A}
     \draw[<->] (\source) -- (\dest);

\node at (0,2.6) {\scriptsize{$\textsf{G}_d$ only}};
\draw[densely dotted, ->] (A) edge [loop above, min distance =22mm, out=70, in=110] (A);

\end{tikzpicture}   
    \caption{$\mathsf{G}_d$ and $\mathsf{G}_a$}
    \label{fig:physical-graph}
    \end{subfigure}
    \begin{subfigure}[b]{0.3 \linewidth}
    \centering
    \begin{tikzpicture}[auto,node distance=2.5cm,
                    thick,main node/.style={circle,fill=\gcolor,draw,font=\sffamily\small\bfseries,inner sep=2}, 
                    gray node/.style={circle,fill=\ncolor,draw,font=\sffamily\small\bfseries,inner sep=2},
                    scale=0.25,minimum size=10pt]

  \node[main node] (s) at (1,0) {\supertiny{}};
  
  \node[main node] (A1) at (3,2) {\supertiny{A}};
  \node[gray node] (B1) at (3,0) {\supertiny{B}};
  \node[gray node] (C1) at (3,-2) {\supertiny{C}};
  
  \node[main node] (A2) at (6,2) {\supertiny{A}};
  \node[main node] (B2) at (6,0) {\supertiny{B}};
  \node[main node] (C2) at (6,-2) {\supertiny{C}};
  
  \node[main node] (A3) at (9,2) {\supertiny{A}};
  \node[main node] (B3) at (9,0) {\supertiny{B}};
  \node[main node] (C3) at (9,-2) {\supertiny{C}};
  

\draw (s) -- (A1.west);

\draw  (A1.east) -- (B2.west);
\draw  (A1.east) -- (C2.west);
\draw  (B1.east) -- (C2.west);
\draw  (B1.east) -- (A2.west);
\draw  (C1.east) -- (B2.west);
\draw  (C1.east) -- (A2.west);
\draw  (A1.east) -- (A2.west);

\draw  (A2.east) -- (B3.west);
\draw  (A2.east) -- (C3.west);
\draw  (B2.east) -- (C3.west);
\draw  (B2.east) -- (A3.west);
\draw  (C2.east) -- (B3.west);
\draw  (C2.east) -- (A3.west);
\draw  (A2.east) -- (A3.west);



\end{tikzpicture}
    \caption{$\mathcal{G}_d$}
    \label{fig:unrolled-defender}
    \end{subfigure}
    \begin{subfigure}[b]{0.3 \linewidth}
    \centering
    \begin{tikzpicture}[auto,node distance=2.5cm,
                    thick,main node/.style={circle,fill=\gcolor,draw,font=\sffamily\small\bfseries,inner sep=2}, 
                    gray node/.style={circle,fill=\ncolor,draw,font=\sffamily\small\bfseries,inner sep=2},
                    scale=0.25,minimum size=10pt]

  \node[main node] (s) at (1,0) {\supertiny{}};
  
  \node[gray node] (A1) at (3,2) {\supertiny{A}};
  \node[main node] (B1) at (3,0) {\supertiny{B}};
  \node[gray node] (C1) at (3,-2) {\supertiny{C}};
  
  \node[main node] (A2) at (6,2) {\supertiny{A}};
  \node[main node] (B2) at (6,0) {\supertiny{B}};
  \node[main node] (C2) at (6,-2) {\supertiny{C}};
  
  \node[main node] (A3) at (9,2) {\supertiny{A}};
  \node[main node] (B3) at (9,0) {\supertiny{B}};
  \node[main node] (C3) at (9,-2) {\supertiny{C}};
  

\draw (s) -- (B1.west);

\draw  (A1.east) -- (B2.west);
\draw  (A1.east) -- (C2.west);
\draw  (B1.east) -- (C2.west);
\draw  (B1.east) -- (A2.west);
\draw  (C1.east) -- (B2.west);
\draw  (C1.east) -- (A2.west);

\draw  (A2.east) -- (B3.west);
\draw  (A2.east) -- (C3.west);
\draw  (B2.east) -- (C3.west);
\draw  (B2.east) -- (A3.west);
\draw  (C2.east) -- (B3.west);
\draw  (C2.east) -- (A3.west);



\end{tikzpicture}
    \caption{$\mathcal{G}_a$}
    \label{fig:unrolled-attacker}
    \end{subfigure}
    \caption{Physical graph and the layered graphs $\mathcal{G}_d, \mathcal{G}_a$ obtained by unrolling over 3 steps. 
    The defender and attacker starts at A and B.
    Note $\mathsf{G}_d$ has an extra loop at A. 
    Unreachable vertices are in white.
    }
    \label{fig:unrolled}
\end{figure}

For $i \in \{a, d \}$, the $(T+1)$-layered graph $\mathcal{G}_i$ is obtained from $\mathsf{G}_i$ by first fixing a source vertex $\mathsf{v}_{i, \text{source}} \in \mathsf{G}_i$ (or a set of possible sources the player may choose from). We then ``unroll'' $\mathsf{G}_i$ over time. 
Each layer of $\mathcal{G}_i$ contains vertices which represent where the player is at a given timestep, while edges between layers represent an action taken in the physical world. 
Figure~\ref{fig:unrolled} shows a simple example of how a graph is unrolled over 3 timesteps. 
Crucially, this unrolling process \textit{does not} result in a tree (whose size is exponential in $T$), but rather, a compact layered DAG. 
Note that in general, layers in $\mathcal{G}_i$ need not have the same edges across layers; in fact, vertices in each layer $\mathcal{V}^\ell$ need not have a 1-1 mapping with $\mathsf{V}$.
Many interesting domains may be described by slightly modifying this unrolling process and/or adding application-specific auxiliary vertices/edges to this DAG. The detailed conversion of $\mathsf{G}_i$ to $\mathcal{G}_i$ for each application is deferred to the appendix.
\footnote{Full paper and source code: \url{https://arxiv.org/abs/2405.03070}.}

\textit{Remark.} The graphs $\mathsf{G}_a, \mathsf{G}_d$ are just a convenient way to generate $\mathcal{G}_a$ and $\mathcal{G}_d$. Our framework and algorithms do not exploit properties of $\mathsf{G}_a, \mathsf{G}_d$. 
In fact, depending on the application, one may choose to sidestep $\mathsf{G}_a$ and $\mathsf{G}_d$ entirely.

\paragraph{Pursuit-Evasion (PE).} The simplest problem described by our formulation is finite-horizon pursuit-evasion games played on graphs \cite{parsons2006pursuit}. The pursuer plays the role of the defender, while the evader is the attacker.
Players start at some vertex $\mathsf{v}_{a,\text{source}}, \mathsf{v}_{d,\text{source}} \in \mathsf{V}$. At each timestep $t < T$, players select an edge to traverse (loops are allowed in $\mathsf{V}$). We define $\edgeadj$ such that the attacker is interdicted if both players share the same vertex $\mathsf{v}$ at the same time. We reiterate that our framework admits many modifications, e.g., allowing the attacker to be interdicted when traversing the same \emph{edge} as the defender, or if both players are physically ``close'' to each other.
Note that the game is one-shot: no information is revealed to either player after each step. 
Nonetheless, this captures a wide variety of pursuit-evasion games. For example,
when the goal of the attacker is simply to evade capture, we set $r^\odot(p_a) = 1$ and use binary utilities $u_\textsc{Bin}$ (we adjust $r^\odot$ appropriately if final attacker locations matter). 

\paragraph{Anti-Terrorism (AT).} An extension to PE games has the attacker playing the role of a terrorist, which seeks to plant an explosive at some target node $\mathsf{v} \in \mathsf{V}$ while evading capture. 
The explosive device requires $T_{\text{setup}} \geq 0$ time to setup, during which the attacker must remain at that same vertex. Once the setup is complete, the explosive detonates and the attacker receives utility equal to $\targetval(\mathsf{v}) \geq 0$. The attacker gets $0$ utility if the explosive was not planted by the time limit, or the attacker was interdicted while moving around or planting the explosive. 
As with PE games, the interdiction function $\edgeadj$ may be defined in various meaningful ways.
This game can be formulated using the same layered graph formulation by introducing additional ``waiting'' vertices for each target which represent how long an attacker has been setting up up the explosive at each vertex. This results in a layered graph of $L=T+1$ layers, each roughly of size $\mathcal{O}(|\mathsf{V}| \cdot T)$. 
Unlike PE games, the representation of the interdiction function in terms of $\mathcal{G}_a, \mathcal{G}_d$ is slightly more complicated. 

\paragraph{Logistical Interdiction (LI) and Persistent Threats (PT).} PE games may be modified to contain select exit vertices which end the game when the attacker reaches them.  We introduce a \textit{delay factor} $\gamma \geq 0$.  If the attacker reaches one of these exit vertices at time $t_\text{exit}$ it obtains a payoff of $\gamma^{t_\text{exit}}$. If it does not, or gets captured, it gets a payoff of $0$. 
When $\gamma=1$, this reduces to PE games with the introduction of a special `exit' vertex with a self-loop only reachable by the attacker. When $\gamma < 1$, the delay factor encourages the attacker to exit as soon as possible. We call these logistical interdiction (LI) games, which model supply lines in warfare where delays in shipments result in casualties. When $\gamma > 1$, the attacker delays exiting as long as possible. We call these persistent threat (PT) games. A raiding party seeks to cause damage for as long as possible (without being captured). Similarly, Advanced Persistent Threats (APTs) in cybersecurity procure classified information for as long as possible (without being evicted) before leaving \cite{rass2017defending}. 

\section{Computing Equilibrium in \lgsgs}\label{sec:equilibria}

This expressiveness of \lgsgs comes at a cost: finding a NE is intractable. This is unsurprising since layered graph games generalize the EIGs of \citet{zhang2017optimal}.
\begin{proposition}\label{prop:np-hard-zhang}
    It is NP-hard to find a NE for a layered graph security game with general utilities given in Equation~\ref{eq:general-min-max}.
\end{proposition}
The proof is essentially identical to the reduction from 3-SAT in \citet{zhang2017optimal}. Their reduction uses multiple defenders, but may be adapted to \lgsgs with $u=u_\textsc{bin}$, $\targetval^\odot=1$ and suitably designed $\edgeadj$ (see \href{https://arxiv.org/abs/2405.03070}{appendix} for details).
We now delve deeper and discuss subclasses of layered games and their respective computational complexities.

\subsection{\lgsgs with Linear Utility Models}
\label{sec:theory-linear-util}
In the case of linear utilities $u_\textsc{lin}$, equilibrium computation is greatly simplified.
This arises from the use of network \textit{flows} as a polynomial-size representation of strategies which is payoff equivalent to $\Delta_d, \Delta_a$. 
Given a layered graph (or any single-source DAG), the unit-flow polytopes $\Gamma_d, \Gamma_a$ are given by flow conservation constraints,
\begin{align*}
    \Gamma_i = 
    \left\{ 
    f_i  \geq 0 \middle \vert 
    \begin{array}{ccc}
     \sum\limits_{e \in \mathcal{E}_i^-(v)} f_i(e) = 1 & v \in \mathcal{V}_i^1 \\
    \sum\limits_{e \in \mathcal{E}_i^-(v)} f_i(e) = \sum\limits_{e \in \mathcal{E}_i^+(v)}  f_i(e) & v \in {\mathcal{V}_i \backslash \mathcal{V}_i^1}
    \end{array}
    \right\}
\end{align*}
Flows describe for each player $i\in\{a, d \}$ the marginal probability that a particular edge is traversed. 
Crucially, every flow is the convex combination of \textit{some} set of paths. In particular, flows embody \textit{Markovian strategies}. For internal vertices $v \in \mathcal{V}_i \backslash \mathcal{V}_i^1$, the conditional probability of taking edge $e = (v, v')$ is $f_i(e | v) = f_i(e)/\sum_{e' \in \mathcal{E}_i^+(v)} f_i(e')$, where by convention we set $0/0 = 0$, while $f_i(e | v) = f_i(e)$ for $v \in \mathcal{V}_i^1$. The probability $x_i(p)$ of choosing a path $p \in \mathcal{P}_i$ is $\prod_{e: (v, v') \in p} f_i(e, v)$. 
Conversely, any distribution over paths $x_i \in \Delta_i$ (not necessarily Markovian), maps to a flow $f_i: \mathcal{E}_i \mapsto [0, 1]$ where $f_i(e_i) = \sum_{p_i\in P_i(e_i)} x_i(p_i)$. In fact, the vertices of $\Gamma_i$ correspond precisely to paths, simplifying linear utility models.

\begin{proposition}[Kuhn's theorem for $u_{\textsc{lin}}$]
    Suppose $x_d \in \Delta_d, x_a \in \Delta_a$ in a layered graph security game. Then $u_{\textsc{Lin}}$ is bilinear in their flows $f_d(x_d)$ and $f_a(x_a)$. Specifically,
    \begin{align*}
    u_\textsc{Lin}(x_d, x_a) = \sum_{e_d\in\mathcal{E}_d}\sum_{e_a\in\mathcal{E}_a} Q(e_d,e_a)  f_d(e_d) f_a(e_a).
    \end{align*}
    \label{thm:kuhn-flows}
\end{proposition}
Proposition~\ref{thm:kuhn-flows} implies that rather than optimizing over $\Delta_i$, we can optimize over $\Gamma_i$ instead. Computing a Nash equilibrium \eqref{eq:general-min-max} may be expressed as a \textit{bilinear saddle point problem}
\begin{align}
    \min_{f_d \in \Gamma_d} \max_{f_a \in \Gamma_a} \sum_{e_d\in\mathcal{E}_d}\sum_{e_a\in\mathcal{E}_a} Q(e_d,e_a)  f_d(e_d) f_a(e_a).
\label{eq:min-max-flow}
\end{align}
We remark that, since Markovian strategies do not capture every distribution over paths (i.e., there may be multiple $x_i \in \Delta_i$ mapping onto the same flow), the solutions to the above optimization will not necessarily capture \textit{all} equilibria. 
Nonetheless, it will provide \textit{some} Markovian equilibrium. In fact, we argue that this is desirable since Markovian strategies are compactly represented.
The optimization problem in \eqref{eq:min-max-flow} lends itself well to computation. Since $\Gamma_d$ and $\Gamma_a$ are compact and convex sets (being polytopes), the minimax theorem \cite{v1928theorie} holds. 
In particular, \eqref{eq:min-max-flow} resembles the classic min-max formulation for solving zero-sum normal-form games, except that we optimize over $\Gamma_i$ instead of over the probability simplex. Taking the dual of the inner maximization problem yields the following linear program.
\begin{align*}
    \min_{f_d \in \Gamma_d,g \in \mathbb{R}^{|\mathcal{V}|}}~&g(v_\text{source}) \\
    g(e_a^+) - g(e_a^-) &\geq \sum_{e_{d} \in \mathcal{E}_{d}} f_{d}(e_{d}) \cdot Q(e_d, e_{a}) &&\forall e_a\in{\cal E}_a
\end{align*}
Here, the $g$ variables are dual variables corresponding to values of \textit{vertices}. 
We remark that other methods such as those involving regret minimization over $\Gamma_i$ \cite{takimoto2003path,farina2019efficient,farina2022kernelized} may be more efficient in practice than this LP. 
Regardless, since the number of variables and constraints is linear in the sizes of $\mathcal{G}_i$ and LPs are solvable in polynomial time, we have:

\begin{proposition}
    A NE for a \lgsg with linear utilities as in Equation~\eqref{eq:linear-utils} may be found in polynomial time.
\end{proposition}
Since \lgsgs with linear utilities are less suited for security applications and also computationally uninteresting, we focus on binary utility models for the rest of the paper.

\subsection{The Role of Flows in Binary Utility Models}
\label{sec:non-markov}
In games with linear utilities, restricting the space of strategies to flows resulted in polynomial-time algorithms. Such optimism is perhaps unwarranted in binary utilities 
given the intractability result of Proposition~\ref{prop:np-hard-zhang}. It turns out that, not only are polynomial-time algorithms unlikely, even the restriction of strategies from $\mathcal{P}_i$ to $\Gamma_i$ itself may not contain any NE. 

\begin{proposition}
    There exist \lgsg with $u_\textsc{bin}$, $\targetval^\odot=1$, $\edgeadj(e_d, e_a)=\indicator[e_d=e_a]$ where no NE is Markovian.
    \label{prop:no-markovian-NE}
\end{proposition}
Proposition~\ref{prop:no-markovian-NE} follows directly from Example~\ref{eg:hexa-eight}, whose unique attacker Nash strategy $x^*_a(UU)=x^*_a(DD)$ is clearly not Markovian since the decision at the second vertex depends on the previous action taken. This dependency is intuitive: imagine the attacker has already taken the top path and is deciding where to go in the second decision vertex. The fact that it was not interdicted implies that the defender did not play $U$, implying that it should continue to play $UU$. 

Proposition~\ref{prop:no-markovian-NE} seems trivial in hindsight, but has important ramifications. For example, one may try running deep reinforcement learning methods with self-play to arrive at an equilibrium \cite{wang2019deep}. Proposition~\ref{prop:no-markovian-NE} implies that the state features fed into such a policy or $Q$-function \textit{must} describe the player's history and not just features of that vertex. This stands in contrast to Markov games, and is analogous to the importance of perfect recall in EFGs.

There exists similar examples to Example~\ref{eg:hexa-eight} where $\mathcal{E}_d = \mathcal{E}_a$ but with $\targetval^\odot(v)$ varying over targets. These negative examples lead us to believe that searching purely in the space of $\Gamma_i$ is unlikely to yield fruitful results. Nevertheless, our experimentation seem to substantiate the following special case.

\begin{conjecture}
    \lgsgs with $u=u_\textsc{bin}$, $\targetval^{\odot}=1$, $R(e_d, e_a)=\indicator[e_d = e_a]$ and $\mathcal{E}_d=\mathcal{E}_a$ have a Markovian NE.
\end{conjecture}

\subsection{Best-Responses in Binary Utility Models}
\label{sec:best-response-binary}
Recall that defender's and attacker's (pure) best-responses to fixed strategies $x_a$ and $x_d$ are defined as $p_d^{\text{BR}} = \argmin_{p_d \in \mathcal{P}_d} u(p_d, x_a)$ and $p_a^{\text{BR}} = \argmax_{p_a \in \mathcal{P}_a} u(x_d, p_a)$, where $u(x_d, p_a)=\mathbb{E}_{p_a \sim x_a} u(p_d, p_a)$ and $u(p_d, x_a)=\mathbb{E}_{p_d \sim x_d} u(p_d, p_a)$.
One may hope that with $u_\textsc{bin}$, computing best-responses is tractable even if equilibrium solving is not. Unfortunately, it turns out that this too is intractable. 

\begin{proposition}\label{prop:sat}
    Let $\widetilde{\mathcal{P}}_i \subseteq \mathcal{P}_i$ be of size $k$ (possibly much smaller than $|\mathcal{P}_i|$) and $\widetilde{x}_i$ be a distribution with support $\widetilde{\mathcal{P}}_i$.
    Finding a best response of player $-i$ against $\widetilde{x}_i$ in a \lgsg with $u=u_\textsc{bin}$ is NP-hard in terms of $|\mathcal{G}_a|, |\mathcal{G}_d|$ and $k$.
\end{proposition}
We stress that while best-responses are closely related to NE computation, these are generally distinct problems: it is possible to devise classes of games where best-responses are difficult to compute but finding NE is easy and vice versa \cite{xu2016mysteries}.
Proposition~\ref{prop:sat} suggests that even \textit{verifying} that a given pair $(x^*_d, x^*_a)$ is a NE may be difficult. Thankfully, it turns out that in many games of interest a best-response can be formulated as a mixed-integer linear program (MILP) which can be practically tackled by commercial solvers, at least when the opponent distribution $\widetilde{x}_i$ has small support.

Let $\widetilde{\mathcal{P}}_a \subseteq \mathcal{P}_a$ be the support of an attacker strategy  $\widetilde{x}_a(p_a)$. The following Mixed-integer linear program (MILP) solves for the best defender response in the form of a path $p_d \in \mathcal{P}_d$.
\begin{align*}
    \max_{f_d \in \Gamma_d}&\smashoperator{\sum_{p_a \in \widetilde{\mathcal{P}}_a}}-\targetval^{\odot}(p_a)\cdot (1-y(p_a))\cdot \widetilde{x}_a(p_a) \\
    &y(p_a) \leq \smashoperator{\sum_{e_d\in \{e_d:\exists e_a\in p_a R(e_d,e_a)=1\}}}f_d(e_d) && \quad\forall p_a \in \widetilde{\mathcal{P}}_a \\
     &f_d(e_d) \in \{0,1\}  &&\quad\forall e_d\in \mathcal{E}_d\\
    &0 \leq y(p_a) \leq 1  &&\quad\forall p_a\in \widetilde{\mathcal{P}}_a.
\end{align*}

In the above, $y(p_a)$ is the number of times the attacker is interdicted when playing $p_a \in \widetilde{\mathcal{P}}_a$; note that this will be integeral as elements of $f_d$ are binary. Observe that $f_d$ lies in the intersection of the unit-flow polytope $\Gamma_d$ and the $\{0, 1\}^{|\mathcal{E}_d|}$ integer lattice, which is precisely the set of all paths in $\mathcal{G}_d$. 
The objective minimizes the attacker utility by trying to achieve large $y(p_a)$. The first constraint ensures that $y(p_a)$ can only be set to $1$ if at least one edge is shared between $p_a$ and the path given by $f_d$, while the last constraint limits the defender to a maximum of one interdiction. 
This MILP grows in size with $|\widetilde{\mathcal{P}}_a|$, and is likely to be hard when $|\widetilde{\mathcal{P}}_a|$ is large.

This MILP serves as the defender's \textit{best-response oracle} to a given distribution of attacker paths $\widetilde{x}_a$. We denote this by $\textsc{DefenderBR}(\widetilde{x}_a)$. 
Similarly, the attacker's best response to a distribution of defender paths $\widetilde{x}_d$ with support in $\widetilde{\mathcal{P}}_d \subseteq \mathcal{P}_d$ is given by $\textsc{AttackerBR}(\widetilde{x}_d)$. 
$\textsc{AttackerBR}$ can also be written as a MILP, the details of which are deferred to the \href{https://arxiv.org/abs/2405.03070}{appendix}. These best-responses are similar in spirit to those in \citet{zhang2017optimal}, except that we account for a wider class of interdiction functions $\edgeadj$ and target values $\targetval^\odot$.

\subsection{Approximating NE using Strategy Generation in LGSGs with Binary Utility Models}
\label{sec:do-desc}
The negative results of Proposition~\ref{prop:np-hard-zhang} compels us to work directly in the space of path distributions instead. Fortunately, despite the number of paths being exponential, we find that in practical applications such as those in Section~\ref{sec:applications}, equilibria exhibit relatively small supports (in path space). This motivates our adoption of the \textit{double oracle} framework.

The double oracle (DO) algorithm is a variant of concurrent column and row generation. It is commonly used to solve large saddle-point problems which admit efficient (in a practical sense) best-response oracles. The DO algorithm is an iterative algorithm that incrementally builds a subgame --- a subset of pure strategies for each player --- with the hope that ``weak'' strategies outside the equilibrium's support are never included in the subgame. DO guarantees that at termination, the subgame (ideally a small fraction of the entire game) has a NE which mirrors the NE of the original game. In our setting, pure strategies are paths $p_i \in \mathcal{P}_i$ and subgames are specified via subsets $\widetilde{\mathcal{P}}_i \subseteq \mathcal{P}_i$ for each $i$. An outline of our proposed DO implementation is shown in Algorithm~\ref{alg:do}.

The algorithm starts from a small subgame for each player $\widetilde{\mathcal{P}}_d, \widetilde{\mathcal{P}}_a$. At each iteration, it computes the equilibrium $(\widetilde{x}_d^*, \widetilde{x}_a^*)$ within the current subgame (i.e., player $i$ may only choose distributions of paths in $\widetilde{\mathcal{P}}_i$) as a normal-form game. 
For each player $i \in \{ a, d \}$ we compute best-responses $p_i^{\text{BR}}$ (of the full game) against their opponent's subgame equilibrium strategy $\widetilde{x}_{-i}^*$. This is done using the best-response oracles.
These best-responses give paths which are added into the subgame, and the procedure repeats.

The DO algorithm terminates when the best-response oracle for \emph{both} players returns responses that do not improve the player's return over the subgame value. 
This implies that the current subgame equilibrium constitutes an equilibrium in the full game, and further addition of strategies will not result in less exploitable strategies for either player. In practice, rather than converging to an exact equilibrium, we compute the \textit{equilibrium gap} $\nabla = u(\widetilde{x}_d^*, p_a^\text{BR}) - u(p_d^\text{BR}, \widetilde{x}_a^*)$ and terminate when $\nabla \leq \epsilon$ for some pre-specified threshold $\epsilon > 0$.

\begin{algorithm}[t]
\caption{Double Oracle for LGSGs}\label{alg:do}
\begin{algorithmic}[1]
\REQUIRE $\mathcal{G}_d, \mathcal{G}_a, \targetval^\odot, \edgeadj, u, \epsilon > 0$
\STATE $\widetilde{\mathcal{P}}_d, \widetilde{\mathcal{P}}_a \gets \textsc{InitialSubgame}(\mathcal{G}_d, \mathcal{G}_a)$
\REPEAT
\STATE $\widetilde{x}_d^*, \widetilde{x}_a^* \gets \textsc{NashEquilibrium}(\widetilde{\mathcal{P}}_d, \widetilde{\mathcal{P}}_a)$
\STATE $p^\text{BR}_d, p^\text{BR}_a \gets \textsc{DefenderBR}(\widetilde{x}_a^*), \textsc{AttackerBR}(\widetilde{x}_d^*)$ 
\STATE $\widetilde{\mathcal{P}}_d, \widetilde{\mathcal{P}}_a \gets \widetilde{\mathcal{P}}_d\cup \{p^\text{BR}_d\}, \widetilde{\mathcal{P}}_a\cup \{p^\text{BR}_a\}$
\UNTIL{$\textsc{EquilibriumGap}(\widetilde{x}_d^*, \widetilde{x}_a^*, p^\text{BR}_d, p^\text{BR}_a) < \epsilon$}
\end{algorithmic}
\end{algorithm}

\textbf{Speeding up DO.} The efficiency of DO hinges on three factors, (i) the time needed to compute a NE of the subgame, (ii) the number of outer iterations of DO, i.e., the number of strategies added to the subgame and (iii) the best response oracles for each player. The first factor is typically negligible, since computing the NE of a zero-sum matrix game may be done in polynomial time. The second factor depends on the underlying game (e.g., does it have a sparse equilibrium). The third factor depends on how hard the MILP is. 

The component most within our control is (iii), since MILP solvers can be tuned via parameters and reformulation. We consider the following speedups, (a) admitting approximate best-responses (or better responses) rather than solving MILPs to completion, (b) strategy management by periodically removing ``weak'' strategies in $\widetilde{\mathcal{P}}_i$ that absent in $\widetilde{x}_i^*$, (c) tightening of MILPs by adding cuts/implied constraints or lifting, and (d) tuning the MILP solver, using warm-starts, or heuristics. Unfortunately, we find that only (a) yielded consistently better results. We discuss implementation details and other speedups in the \href{https://arxiv.org/abs/2405.03070}{appendix}.
\section{Empirical Evaluation}\label{sec:experiments}

We seek to answer the following questions. (i) Performance: how does DO compare with the full LP solver? (ii) Sparsity: how do sizes of the (approximate) NE's support and DO subgames compare to the game size itself? (iii) Factors: how does performance scale with other game parameters?

The experiments were conducted on an Intel Xeon Gold 6226, 2.9Ghz on a Linux 64-bit platform. We used Gurobi 10.0.3 \cite{gurobi} for MILPs and LPs. Each run was restricted to 8 threads and 32GB of RAM. Our DO algorithm was implemented in Python 3.7.9 and configured with a tolerance of $\epsilon=10^{-3}$. 
The real-world physical graphs were obtained using OSMnx \cite{boeing2017osmnx}. For experiments with randomness, 20 instances were generated and solved. We report their standard errors in plots, \textit{noting that in almost all cases these are negligible}. 
We limit solvers to 6 hours for each instance .
Game sizes are defined as $|\mathcal{P}_d| + |\mathcal{P}_a|$ when reporting runtimes. 
We allow loops in $\mathsf{G}_i$ and set $\edgeadj$ such that the attacker is interdicted when players share a vertex. Application specific instantiations of $\edgeadj$ and $\targetval^\odot$, as well as additional setup and results are deferred to the \href{https://arxiv.org/abs/2405.03070}{appendix}. A link to the source code is included in the full paper.

\begin{figure}[t]
    \centering
\begin{subfigure}[t]{0.45\linewidth}
    \centering
    \begin{tikzpicture}[scale=\fscale]
\pgfplotsset{compat=1.3}
\begin{loglogaxis}[
    title={\titfont \bf Pursuit-evasion on grid world},
    xlabel={\gamesizelabel},
    ylabel={\computationlabel},
    label style={font=\labfont},
    tick label style={font=\tickfont},
    ylabel style={align=center},
    ymajorgrids=true,
    grid style=dashed,
    scaled ticks=false, 
    ytick pos=left,
    ymin = 0.7,
    ymax = 90000,
    mark size=\fmarksize,  
]

\addplot[color=black,mark=\lpmark,error bars/.cd,y dir=both, y explicit]
coordinates {
(1218.25, 8.12875) += (0, 0.8612914669677325) -= (0, 0.8612914669677325)
(4690.7, 36.559) += (0, 3.564506027799956) -= (0, 3.564506027799956)
(18341.333333333332, 563.8659999999999) += (0, 70.9325523081778) -= (0, 70.9325523081778)
(75720.26315789473, 10477.111578947368) += (0, 1214.5436230572066) -= (0, 1214.5436230572066)
};\label{bipu:grid:lp}
\addplot[color=black,mark=\domark,error bars/.cd,y dir=both, y explicit]
coordinates {
(1218.25, 2.7549999999999994) += (0, 0.32528009907945926) -= (0, 0.32528009907945926)
(4690.7, 2.978) += (0, 0.45839284462129204) -= (0, 0.45839284462129204)
(18341.333333333332, 6.319333333333333) += (0, 1.8437052410493973) -= (0, 1.8437052410493973)
(75720.26315789473, 25.18157894736842) += (0, 4.4508397435651785) -= (0, 4.4508397435651785)
(316726.05263157893, 128.48105263157896) += (0, 21.66028320127636) -= (0, 21.66028320127636)
(1332211.6315789474, 622.1368421052633) += (0, 94.46723676316675) -= (0, 94.46723676316675)
(5626509.2105263155, 5489.717894736842) += (0, 1301.3540782532261) -= (0, 1301.3540782532261)
(20802633.545454547, 14144.370909090909) += (0, 3766.157478775816) -= (0, 3766.157478775816)
}; \label{bipu:grid:do}



\addplot[color=black,mark=\dolmark,error bars/.cd,y dir=both, y explicit]
coordinates {
(1095.5, 4.815) += (0, 3.01281417323144) -= (0, 3.01281417323144)
(4496.333333333333, 2.898888888888889) += (0, 0.17899663868186047) -= (0, 0.17899663868186047)
(18579.428571428572, 5.039285714285714) += (0, 0.6831178234781553) -= (0, 0.6831178234781553)
(74154.05, 27.679000000000002) += (0, 3.252475243645417) -= (0, 3.252475243645417)
(309773.4, 114.98049999999998) += (0, 12.445302611511915) -= (0, 12.445302611511915)
(1301483.7, 353.96000000000004) += (0, 46.857540756952595) -= (0, 46.857540756952595)
(5491216.95, 1751.8444999999997) += (0, 312.01932666901723) -= (0, 312.01932666901723)
(23245349.8, 11646.032) += (0, 2641.2553624420225) -= (0, 2641.2553624420225)
}; \label{bipu:grid:dol}

\end{loglogaxis}

\begin{loglogaxis}[
  axis y line*=right,
  axis x line=none,
  ymin = 0.000007,
  ymax = 0.9,
  legend pos=north west,
  legend columns=\legcols,
  legend style={font=\legfont},
  mark size=\fmarksize,
  label style={font=\labfont},
  tick label style={font=\tickfont},
]
\addlegendimage{/pgfplots/refstyle=bipu:grid:lp}\addlegendentry{\lpname}
\addlegendimage{/pgfplots/refstyle=bipu:grid:do}\addlegendentry{\doname}
\addlegendimage{/pgfplots/refstyle=bipu:grid:dol}\addlegendentry{\dolname}

\addplot[color=\sizescolor,mark=\sgmark,error bars/.cd,y dir=both, y explicit]
coordinates {
(1218.25, 0.023136783116676617) += (0, 0.0014232123707259773) -= (0, 0.0014232123707259773)
(4690.7, 0.008360009773112333) += (0, 0.0008981649189390862) -= (0, 0.0008981649189390862)
(18341.333333333332, 0.003241546084559402) += (0, 0.00036826221921117455) -= (0, 0.00036826221921117455)
(75720.26315789473, 0.0014574368466705903) += (0, 0.0001570201588751909) -= (0, 0.0001570201588751909)
(316726.05263157893, 0.0006287133286441903) += (0, 5.257340796489231e-05) -= (0, 5.257340796489231e-05)
(1332211.6315789474, 0.00027884792106495465) += (0, 2.7726609816014805e-05) -= (0, 2.7726609816014805e-05)
(5626509.2105263155, 0.0001175251534375624) += (0, 1.1680336480270165e-05) -= (0, 1.1680336480270165e-05)
(20802633.545454547, 4.0705678912562396e-05) += (0, 5.06259993827696e-06) -= (0, 5.06259993827696e-06)
}; \addlegendentry{\subgamename}

\addplot[color=\sizescolor,mark=\spmark,error bars/.cd,y dir=both, y explicit]
coordinates {
(1218.25, 0.003903423293885401) += (0, 0.0006376284522985141) -= (0, 0.0006376284522985141)
(4690.7, 0.0020861617197999774) += (0, 0.00044439423347570397) -= (0, 0.00044439423347570397)
(18341.333333333332, 0.000781361022546474) += (0, 0.00011951219912234135) -= (0, 0.00011951219912234135)
(75720.26315789473, 0.00038921278593182575) += (0, 5.7639044056397265e-05) -= (0, 5.7639044056397265e-05)
(316726.05263157893, 0.0001960667960398447) += (0, 2.4977721702925037e-05) -= (0, 2.4977721702925037e-05)
(1332211.6315789474, 9.732805372611934e-05) += (0, 1.0644499264695235e-05) -= (0, 1.0644499264695235e-05)
(5626509.2105263155, 4.2371483871208446e-05) += (0, 4.750220898037465e-06) -= (0, 4.750220898037465e-06)
(20802633.545454547, 1.4569235083147282e-05) += (0, 1.926230683596864e-06) -= (0, 1.926230683596864e-06)
}; \addlegendentry{\supportname}

\end{loglogaxis}

\end{tikzpicture}
\end{subfigure}
\hfill
\begin{subfigure}[t]{0.475\linewidth}
    \centering
    \begin{tikzpicture}[scale=\fscale]
\pgfplotsset{compat=1.3}
\begin{loglogaxis}[
    title={\titfont \bf Anti-terrorism on grid world},
    xlabel={\gamesizelabel},
    ylabel style={align=center},
    ymajorgrids=true,
    grid style=dashed,
    ytick pos=left,
    ymin = 1.0,
    ymax = 200000,
    mark size=\fmarksize,
    label style={font=\labfont},
    tick label style={font=\tickfont},
    xtick={1000,10000,100000,1000000,10000000,100000000,1000000000,10000000000,100000000000,1000000000000},
    xticklabels={$10^3$,,$10^5$,,$10^7$,,$10^9$,,$10^{11}$}
]

\addplot[color=black,mark=\lpmark,error bars/.cd,y dir=both, y explicit]
coordinates {
(1672.8947368421052, 7.1478947368421055) += (0, 0.4632074667782923) -= (0, 0.4632074667782923)
(6811.526315789473, 93.89210526315789) += (0, 10.478180184921833) -= (0, 10.478180184921833)
(28105.473684210527, 1839.596842105263) += (0, 252.4443773186419) -= (0, 252.4443773186419)
(104134.75, 26303.006875000003) += (0, 2576.93869501047) -= (0, 2576.93869501047)
};\label{bopl:grid:lp}

\addplot[color=black,mark=\domark,error bars/.cd,y dir=both, y explicit]
coordinates {
(1694.25, 3.4314999999999998) += (0, 0.27700586636387325) -= (0, 0.27700586636387325)
(6918.1, 5.91) += (0, 0.7626491226052295) -= (0, 0.7626491226052295)
(28627.2, 15.505) += (0, 2.421542516226947) -= (0, 2.421542516226947)
(119614.85, 32.827) += (0, 7.486356224914481) -= (0, 7.486356224914481)
(503341.5, 82.69500000000001) += (0, 12.335229618326611) -= (0, 12.335229618326611)
(2129194.45, 327.2205000000001) += (0, 44.97693910553087) -= (0, 44.97693910553087)
(9042422.55, 526.63) += (0, 92.59087704577198) -= (0, 92.59087704577198)
(38519721.85, 1061.472) += (0, 166.2871137546205) -= (0, 166.2871137546205)
(164489212.7, 1838.6414999999997) += (0, 330.88263810544436) -= (0, 330.88263810544436)
(703808134.5, 3435.1205) += (0, 958.3847773891772) -= (0, 958.3847773891772)
(3016439214.4, 5738.2515) += (0, 967.3937075688619) -= (0, 967.3937075688619)
(11990482216.157894, 11406.10105263158) += (0, 2494.304856712551) -= (0, 2494.304856712551)
(52404061261.76471, 16410.94941176471) += (0, 2994.693013724871) -= (0, 2994.693013724871)
(196994871431.06668, 20629.17666666667) += (0, 3063.1466636570008) -= (0, 3063.1466636570008)
}; \label{bopl:grid:do}

\addplot[color=black,mark=\dolmark,error bars/.cd,y dir=both, y explicit]
coordinates {
(1694.25, 3.188) += (0, 0.2286442881996307) -= (0, 0.2286442881996307)
(6918.1, 5.2845) += (0, 0.5057727152325283) -= (0, 0.5057727152325283)
(28627.2, 12.371500000000001) += (0, 2.3667127185413235) -= (0, 2.3667127185413235)
(119614.85, 40.77550000000001) += (0, 6.182523734178803) -= (0, 6.182523734178803)
(503341.5, 100.58150000000002) += (0, 13.98401742356722) -= (0, 13.98401742356722)
(2129194.45, 194.87400000000005) += (0, 18.40040433480228) -= (0, 18.40040433480228)
(9042422.55, 330.92) += (0, 27.460822073407932) -= (0, 27.460822073407932)
(38519721.85, 564.6185) += (0, 50.40060344393863) -= (0, 50.40060344393863)
(164489212.7, 982.473) += (0, 98.79662581362733) -= (0, 98.79662581362733)
(703808134.5, 1696.2275000000002) += (0, 182.99837814765203) -= (0, 182.99837814765203)
(3016439214.4, 2914.6870000000004) += (0, 326.69430667159105) -= (0, 326.69430667159105)
(12946629534.6, 4920.823) += (0, 702.8178686983075) -= (0, 702.8178686983075)
(55637189639.0, 8231.3035) += (0, 1297.32758775173) -= (0, 1297.32758775173)
(239366196479.3, 13870.592999999999) += (0, 2147.2263109075707) -= (0, 2147.2263109075707)
}; \label{bopl:grid:dol}

\end{loglogaxis}

\begin{loglogaxis}[
  axis y line*=right,
  axis x line=none,
  ymajorgrids=true,
  grid style=dashed,
  scaled ticks=false,
  ymin = 0.0000000001,
  ymax = 2.0,
  ylabel = {\textcolor{\sizescolor}{\sparsitylabel}},
  legend pos=north west,
  legend columns=\legcols,
  legend style={font=\legfont},
  mark size=\fmarksize,
  label style={font=\labfont},
  tick label style={font=\tickfont},
]
\addlegendimage{/pgfplots/refstyle=bopl:grid:lp}\addlegendentry{\lpname}
\addlegendimage{/pgfplots/refstyle=bopl:grid:do}\addlegendentry{\doname}
\addlegendimage{/pgfplots/refstyle=bopl:grid:dol}\addlegendentry{\dolname}

\addplot[color=\sizescolor,mark=\sgmark,error bars/.cd,y dir=both, y explicit]
coordinates {
(1694.25, 0.023118818610948326) += (0, 0.0010386562040405145) -= (0, 0.0010386562040405145)
(6918.1, 0.007941658554121716) += (0, 0.0002712276869363806) -= (0, 0.0002712276869363806)
(28627.2, 0.002883053733911604) += (0, 0.00017688754084429494) -= (0, 0.00017688754084429494)
(119614.85, 0.0009036988536371222) += (0, 5.500663977809869e-05) -= (0, 5.500663977809869e-05)
(503341.5, 0.00035331981940501524) += (0, 3.246914017110816e-05) -= (0, 3.246914017110816e-05)
(2129194.45, 0.00013110976617837334) += (0, 1.2367080630783937e-05) -= (0, 1.2367080630783937e-05)
(9042422.55, 3.850534877357736e-05) += (0, 3.230680814496195e-06) -= (0, 3.230680814496195e-06)
(38519721.85, 1.1312707749334538e-05) += (0, 9.275813093323522e-07) -= (0, 9.275813093323522e-07)
(164489212.7, 3.417421746637419e-06) += (0, 4.273453337878623e-07) -= (0, 4.273453337878623e-07)
(703808134.5, 9.208002561534517e-07) += (0, 8.424400958319839e-08) -= (0, 8.424400958319839e-08)
(3016439214.4, 2.8413015945839076e-07) += (0, 3.315637453107939e-08) -= (0, 3.315637453107939e-08)
(12233356088.705883, 7.795989576963934e-08) += (0, 7.788393548319634e-09) -= (0, 7.788393548319634e-09)
}; \addlegendentry{\subgamename}

\addplot[color=\sizescolor,mark=\spmark,error bars/.cd,y dir=both, y explicit]
coordinates {
(1694.25, 0.005067420692117928) += (0, 0.00038708952246671537) -= (0, 0.00038708952246671537)
(6918.1, 0.0020577594489970576) += (0, 0.00010281529532346044) -= (0, 0.00010281529532346044)
(28627.2, 0.0008316113887059157) += (0, 5.132311597970281e-05) -= (0, 5.132311597970281e-05)
(119614.85, 0.00027348875479717394) += (0, 2.8644040623449414e-05) -= (0, 2.8644040623449414e-05)
(503341.5, 0.0001006818411377886) += (0, 1.4428811576220048e-05) -= (0, 1.4428811576220048e-05)
(2129194.45, 4.23013381267675e-05) += (0, 4.972061875648779e-06) -= (0, 4.972061875648779e-06)
(9042422.55, 1.1459476124939913e-05) += (0, 1.1851203296476838e-06) -= (0, 1.1851203296476838e-06)
(38519721.85, 3.295188961067028e-06) += (0, 3.7147610807971533e-07) -= (0, 3.7147610807971533e-07)
(164489212.7, 9.192185084731103e-07) += (0, 1.2748037805401095e-07) -= (0, 1.2748037805401095e-07)
(703808134.5, 2.2838122111096713e-07) += (0, 2.6992280832965928e-08) -= (0, 2.6992280832965928e-08)
(3016439214.4, 7.144058975711671e-08) += (0, 9.911439167565622e-09) -= (0, 9.911439167565622e-09)
(12233356088.705883, 1.7169437185211586e-08) += (0, 2.103865405518254e-09) -= (0, 2.103865405518254e-09)
}; \addlegendentry{\supportname}

\end{loglogaxis}

\end{tikzpicture}
\end{subfigure}
\\
\begin{subfigure}[t]{0.45\linewidth}
    \centering
    \begin{tikzpicture}[scale=\fscale]
\pgfplotsset{compat=1.3}
\begin{loglogaxis}[
    title={\titfont \bf Pursuit-evasion on Manhattan},
    xlabel={\gamesizelabel},
    ylabel={\computationlabel},
    ylabel style={align=center},
    ymajorgrids=true,
    grid style=dashed,
    scaled ticks=false, 
    ytick pos=left,
    ymin = 1.5,
    ymax = 180000,
    label style={font=\labfont},
    tick label style={font=\tickfont},
    mark size=\fmarksize,
]

\addplot[color=black,mark=\lpmark,error bars/.cd,y dir=both, y explicit]
coordinates {
(1342.0, 7.378) += (0, 0.26521351874728366) -= (0, 0.26521351874728366)
(3206.0, 23.350000000000005) += (0, 0.13371611720357424) -= (0, 0.13371611720357424)
(7824.0, 119.92050000000002) += (0, 0.5224292825475594) -= (0, 0.5224292825475594)
(19568.0, 799.2170000000001) += (0, 6.239065344137777) -= (0, 6.239065344137777)
(50088.0, 5723.7495) += (0, 36.632341358530375) -= (0, 36.632341358530375)

};\label{bipu:osm:lp}

\addplot[color=black,mark=\domark,error bars/.cd,y dir=both, y explicit]
coordinates {
(1342.0, 15.148499999999999) += (0, 5.396439239028482) -= (0, 5.396439239028482)
(3206.0, 7.487499999999999) += (0, 0.5535891859873214) -= (0, 0.5535891859873214)
(7824.0, 12.411000000000001) += (0, 1.0849717192530628) -= (0, 1.0849717192530628)
(19568.0, 31.002499999999998) += (0, 4.696211624109485) -= (0, 4.696211624109485)
(50088.0, 56.55000000000001) += (0, 7.188269318095537) -= (0, 7.188269318095537)
(130714.0, 125.86549999999997) += (0, 13.811735770236922) -= (0, 13.811735770236922)
(346464.0, 272.38200000000006) += (0, 26.90784215174375) -= (0, 26.90784215174375)
(930446.0, 536.2605000000001) += (0, 55.11270811745212) -= (0, 55.11270811745212)
(2529336.0, 1182.9540000000002) += (0, 158.97642549397727) -= (0, 158.97642549397727)
(6959182.0, 3290.0960000000005) += (0, 509.95668550759757) -= (0, 509.95668550759757)
(19380428.0, 10142.5805) += (0, 1610.8069208061816) -= (0, 1610.8069208061816)
(54615200.0, 17736.013684210528) += (0, 2070.378687199109) -= (0, 2070.378687199109)
}; \label{bipu:osm:do}

\addplot[color=black,mark=\dolmark,error bars/.cd,y dir=both, y explicit]
coordinates {
(1342.0, 8.303) += (0, 3.2152546023440722) -= (0, 3.2152546023440722)
(3206.0, 8.093500000000002) += (0, 0.6056476633367481) -= (0, 0.6056476633367481)
(7824.0, 12.372) += (0, 1.0036474533053503) -= (0, 1.0036474533053503)
(19568.0, 26.127500000000005) += (0, 3.3529635654227787) -= (0, 3.3529635654227787)
(50088.0, 47.90599999999999) += (0, 5.254391015842694) -= (0, 5.254391015842694)
(130714.0, 83.84700000000001) += (0, 9.8642044814892) -= (0, 9.8642044814892)
(346464.0, 152.91699999999997) += (0, 16.367301679359304) -= (0, 16.367301679359304)
(930446.0, 253.88100000000003) += (0, 24.411531549227863) -= (0, 24.411531549227863)
(2529336.0, 444.38149999999996) += (0, 44.82204478340751) -= (0, 44.82204478340751)
(6959182.0, 826.7955000000002) += (0, 91.31268099654241) -= (0, 91.31268099654241)
(19380428.0, 1620.4299999999998) += (0, 236.6039331256484) -= (0, 236.6039331256484)
(54615200.0, 2526.3744999999994) += (0, 238.65211163232098) -= (0, 238.65211163232098)
(155646044.0, 5764.741) += (0, 738.4340716361245) -= (0, 738.4340716361245)
(448200828.0, 8554.13625) += (0, 805.4288223807008) -= (0, 805.4288223807008)
}; \label{bipu:osm:dol}

\end{loglogaxis}

\begin{loglogaxis}[
  axis y line*=right,
  axis x line=none,
  scaled ticks=false,
  ymin = 0.0000015,
  ymax = 0.18,
  legend pos=north west,
  legend columns=\legcols,
  legend style={font=\legfont},
  mark size=\fmarksize,
  label style={font=\labfont},
  tick label style={font=\tickfont}  
]
\addlegendimage{/pgfplots/refstyle=bipu:osm:lp}\addlegendentry{\lpname}
\addlegendimage{/pgfplots/refstyle=bipu:osm:do}\addlegendentry{\doname}
\addlegendimage{/pgfplots/refstyle=bipu:osm:dol}\addlegendentry{\dolname}

\addplot[color=\sizescolor,mark=\sgmark,error bars/.cd,y dir=both, y explicit]
coordinates {
(1342.0, 0.02738450074515648) += (0, 0.0010715140763854296) -= (0, 0.0010715140763854296)
(3206.0, 0.015018714909544604) += (0, 0.0006300682149836946) -= (0, 0.0006300682149836946)
(7824.0, 0.007866820040899798) += (0, 0.00030295157461421855) -= (0, 0.00030295157461421855)
(19568.0, 0.004400040883074408) += (0, 0.00027985570929567875) -= (0, 0.00027985570929567875)
(50088.0, 0.0022310733109726882) += (0, 0.00014152075175685725) -= (0, 0.00014152075175685725)
(130714.0, 0.0011601664703092247) += (0, 5.7823117806696214e-05) -= (0, 5.7823117806696214e-05)
(346464.0, 0.0005514281426064469) += (0, 2.6647834929202172e-05) -= (0, 2.6647834929202172e-05)
(930446.0, 0.00026170245237230313) += (0, 1.3799401006276844e-05) -= (0, 1.3799401006276844e-05)
(2529336.0, 0.00012305601153820607) += (0, 6.8222632847823985e-06) -= (0, 6.8222632847823985e-06)
(6959182.0, 5.663021889641628e-05) += (0, 3.484839229517093e-06) -= (0, 3.484839229517093e-06)
(19380428.0, 2.6150093279673704e-05) += (0, 1.6097888753921267e-06) -= (0, 1.6097888753921267e-06)
(54615200.0, 1.0810562858537163e-05) += (0, 4.97475813662821e-07) -= (0, 4.97475813662821e-07)
}; \addlegendentry{\subgamename}

\addplot[color=\sizescolor,mark=\spmark,error bars/.cd,y dir=both, y explicit]
coordinates {
(1342.0, 0.008941877794336809) += (0, 0.00024773098591817017) -= (0, 0.00024773098591817017)
(3206.0, 0.0044915782907049276) += (0, 0.00020541080631207342) -= (0, 0.00020541080631207342)
(7824.0, 0.0021344580777096114) += (0, 0.0001288503179749068) -= (0, 0.0001288503179749068)
(19568.0, 0.0012699304987735078) += (0, 9.761791548716775e-05) -= (0, 9.761791548716775e-05)
(50088.0, 0.0006338843635202044) += (0, 4.5293009706250616e-05) -= (0, 4.5293009706250616e-05)
(130714.0, 0.0002930061049313768) += (0, 2.1928880283399275e-05) -= (0, 2.1928880283399275e-05)
(346464.0, 0.00015008774360395307) += (0, 1.271112034542713e-05) -= (0, 1.271112034542713e-05)
(930446.0, 6.711834969466257e-05) += (0, 5.836090125134776e-06) -= (0, 5.836090125134776e-06)
(2529336.0, 3.085790104596622e-05) += (0, 2.8216146606428676e-06) -= (0, 2.8216146606428676e-06)
(6959182.0, 1.477185105950671e-05) += (0, 1.2855265960509324e-06) -= (0, 1.2855265960509324e-06)
(19380428.0, 6.7155379643834505e-06) += (0, 4.4995370814583804e-07) -= (0, 4.4995370814583804e-07)
(54615200.0, 2.568207346942552e-06) += (0, 1.3326030083885746e-07) -= (0, 1.3326030083885746e-07)
}; \addlegendentry{\supportname}

\end{loglogaxis}

\end{tikzpicture}
\end{subfigure}
\hfill
\begin{subfigure}[t]{0.475\linewidth}
    \centering
    \begin{tikzpicture}[scale=\fscale]
\pgfplotsset{compat=1.3}
\begin{loglogaxis}[
    title={\titfont \bf Anti-terrorism on Manhattan},
    xlabel={\gamesizelabel},
    ylabel style={align=center},
    ymajorgrids=true,
    grid style=dashed,
    scaled ticks=false, 
    ytick pos=left,
    ymin = 1.0,
    ymax = 200000,
    mark size=\fmarksize,
    label style={font=\labfont},
    tick label style={font=\tickfont},
    xtick={1000,10000,100000,1000000,10000000,100000000,1000000000,10000000000,100000000000,1000000000000},
    xticklabels={$10^3$,,$10^5$,,$10^7$,,$10^9$,,$10^{11}$}
]

\addplot[color=black,mark=\lpmark,error bars/.cd,y dir=both, y explicit]
coordinates {
(1443.0, 6.2945) += (0, 0.21606952462965295) -= (0, 0.21606952462965295)
(3432.0, 18.110500000000005) += (0, 0.0962657369226619) -= (0, 0.0962657369226619)
(8276.0, 97.7225) += (0, 0.25929547766369204) -= (0, 0.25929547766369204)
(20369.0, 622.7529999999999) += (0, 4.202460062742702) -= (0, 4.202460062742702)
(51285.0, 4272.901999999999) += (0, 31.40773078315858) -= (0, 31.40773078315858)
(131902.0, 28977.0545) += (0, 103.59028190163643) -= (0, 103.59028190163643)
};\label{bopl:osm:lp}

\addplot[color=black,mark=\domark,error bars/.cd,y dir=both, y explicit]
coordinates {
(1443.0, 4.031499999999999) += (0, 0.17126814156692852) -= (0, 0.17126814156692852)
(3432.0, 4.743) += (0, 0.16984218681191263) -= (0, 0.16984218681191263)
(8276.0, 5.563000000000001) += (0, 0.23314678006061243) -= (0, 0.23314678006061243)
(20369.0, 8.218) += (0, 0.40610252788907225) -= (0, 0.40610252788907225)
(51285.0, 11.3095) += (0, 0.7806978912283682) -= (0, 0.7806978912283682)
(131902.0, 15.993500000000003) += (0, 2.5732773773004567) -= (0, 2.5732773773004567)
(345447.0, 18.759999999999998) += (0, 3.0979718322935423) -= (0, 3.0979718322935423)
(918545.0, 29.940000000000005) += (0, 4.959562692946139) -= (0, 4.959562692946139)
(2475204.0, 73.92600000000002) += (0, 16.267815261073657) -= (0, 16.267815261073657)
(6754240.0, 153.4905) += (0, 34.095792883510754) -= (0, 34.095792883510754)
(18659377.0, 301.9995) += (0, 59.05579049319648) -= (0, 59.05579049319648)
(52177600.0, 752.6775) += (0, 151.4776189632525) -= (0, 151.4776189632525)
(147618008.0, 1419.3440000000003) += (0, 197.40031195037474) -= (0, 197.40031195037474)
(422233844.0, 2508.8995000000004) += (0, 346.0467401221941) -= (0, 346.0467401221941)
(1220032270.0, 4837.1625) += (0, 640.7135292805141) -= (0, 640.7135292805141)
(3558575559.0, 9103.565999999999) += (0, 1146.5403462452184) -= (0, 1146.5403462452184)
(10472256655.0, 16005.706842105263) += (0, 2161.247041842946) -= (0, 2161.247041842946)
(31084654512.0, 25014.306000000004) += (0, 2420.862789670697) -= (0, 2420.862789670697)
}; \label{bopl:osm:do}

\addplot[color=black,mark=\dolmark,error bars/.cd,y dir=both, y explicit]
coordinates {
(1443.0, 3.941499999999999) += (0, 0.2889156451284699) -= (0, 0.2889156451284699)
(3432.0, 4.1795) += (0, 0.11076403320858073) -= (0, 0.11076403320858073)
(8276.0, 5.282000000000001) += (0, 0.2246747356560262) -= (0, 0.2246747356560262)
(20369.0, 7.717999999999999) += (0, 0.37259304736853105) -= (0, 0.37259304736853105)
(51285.0, 10.916999999999998) += (0, 0.7997881627425839) -= (0, 0.7997881627425839)
(131902.0, 15.741499999999997) += (0, 2.1555164572251404) -= (0, 2.1555164572251404)
(345447.0, 17.2885) += (0, 2.3993714390714036) -= (0, 2.3993714390714036)
(918545.0, 25.0825) += (0, 3.723367736691016) -= (0, 3.723367736691016)
(2475204.0, 49.7445) += (0, 8.097393390695812) -= (0, 8.097393390695812)
(6754240.0, 84.10650000000001) += (0, 14.943138247070213) -= (0, 14.943138247070213)
(18659377.0, 148.57649999999998) += (0, 21.629169172487998) -= (0, 21.629169172487998)
(52177600.0, 307.553) += (0, 40.455501363584) -= (0, 40.455501363584)
(147618008.0, 526.8950000000001) += (0, 65.15321240010067) -= (0, 65.15321240010067)
(422233844.0, 877.8165000000001) += (0, 116.11059281037332) -= (0, 116.11059281037332)
(1220032270.0, 1415.0840000000003) += (0, 147.55942834989355) -= (0, 147.55942834989355)
(3558575559.0, 2376.2290000000003) += (0, 287.4248035609533) -= (0, 287.4248035609533)
(10472256655.0, 3827.8229999999994) += (0, 568.5413833061157) -= (0, 568.5413833061157)
(31084654512.0, 6836.142666666667) += (0, 1038.6340700845724) -= (0, 1038.6340700845724)
(93063067607.0, 10902.193684210526) += (0, 1405.2666701710712) -= (0, 1405.2666701710712)
(281058342216.0, 17730.841176470585) += (0, 2478.1571717958977) -= (0, 2478.1571717958977)
}; \label{bopl:osm:dol}

\end{loglogaxis}

\begin{loglogaxis}[
  axis y line*=right,
  axis x line=none,
  ymajorgrids=true,
  grid style=dashed,
  scaled ticks=false,
  ymin = 0.0000000001,
  ymax = 2.0,
  ylabel = {\textcolor{\sizescolor}{\sparsitylabel}},
  legend pos=north west,
  legend columns=\legcols,
  legend style={font=\legfont},
  mark size=\fmarksize,
  label style={font=\labfont},
  tick label style={font=\tickfont},
]
\addlegendimage{/pgfplots/refstyle=bopl:osm:lp}\addlegendentry{\lpname}
\addlegendimage{/pgfplots/refstyle=bopl:osm:do}\addlegendentry{\doname}
\addlegendimage{/pgfplots/refstyle=bopl:osm:dol}\addlegendentry{\dolname}

\addplot[color=\sizescolor,mark=\sgmark,error bars/.cd,y dir=both, y explicit]
coordinates {
(1443.0, 0.023804573804573806) += (0, 0.0007612638718587774) -= (0, 0.0007612638718587774)
(3432.0, 0.011305361305361304) += (0, 0.0003167968074603098) -= (0, 0.0003167968074603098)
(8276.0, 0.005860318994683421) += (0, 0.00021265517270380036) -= (0, 0.00021265517270380036)
(20369.0, 0.003232853846531494) += (0, 0.000138276933326781) -= (0, 0.000138276933326781)
(51285.0, 0.001544311202105879) += (0, 6.856931428892277e-05) -= (0, 6.856931428892277e-05)
(131902.0, 0.0006364573698655063) += (0, 3.911458178826368e-05) -= (0, 3.911458178826368e-05)
(345447.0, 0.00027254542665010843) += (0, 1.7117206475055636e-05) -= (0, 1.7117206475055636e-05)
(918545.0, 0.00012193196849365029) += (0, 7.679863987715358e-06) -= (0, 7.679863987715358e-06)
(2475204.0, 6.0621265964340715e-05) += (0, 4.356367989612997e-06) -= (0, 4.356367989612997e-06)
(6754240.0, 2.8174894584734928e-05) += (0, 2.2965785175805696e-06) -= (0, 2.2965785175805696e-06)
(18659377.0, 1.3457040929072816e-05) += (0, 8.714486444161051e-07) -= (0, 8.714486444161051e-07)
(52177600.0, 6.533454969182177e-06) += (0, 4.7421924831925317e-07) -= (0, 4.7421924831925317e-07)
(147618008.0, 2.98540812175165e-06) += (0, 1.8455111306020542e-07) -= (0, 1.8455111306020542e-07)
(422233844.0, 1.2992800264490405e-06) += (0, 7.520757128781134e-08) -= (0, 7.520757128781134e-08)
(1220032270.0, 5.695750982062139e-07) += (0, 3.246869590146081e-08) -= (0, 3.246869590146081e-08)
(3558575559.0, 2.458286990106313e-07) += (0, 1.274616748814802e-08) -= (0, 1.274616748814802e-08)
(10472256655.0, 9.95462313953884e-08) += (0, 4.939867380059127e-09) -= (0, 4.939867380059127e-09)
(31084654512.0, 3.849058918138465e-08) += (0, 1.7975935975312533e-09) -= (0, 1.7975935975312533e-09)
}; \addlegendentry{\subgamename}

\addplot[color=\sizescolor,mark=\spmark,error bars/.cd,y dir=both, y explicit]
coordinates {
(1443.0, 0.008974358974358974) += (0, 0.00041905511511455997) -= (0, 0.00041905511511455997)
(3432.0, 0.004487179487179488) += (0, 0.00018233178686911851) -= (0, 0.00018233178686911851)
(8276.0, 0.002066215563073949) += (0, 9.63864240005251e-05) -= (0, 9.63864240005251e-05)
(20369.0, 0.0010015219205655652) += (0, 4.5944142716135445e-05) -= (0, 4.5944142716135445e-05)
(51285.0, 0.0004835721945988106) += (0, 2.309739025138221e-05) -= (0, 2.309739025138221e-05)
(131902.0, 0.00019939045655107578) += (0, 1.3212875900356606e-05) -= (0, 1.3212875900356606e-05)
(345447.0, 7.454110181880288e-05) += (0, 4.4195938110861e-06) -= (0, 4.4195938110861e-06)
(918545.0, 3.587194965951587e-05) += (0, 2.694022943130219e-06) -= (0, 2.694022943130219e-06)
(2475204.0, 1.6301686648858038e-05) += (0, 1.3219560973010934e-06) -= (0, 1.3219560973010934e-06)
(6754240.0, 7.2324939593499775e-06) += (0, 5.815338417170923e-07) -= (0, 5.815338417170923e-07)
(18659377.0, 3.341483480397014e-06) += (0, 2.4501407928658584e-07) -= (0, 2.4501407928658584e-07)
(52177600.0, 1.7152954524546936e-06) += (0, 1.419602375063506e-07) -= (0, 1.419602375063506e-07)
(147618008.0, 7.912313787624068e-07) += (0, 6.540342567487868e-08) -= (0, 6.540342567487868e-08)
(422233844.0, 3.108467070204822e-07) += (0, 2.273522620232468e-08) -= (0, 2.273522620232468e-08)
(1220032270.0, 1.3634885247748403e-07) += (0, 9.560400038942898e-09) -= (0, 9.560400038942898e-09)
(3558575559.0, 5.84925053715854e-08) += (0, 4.066177726244252e-09) -= (0, 4.066177726244252e-09)
(10472256655.0, 2.1440108200773814e-08) += (0, 1.2037862758493731e-09) -= (0, 1.2037862758493731e-09)
(31084654512.0, 8.432885533025264e-09) += (0, 5.092064164175095e-10) -= (0, 5.092064164175095e-10)
}; \addlegendentry{\supportname}

\end{loglogaxis}

\end{tikzpicture}
\end{subfigure}
    \caption{Computation times (black lines) and sparsity metrics (orange lines, for vanilla DO with exact best-responses) for PE and AT domains on grid world and Lower Manhattan.}
    \label{fig:runtimes}
\end{figure}
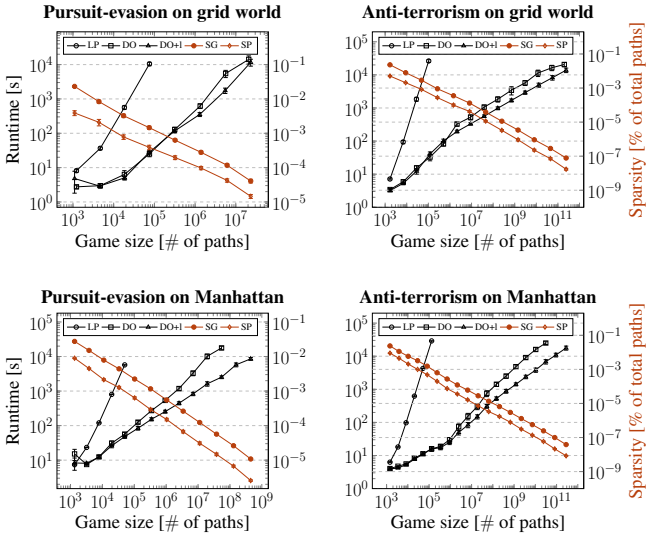

\begin{figure}[t]
    \centering
\begin{subfigure}[t]{0.45\linewidth}
    \centering
    \begin{tikzpicture}[scale=\fscale]
\pgfplotsset{compat=1.3}
\begin{loglogaxis}[
    title={\titfont \bf Anti-poaching on Minnewaska},
    ylabel style={align=center},
    xlabel={\gamesizelabel},
    ylabel={\computationlabel},
    ymajorgrids=true,
    grid style=dashed,
    scaled ticks=false, 
    ytick pos=left,
    ymin = 5.0,
    ymax = 50000,
    mark size=\fmarksize,
    label style={font=\labfont},
    tick label style={font=\tickfont},
]

\addplot[color=black,mark=\lpmark,error bars/.cd,y dir=both, y explicit]
coordinates {
(8917.0, 14.290499999999998) += (0, 0.47686336727896345) -= (0, 0.47686336727896345)
(19885.0, 13.985499999999998) += (0, 0.1586198285209009) -= (0, 0.1586198285209009)
(46100.0, 16.073) += (0, 0.33887282197556684) -= (0, 0.33887282197556684)
(111559.0, 19.865000000000002) += (0, 0.6649877323759636) -= (0, 0.6649877323759636)
(282033.0, 38.4555) += (0, 4.062518824247886) -= (0, 4.062518824247886)
(742608.0, 34.483999999999995) += (0, 4.539047453979049) -= (0, 4.539047453979049)
(2023907.0, 38.518) += (0, 7.301165754071129) -= (0, 7.301165754071129)
(5666415.0, 49.7175) += (0, 9.889790000298287) -= (0, 9.889790000298287)
(16180286.0, 58.68349999999997) += (0, 12.349232924718336) -= (0, 12.349232924718336)
(46845883.0, 62.327999999999996) += (0, 14.529318193153275) -= (0, 14.529318193153275)
(136921875.0, 72.03049999999999) += (0, 17.886670256970103) -= (0, 17.886670256970103)
(402777130.0, 120.55850000000001) += (0, 32.78718044790806) -= (0, 32.78718044790806)
(1189983853.0, 119.93550000000002) += (0, 38.21831532872412) -= (0, 38.21831532872412)
(3526023792.0, 83.4125) += (0, 36.725379736311986) -= (0, 36.725379736311986)
(10468306005.0, 121.95750000000001) += (0, 53.91631721022086) -= (0, 53.91631721022086)
(31119019396.0, 354.18750000000006) += (0, 205.2383250348046) -= (0, 205.2383250348046)
(92583944059.0, 481.12600000000003) += (0, 344.86889236071323) -= (0, 344.86889236071323)
(275594455538.0, 429.7404999999999) += (0, 281.34137302205346) -= (0, 281.34137302205346)
(820613423424.0, 734.8409999999999) += (0, 503.9780435160476) -= (0, 503.9780435160476)
(2443872086675.0, 170.19400000000002) += (0, 117.44476338710308) -= (0, 117.44476338710308)
(7278617927795.0, 61.239) += (0, 12.744199648632726) -= (0, 12.744199648632726)
};\label{bopl:park:lin}

\addplot[color=\minnewaskaexpcolor,,mark=\domark,error bars/.cd,y dir=both, y explicit]
coordinates {
(8917.0, 17.698) += (0, 4.584784021774177) -= (0, 4.584784021774177)
(19885.0, 13.497) += (0, 0.14015048303426153) -= (0, 0.14015048303426153)
(46100.0, 15.123500000000002) += (0, 0.17208974861666618) -= (0, 0.17208974861666618)
(111559.0, 20.722) += (0, 0.556026125853434) -= (0, 0.556026125853434)
(282033.0, 31.059500000000003) += (0, 1.62617883961791) -= (0, 1.62617883961791)
(742608.0, 37.393) += (0, 1.0189680280093498) -= (0, 1.0189680280093498)
(2023907.0, 42.1775) += (0, 1.7190281640937033) -= (0, 1.7190281640937033)
(5666415.0, 70.85549999999998) += (0, 3.7754661946363783) -= (0, 3.7754661946363783)
(16180286.0, 117.102) += (0, 8.73802264757157) -= (0, 8.73802264757157)
(46845883.0, 126.00499999999997) += (0, 11.811154261967793) -= (0, 11.811154261967793)
(136921875.0, 155.2765) += (0, 16.853181056515243) -= (0, 16.853181056515243)
(402777130.0, 226.353) += (0, 29.89080725697592) -= (0, 29.89080725697592)
(1189983853.0, 338.651) += (0, 28.75277093191468) -= (0, 28.75277093191468)
(3526023792.0, 495.7205000000001) += (0, 43.316425968344305) -= (0, 43.316425968344305)
(10468306005.0, 666.2090000000001) += (0, 82.51859012527514) -= (0, 82.51859012527514)
(31119019396.0, 1018.854736842105) += (0, 94.04956389702878) -= (0, 94.04956389702878)
(92583944059.0, 2178.3650000000002) += (0, 238.1927686841501) -= (0, 238.1927686841501)
(275594455538.0, 3929.9010000000003) += (0, 591.1302373526319) -= (0, 591.1302373526319)
(820613423424.0, 5862.337) += (0, 926.059796279547) -= (0, 926.059796279547)
(2443872086675.0, 9397.674499999997) += (0, 1287.2883178390873) -= (0, 1287.2883178390873)
(7278617927795.0, 11837.223999999998) += (0, 1669.0014278317951) -= (0, 1669.0014278317951)
}; \label{bopl:park:exp}

\end{loglogaxis}

\begin{loglogaxis}[
  axis y line*=right,
  axis x line=none,
  ymajorgrids=true,
  grid style=dashed,
  scaled ticks=false,
  ymin = 0.00000000001,
  ymax = 1.0,
  ylabel = {\textcolor{\sizescolor}{\sparsitylabel}},
  legend pos=north west,
  legend columns=\legcols,
  legend style={font=\legfont},
  mark size=\fmarksize,
  label style={font=\labfont},
    tick label style={font=\tickfont},
]
\addlegendimage{/pgfplots/refstyle=bopl:park:lin}\addlegendentry{\APlin}
\addlegendimage{/pgfplots/refstyle=bopl:park:exp}\addlegendentry{\APexp}

\addplot[color=\sizescolor,mark=\sgmark,error bars/.cd,y dir=both, y explicit]
coordinates {
(8917.0, 0.003953123247729057) += (0, 0.00010190167020191798) -= (0, 0.00010190167020191798)
(19885.0, 0.002069399044505909) += (0, 6.867012810160942e-05) -= (0, 6.867012810160942e-05)
(46100.0, 0.0012754880694143167) += (0, 2.891501973883179e-05) -= (0, 2.891501973883179e-05)
(111559.0, 0.0007063526922973494) += (0, 2.491279408559546e-05) -= (0, 2.491279408559546e-05)
(282033.0, 0.0003485407735974159) += (0, 1.5036685512790484e-05) -= (0, 1.5036685512790484e-05)
(742608.0, 0.0001448947493159244) += (0, 3.403028382814005e-06) -= (0, 3.403028382814005e-06)
(2023907.0, 5.570908149435719e-05) += (0, 1.4297189569039e-06) -= (0, 1.4297189569039e-06)
(5666415.0, 2.6171750568922326e-05) += (0, 9.337570604500485e-07) -= (0, 9.337570604500485e-07)
(16180286.0, 1.0929967492539994e-05) += (0, 4.346251614712028e-07) -= (0, 4.346251614712028e-07)
(46845883.0, 3.825309472766262e-06) += (0, 1.8563463034305956e-07) -= (0, 1.8563463034305956e-07)
(136921875.0, 1.4070067328540454e-06) += (0, 6.888880265755222e-08) -= (0, 6.888880265755222e-08)
(402777130.0, 5.582491736807399e-07) += (0, 2.7561828615936466e-08) -= (0, 2.7561828615936466e-08)
(1189983853.0, 2.1735589045845653e-07) += (0, 9.783212782908767e-09) -= (0, 9.783212782908767e-09)
(3526023792.0, 8.537945792737862e-08) += (0, 3.954264214009235e-09) -= (0, 3.954264214009235e-09)
(10468306005.0, 3.1647909398307655e-08) += (0, 1.7641736466218645e-09) -= (0, 1.7641736466218645e-09)
(31119019396.0, 1.2564663269892709e-08) += (0, 5.176257369156896e-10) -= (0, 5.176257369156896e-10)
(92583944059.0, 5.421026346427404e-09) += (0, 2.1577118412499646e-10) -= (0, 2.1577118412499646e-10)
(275594455538.0, 2.082589066893843e-09) += (0, 1.1436413440340121e-10) -= (0, 1.1436413440340121e-10)
(820613423424.0, 8.150000730056787e-10) += (0, 5.1297272481857376e-11) -= (0, 5.1297272481857376e-11)
(2443872086675.0, 3.1658776423632474e-10) += (0, 1.9304633838744167e-11) -= (0, 1.9304633838744167e-11)
(7278617927795.0, 1.1625833480949723e-10) += (0, 7.357324011105098e-12) -= (0, 7.357324011105098e-12)
}; \addlegendentry{\subgamename}

\addplot[color=\sizescolor,mark=\spmark,error bars/.cd,y dir=both, y explicit]
coordinates {
(8917.0, 0.0007345519793652576) += (0, 1.5166376651965305e-05) -= (0, 1.5166376651965305e-05)
(19885.0, 0.0003997988433492583) += (0, 1.4352016766175922e-05) -= (0, 1.4352016766175922e-05)
(46100.0, 0.0002754880694143168) += (0, 8.039749693553262e-06) -= (0, 8.039749693553262e-06)
(111559.0, 0.00013625077313349888) += (0, 5.267087177170961e-06) -= (0, 5.267087177170961e-06)
(282033.0, 6.701343459807895e-05) += (0, 3.149372920264265e-06) -= (0, 3.149372920264265e-06)
(742608.0, 3.743563225820352e-05) += (0, 3.1806545909206563e-07) -= (0, 3.1806545909206563e-07)
(2023907.0, 1.3093487003108345e-05) += (0, 4.308901541782212e-07) -= (0, 4.308901541782212e-07)
(5666415.0, 5.117874352655074e-06) += (0, 1.864158139798812e-07) -= (0, 1.864158139798812e-07)
(16180286.0, 2.1662163450015653e-06) += (0, 1.0268320165446053e-07) -= (0, 1.0268320165446053e-07)
(46845883.0, 7.39659448835664e-07) += (0, 4.408268003724626e-08) -= (0, 4.408268003724626e-08)
(136921875.0, 2.848339609722698e-07) += (0, 1.7323642563933554e-08) -= (0, 1.7323642563933554e-08)
(402777130.0, 1.0651051612587836e-07) += (0, 6.760806108439468e-09) -= (0, 6.760806108439468e-09)
(1189983853.0, 4.428631520263158e-08) += (0, 1.7837926889683183e-09) -= (0, 1.7837926889683183e-09)
(3526023792.0, 2.0135995724444047e-08) += (0, 1.1178824490446656e-09) -= (0, 1.1178824490446656e-09)
(10468306005.0, 6.67729811935317e-09) += (0, 4.297795868825725e-10) -= (0, 4.297795868825725e-10)
(31119019396.0, 2.5673924947768385e-09) += (0, 1.1260944136461817e-10) -= (0, 1.1260944136461817e-10)
(92583944059.0, 1.2253744550750929e-09) += (0, 5.31597880706987e-11) -= (0, 5.31597880706987e-11)
(275594455538.0, 4.3959520072164646e-10) += (0, 3.349188534954167e-11) -= (0, 3.349188534954167e-11)
(820613423424.0, 1.7858591611689925e-10) += (0, 1.568225927784844e-11) -= (0, 1.568225927784844e-11)
(2443872086675.0, 7.017552225220333e-11) += (0, 6.378555139997516e-12) -= (0, 6.378555139997516e-12)
(7278617927795.0, 2.496811003629476e-11) += (0, 2.5409916068343037e-12) -= (0, 2.5409916068343037e-12)
}; \addlegendentry{\supportname}

\end{loglogaxis}

\end{tikzpicture}
\end{subfigure}
\hfill
\begin{subfigure}[t]{0.45\linewidth}
    \centering
    \begin{tikzpicture}[scale=\fscale]
\begin{semilogyaxis}[
    title={\titfont \bf Anti-poaching on Minnewaska},
    xlabel={Game size [depth]},
    ylabel style={align=center},
    ylabel={\computationlabel},
    legend pos=north west,
    legend style={font=\legfont} ,
    ymajorgrids=true,
    grid style=dashed,
    scaled ticks=false, 
    mark size=\fmarksize,
    label style={font=\labfont},
    tick label style={font=\tickfont},  
    legend columns=2,
    cycle list name=exotic
]

\addplot[color=black,mark=*,error bars/.cd,y dir=both, y explicit]
coordinates {
( 10, 13.61 )
( 11, 14.42 )
( 12, 15.6 )
( 13, 20.47 )
( 14, 33.3 )
( 15, 76.3 )
( 16, 151.03 )
( 17, 134.8 )
( 18, 154.02 )
( 19, 265.99 )
( 20, 317.34 )
( 21, 548.28 )
( 22, 677.89 )
( 23, 49.43 )
( 24, 137.92 )
( 25, 236.13 )
( 26, 146.44 )
( 27, 169.82 )
( 28, 394.22 )
( 29, 176.43 )
( 30, 24.15 )
}; \addlegendentry{\APlin}

\addplot[forget plot,color=black,mark=*,error bars/.cd,y dir=both, y explicit]
coordinates {
( 10, 13.34 )
( 11, 13.48 )
( 12, 16.22 )
( 13, 19.49 )
( 14, 32.71 )
( 15, 55.15 )
( 16, 87.65 )
( 17, 189.72 )
( 18, 196.01 )
( 19, 175.87 )
( 20, 187.46 )
( 21, 375.71 )
( 22, 38.42 )
( 23, 37.28 )
( 24, 50.71 )
( 25, 54.92 )
( 26, 76.08 )
( 27, 75.98 )
( 28, 172.45 )
( 29, 23.59 )
( 30, 15.62 )
};

\addplot[forget plot,color=black,mark=*,error bars/.cd,y dir=both, y explicit]
coordinates {
( 10, 13.89 )
( 11, 13.84 )
( 12, 17.26 )
( 13, 23.51 )
( 14, 34.98 )
( 15, 33.31 )
( 16, 47.19 )
( 17, 59.23 )
( 18, 107.56 )
( 19, 90.65 )
( 20, 145.43 )
( 21, 172.69 )
( 22, 265.38 )
( 23, 21.0 )
( 24, 15.99 )
( 25, 21.61 )
( 26, 23.0 )
( 27, 36.3 )
( 28, 43.35 )
( 29, 59.89 )
( 30, 90.69 )
};

\addplot[forget plot,color=black,mark=*,error bars/.cd,y dir=both, y explicit]
coordinates {
( 10, 15.84 )
( 11, 14.83 )
( 12, 17.14 )
( 13, 21.22 )
( 14, 33.13 )
( 15, 32.76 )
( 16, 32.01 )
( 17, 38.85 )
( 18, 67.88 )
( 19, 73.0 )
( 20, 87.86 )
( 21, 83.15 )
( 22, 216.9 )
( 23, 166.38 )
( 24, 315.62 )
( 25, 373.18 )
( 26, 422.37 )
( 27, 532.14 )
( 28, 979.25 )
( 29, 30.28 )
( 30, 28.73 )
};

\addplot[forget plot,color=black,mark=*,error bars/.cd,y dir=both, y explicit]
coordinates {
( 10, 12.96 )
( 11, 13.26 )
( 12, 13.44 )
( 13, 15.49 )
( 14, 97.04 )
( 15, 47.33 )
( 16, 32.33 )
( 17, 63.98 )
( 18, 108.56 )
( 19, 51.84 )
( 20, 105.68 )
( 21, 208.38 )
( 22, 386.5 )
( 23, 350.15 )
( 24, 731.84 )
( 25, 2553.24 )
( 26, 6840.9 )
( 27, 5535.96 )
( 28, 10045.72 )
( 29, 58.47 )
( 30, 91.82 )
};

\addplot[forget plot,color=black,mark=*,error bars/.cd,y dir=both, y explicit]
coordinates {
( 10, 13.67 )
( 11, 13.32 )
( 12, 15.69 )
( 13, 20.16 )
( 14, 31.34 )
( 15, 30.81 )
( 16, 52.86 )
( 17, 65.22 )
( 18, 123.57 )
( 19, 76.81 )
( 20, 154.45 )
( 21, 272.45 )
( 22, 279.97 )
( 23, 697.99 )
( 24, 858.39 )
( 25, 3432.86 )
( 26, 1674.83 )
( 27, 1666.61 )
( 28, 2279.68 )
( 29, 2393.74 )
( 30, 17.29 )
};

\addplot[color=\minnewaskaexpcolor,mark=square*,error bars/.cd,y dir=both, y explicit]
coordinates {
( 10, 13.14 )
( 11, 14.92 )
( 12, 14.63 )
( 13, 21.1 )
( 14, 25.32 )
( 15, 39.35 )
( 16, 41.25 )
( 17, 72.81 )
( 18, 110.9 )
( 19, 175.52 )
( 20, 198.66 )
( 21, 281.67 )
( 22, 449.14 )
( 23, 657.8 )
( 24, 815.3 )
( 25, 947.23 )
( 26, 2746.85 )
( 27, 3969.0 )
( 28, 6155.83 )
( 29, 16725.26 )
}; \addlegendentry{\APexp}

\addplot[color=\minnewaskaexpcolor,mark=square*,error bars/.cd,y dir=both, y explicit]
coordinates {
( 10, 13.06 )
( 11, 13.0 )
( 12, 15.53 )
( 13, 21.84 )
( 14, 30.41 )
( 15, 37.14 )
( 16, 40.02 )
( 17, 67.67 )
( 18, 123.33 )
( 19, 98.9 )
( 20, 124.84 )
( 21, 157.42 )
( 22, 457.43 )
( 23, 553.85 )
( 24, 681.04 )
( 25, 795.71 )
( 26, 2771.16 )
( 27, 4068.47 )
( 28, 7154.4 )
( 29, 11262.11 )
};

\addplot[color=\minnewaskaexpcolor,mark=square*,error bars/.cd,y dir=both, y explicit]
coordinates {
( 10, 12.97 )
( 11, 13.18 )
( 12, 14.43 )
( 13, 20.6 )
( 14, 33.49 )
( 15, 32.23 )
( 16, 39.45 )
( 17, 62.63 )
( 18, 87.31 )
( 19, 113.27 )
( 20, 160.72 )
( 21, 189.74 )
( 22, 274.49 )
( 23, 267.17 )
( 24, 292.8 )
( 25, 620.31 )
( 26, 1457.17 )
( 27, 942.92 )
( 28, 1800.74 )
( 29, 2101.48 )
( 30, 4954.08 )
};

\addplot[color=\minnewaskaexpcolor,mark=square*,error bars/.cd,y dir=both, y explicit]
coordinates {
( 10, 12.87 )
( 11, 13.61 )
( 12, 15.74 )
( 13, 18.85 )
( 14, 30.22 )
( 15, 37.98 )
( 16, 48.03 )
( 17, 64.91 )
( 18, 132.98 )
( 19, 123.21 )
( 20, 205.32 )
( 21, 298.82 )
( 22, 305.42 )
( 23, 676.31 )
( 24, 667.37 )
( 26, 1574.28 )
( 27, 7427.21 )
( 28, 6146.39 )
( 29, 10053.01 )
};

\addplot[color=\minnewaskaexpcolor,mark=square*,error bars/.cd,y dir=both, y explicit]
coordinates {
( 10, 12.57 )
( 11, 12.84 )
( 12, 13.05 )
( 13, 15.54 )
( 14, 16.72 )
( 15, 44.69 )
( 16, 31.84 )
( 17, 64.04 )
( 18, 109.28 )
( 19, 88.17 )
( 20, 106.6 )
( 21, 131.3 )
( 22, 272.14 )
( 23, 496.71 )
( 24, 886.24 )
( 25, 1640.45 )
( 26, 3111.69 )
( 27, 9487.04 )
( 28, 17630.36 )
( 29, 19398.71 )
};

\addplot[color=\minnewaskaexpcolor,mark=square*,error bars/.cd,y dir=both, y explicit]
coordinates {
( 10, 12.56 )
( 11, 13.06 )
( 12, 15.21 )
( 13, 21.36 )
( 14, 36.88 )
( 15, 31.06 )
( 16, 39.65 )
( 17, 48.87 )
( 18, 58.81 )
( 19, 82.66 )
( 20, 113.17 )
( 21, 140.82 )
( 22, 252.12 )
( 23, 467.23 )
( 24, 377.87 )
( 25, 534.88 )
( 26, 975.53 )
( 27, 1462.83 )
( 28, 2025.45 )
( 29, 3333.21 )
( 30, 7767.29 )
};

\end{semilogyaxis}
\end{tikzpicture}
\end{subfigure}
\\
\begin{subfigure}[t]{0.45\linewidth}
    \centering
    \begin{tikzpicture}[scale=\fscale]
\pgfplotsset{compat=1.3}
\begin{semilogyaxis}[
    title={\titfont \bf Logistical interdiction},
    xlabel={Delay factor $\gamma$},
    ylabel style={align=center},
    ylabel={\computationlabel},
    legend pos=north west,
    ymax = 100000,
    ymin = 1.0,
    ymajorgrids=true,
    grid style=dashed,
    scaled ticks=false, 
    ytick pos=left, 
    label style={font=\labfont},
    tick label style={font=\tickfont},
]


\addplot[color=black,mark=square,error bars/.cd,y dir=both, y explicit]
coordinates {
(0.8, 5.0525) += (0, 0.4402157814192108) -= (0, 0.4402157814192108)
(0.85, 6.92) += (0, 0.5855752906780062) -= (0, 0.5855752906780062)
(0.9, 19.706) += (0, 1.5051534980298784) -= (0, 1.5051534980298784)
(0.95, 44.5975) += (0, 3.895587103580017) -= (0, 3.895587103580017)
(1.0, 110.3945) += (0, 12.372915889895197) -= (0, 12.372915889895197)
(1.05, 5841.73) += (0, 1379.6820369052236) -= (0, 1379.6820369052236)
(1.1, 11756.090263157894) += (0, 1623.4219296258482) -= (0, 1623.4219296258482)
(1.15, 13639.003333333334) += (0, 1714.367156053179) -= (0, 1714.367156053179)
(1.2, 17446.97125) += (0, 1681.8946141223837) -= (0, 1681.8946141223837)
};\label{starve:gw}

\addplot[color=black,mark=o,error bars/.cd,y dir=both, y explicit]
coordinates {
(0.8, 1655.87) 
(0.85, 2088.40) 
(0.9, 1596.44) 
(0.95, 1515.10) 
(1.0, 3239.61)
};\label{starve:bk}
\end{semilogyaxis}

\begin{loglogaxis}[
  axis y line*=right,
  axis x line=none,
  ymajorgrids=true,
  grid style=dashed,
  scaled ticks=false,
  ymin = 0.0000000001,
  ymax = 0.00001,
  ylabel = {\textcolor{\sizescolor}{\sparsitylabel}},
  legend pos=north west,
  legend columns=\legcols,
  legend style={font=\legfont},
  mark size=\fmarksize,
  label style={font=\labfont},
  tick label style={font=\tickfont},
]
\addlegendimage{/pgfplots/refstyle=starve:gw}\addlegendentry{GW}
\addlegendimage{/pgfplots/refstyle=starve:bk}\addlegendentry{BK}

\addplot[color=\sizescolor,mark=\sgmark,error bars/.cd,y dir=both, y explicit]
coordinates {
(0.8, 5.0037066708818715e-09) += (0, 2.1466354843839897e-10) -= (0, 2.1466354843839897e-10)
(0.85, 5.570026040014285e-09) += (0, 1.8327842787782626e-10) -= (0, 1.8327842787782626e-10)
(0.9, 8.068933940120207e-09) += (0, 2.484346261505853e-10) -= (0, 2.484346261505853e-10)
(0.95, 1.143764405698316e-08) += (0, 4.2548176461300734e-10) -= (0, 4.2548176461300734e-10)
(1.0, 1.875309187858269e-08) += (0, 1.1517895556170058e-09) -= (0, 1.1517895556170058e-09)
(1.05, 1.3010098086801312e-07) += (0, 1.684990023576196e-08) -= (0, 1.684990023576196e-08)
(1.1, 1.993260028747517e-07) += (0, 8.91062098342189e-09) -= (0, 8.91062098342189e-09)
(1.15, 2.226702586856307e-07) += (0, 6.987219043892584e-09) -= (0, 6.987219043892584e-09)
(1.2, 2.3135561244494645e-07) += (0, 5.80326814117092e-09) -= (0, 5.80326814117092e-09)
}; \addlegendentry{\subgamename}

\addplot[color=\sizescolor,mark=\spmark,error bars/.cd,y dir=both, y explicit]
coordinates {
(0.8, 6.881851747952157e-10) += (0, 3.4091269460834944e-11) -= (0, 3.4091269460834944e-11)
(0.85, 8.054494874767175e-10) += (0, 4.5204935018385164e-11) -= (0, 4.5204935018385164e-11)
(0.9, 1.0845164235410405e-09) += (0, 4.8242165929938626e-11) -= (0, 4.8242165929938626e-11)
(0.95, 1.446273981626872e-09) += (0, 9.595842642744273e-11) -= (0, 9.595842642744273e-11)
(1.0, 1.4526114724655908e-09) += (0, 2.2915234127070972e-10) -= (0, 2.2915234127070972e-10)
(1.05, 2.5563965267988623e-08) += (0, 4.411379119316168e-09) -= (0, 4.411379119316168e-09)
(1.1, 4.114829813725704e-08) += (0, 2.017112338110544e-09) -= (0, 2.017112338110544e-09)
(1.15, 4.3046133482677794e-08) += (0, 2.3946801709077964e-09) -= (0, 2.3946801709077964e-09)
(1.2, 5.125126788828273e-08) += (0, 1.891405408766032e-09) -= (0, 1.891405408766032e-09)
}; \addlegendentry{\supportname}

\end{loglogaxis}

\end{tikzpicture}
\end{subfigure}
\hfill
\begin{subfigure}[t]{0.45\linewidth}
    \centering
    \begin{tikzpicture}[scale=\fscale]
\begin{semilogyaxis}[
    title={\titfont \bf Logistical interdiction on grid world},
    xlabel={Delay factor $\gamma$},
    ylabel style={align=center,yshift=.2cm},
    ylabel={\computationlabel},
    legend pos=south east,
    ymajorgrids=true,
    grid style=dashed,
    scaled ticks=false, 
    mark size=\fmarksize,
    label style={font=\labfont},
    tick label style={font=\tickfont},
    cycle list name=exotic
]

\addplot+[mark=square*,error bars/.cd,y dir=both, y explicit]
coordinates {
( 0.8, 5.26 )
( 0.85, 4.15 )
( 0.9, 11.13 )
( 0.95, 64.37 )
( 1.0, 94.94 )
( 1.05, 465.02 )
( 1.1, 3914.49 )
( 1.15, 11687.71 )
( 1.2, 10883.56 )
};

\addplot+[mark=square*,error bars/.cd,y dir=both, y explicit]
coordinates {
( 0.8, 5.27 )
( 0.85, 8.92 )
( 0.9, 16.89 )
( 0.95, 37.76 )
( 1.0, 75.47 )
( 1.05, 932.24 )
( 1.1, 6757.1 )
( 1.15, 14862.19 )
( 1.2, 18068.16 )
};

\addplot+[mark=square*,error bars/.cd,y dir=both, y explicit]
coordinates {
( 0.8, 3.51 )
( 0.85, 5.27 )
( 0.9, 8.99 )
( 0.95, 11.37 )
( 1.0, 35.79 )
( 1.05, 235.26 )
( 1.1, 2531.94 )
( 1.15, 5843.39 )
( 1.2, 8993.39 )
};

\addplot+[mark=square*,error bars/.cd,y dir=both, y explicit]
coordinates {
( 0.8, 4.25 )
( 0.85, 9.46 )
( 0.9, 28.63 )
( 0.95, 51.21 )
( 1.0, 91.19 )
( 1.05, 1408.4 )
( 1.1, 10106.16 )
( 1.15, 11729.7 )
( 1.2, 14509.05 )
};

\addplot+[mark=square*,error bars/.cd,y dir=both, y explicit]
coordinates {
( 0.8, 7.7 )
( 0.85, 7.74 )
( 0.9, 15.54 )
( 0.95, 21.62 )
( 1.0, 67.47 )
( 1.05, 10455.85 )
( 1.1, 16013.36 )
( 1.15, 38500.64 )
( 1.2, 34383.31 )
};

\addplot+[mark=square*,error bars/.cd,y dir=both, y explicit]
coordinates {
( 0.8, 10.46 )
( 0.85, 3.72 )
( 0.9, 26.07 )
( 0.95, 54.97 )
( 1.0, 251.3 )
( 1.05, 8960.93 )
( 1.1, 23435.33 )
};

\addplot+[mark=square*,error bars/.cd,y dir=both, y explicit]
coordinates {
( 0.8, 2.89 )
( 0.85, 6.89 )
( 0.9, 21.99 )
( 0.95, 42.69 )
( 1.0, 130.45 )
( 1.05, 9518.3 )
( 1.1, 22096.18 )
( 1.15, 37791.56 )
( 1.2, 42167.14 )
};

\addplot+[mark=square*,error bars/.cd,y dir=both, y explicit]
coordinates {
( 0.8, 3.7 )
( 0.85, 5.26 )
( 0.9, 16.22 )
( 0.95, 76.41 )
( 1.0, 147.82 )
( 1.05, 13656.0 )
( 1.1, 17400.9 )
};

\addplot+[mark=square*,error bars/.cd,y dir=both, y explicit]
coordinates {
( 0.8, 4.65 )
( 0.85, 7.5 )
( 0.9, 11.24 )
( 0.95, 59.38 )
( 1.0, 162.02 )
( 1.05, 860.62 )
( 1.1, 4579.73 )
( 1.15, 10001.88 )
( 1.2, 19028.24 )
};

\addplot+[mark=square*,error bars/.cd,y dir=both, y explicit]
coordinates {
( 0.8, 3.58 )
( 0.85, 4.19 )
( 0.9, 17.23 )
( 0.95, 46.01 )
( 1.0, 131.42 )
( 1.05, 14731.1 )
( 1.1, 43116.13 )
};

\addplot+[mark=square*,error bars/.cd,y dir=both, y explicit]
coordinates {
( 0.8, 5.72 )
( 0.85, 8.43 )
( 0.9, 19.18 )
( 0.95, 41.02 )
( 1.0, 120.81 )
( 1.05, 1340.16 )
( 1.1, 7868.13 )
( 1.15, 17607.61 )
( 1.2, 23216.51 )
};

\addplot+[mark=square*,error bars/.cd,y dir=both, y explicit]
coordinates {
( 0.8, 3.77 )
( 0.85, 6.29 )
( 0.9, 18.69 )
( 0.95, 46.99 )
( 1.0, 18.34 )
( 1.05, 17510.72 )
( 1.1, 42589.08 )
};

\addplot+[mark=square*,error bars/.cd,y dir=both, y explicit]
coordinates {
( 0.8, 3.03 )
( 0.85, 11.54 )
( 0.9, 18.31 )
( 0.95, 84.38 )
( 1.0, 184.06 )
( 1.05, 14549.45 )
( 1.1, 20688.32 )
};

\addplot+[mark=square*,error bars/.cd,y dir=both, y explicit]
coordinates {
( 0.8, 4.93 )
( 0.85, 13.45 )
( 0.9, 34.49 )
( 0.95, 37.54 )
( 1.0, 183.81 )
( 1.05, 5284.74 )
( 1.1, 19690.33 )
};

\addplot+[mark=square*,error bars/.cd,y dir=both, y explicit]
coordinates {
( 0.8, 6.07 )
( 0.85, 6.95 )
( 0.9, 19.76 )
( 0.95, 31.99 )
( 1.0, 86.17 )
( 1.05, 922.49 )
( 1.1, 6779.51 )
( 1.15, 12769.51 )
( 1.2, 20655.68 )
};

\addplot+[mark=square*,error bars/.cd,y dir=both, y explicit]
coordinates {
( 0.8, 6.21 )
( 0.85, 3.69 )
( 0.9, 10.47 )
( 0.95, 31.74 )
( 1.0, 96.6 )
( 1.05, 12914.21 )
( 1.1, 20214.25 )
};

\addplot+[mark=square*,error bars/.cd,y dir=both, y explicit]
coordinates {
( 0.8, 7.91 )
( 0.85, 6.0 )
( 0.9, 21.23 )
( 0.95, 35.21 )
( 1.0, 69.59 )
( 1.05, 418.31 )
( 1.1, 2429.36 )
( 1.15, 7393.65 )
( 1.2, 10531.67 )
};

\addplot+[mark=square*,error bars/.cd,y dir=both, y explicit]
coordinates {
( 0.8, 2.57 )
( 0.85, 4.9 )
( 0.9, 27.07 )
( 0.95, 34.23 )
( 1.0, 102.23 )
( 1.05, 948.33 )
( 1.1, 5368.42 )
( 1.15, 12233.69 )
( 1.2, 13479.23 )
};

\addplot+[mark=square*,error bars/.cd,y dir=both, y explicit]
coordinates {
( 0.8, 5.88 )
( 0.85, 8.75 )
( 0.9, 24.72 )
( 0.95, 49.27 )
( 1.0, 91.86 )
( 1.05, 1027.36 )
( 1.1, 3649.89 )
( 1.15, 6827.78 )
( 1.2, 14996.3 )
};

\addplot+[mark=square*,error bars/.cd,y dir=both, y explicit]
coordinates {
( 0.8, 3.69 )
( 0.85, 5.3 )
( 0.9, 26.27 )
( 0.95, 33.79 )
( 1.0, 66.55 )
( 1.05, 695.11 )
( 1.1, 6017.61 )
( 1.15, 16922.76 )
( 1.2, 21530.25 )
};

\end{semilogyaxis}
\end{tikzpicture}
\end{subfigure}
    \caption{Computation times and sparsity metrics for AP and LI domains demonstrating the effects of additional factors on the performance of our DO algorithm with approximate best-responses.}
    \label{fig:weird}
\end{figure}
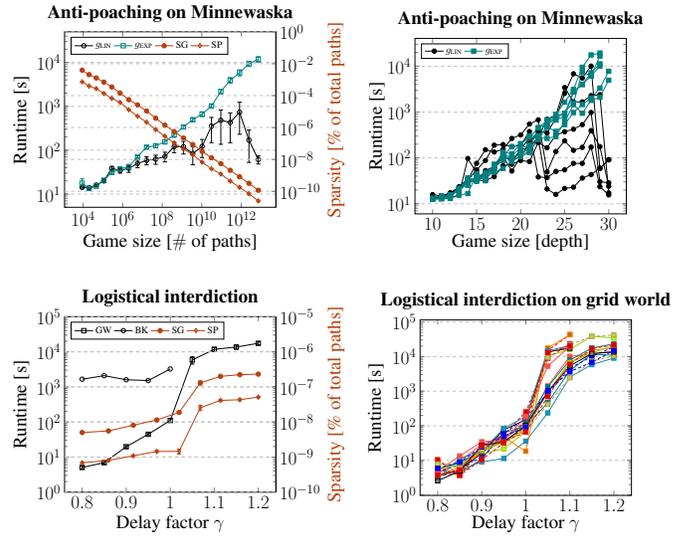

\paragraph{Performance and sparsity.} 
We compare linear programming for the full game (denoted \lpname, in the figures) against DO with (i) exact (denoted \doname) and (ii) approximate (denoted \dolname) best-responses achieved by halting best-response computations after 1s. For sparsity, we report subgame and support sizes (denoted SG and SP) summed over players and as a fraction of the total number of paths $|\mathcal{P}_d| + |\mathcal{P}_a|$.

We experiment on (a) a synthetic 4-connected $5\times 5$ \textit{grid world}, 
with a few edges randomly removed per player (making the grids unique), and (b) the map of \textit{Lower Manhattan} in New York City, USA, with 87 nodes and 191 edges (depicted in Figure~\ref{fig:physical_graphs}). In each domain, we ran pursuit-evasion (PE) and anti-terrorism (AT) scenarios for varying horizons (i.e., different game sizes). For each of the 4 settings, we report in Figure~\ref{fig:runtimes} running times (over 3 algorithms) and the final DO subgame (\subgamename) and support (\supportname) sizes. Note that for purposes of comparison, target values are identical over PE and AT.

As expected, both variants of DO outperform linear programming, which is unable to solve modestly-sized games within the time limit of 6 hours.
On Lower Manhattan, the use of approximate best-responses  yields up to half an order-of-magnitude speedup over exact best-responses, despite final subgame sizes being virtually identical. We report in the \href{https://arxiv.org/abs/2405.03070}{appendix} how equilibrium gaps evolve with running time. 
 
Even though PE and AT utilize identical physical graphs and target values, 
solving PE games appears harder than AT games. 
This stems from more involved best-response computations. 
Furthermore, the size of the NE support remains no more than 2-5 times less than the subgame sizes. This means DO avoids adding too many unnecessary strategies. We discuss  the qualitative behavior of equilibrium in the \href{https://arxiv.org/abs/2405.03070}{appendix}.

\paragraph{Effect of diameter and horizon.}
Consider \textit{Minnewaska State Park} in NY, USA (138 nodes, 201 edges, Figure~\ref{fig:physical_graphs}). Here, we apply AT for anti-poaching, i.e., planting explosives $\approx$ poaching.
To model spatial correlations in animal density, we randomly assign 4 ``animal habitats'' (yellow dots in figure), 
each associated with a positive animal score. Each node's value is the sum of animal scores, attenuated by $g_{\text{LIN}}(z) = 1/z$, where $z$ is the node's euclidean distance to the habitat.
We report results in the top row of Figure~\ref{fig:weird}.

The results on DO+l in Figure~\ref{fig:weird} (black line, top left) exhibit an anomaly where increasing horizon (i.e., larger games) leads to a \textit{decrease} in running time around the $10^{10}$ mark. Furthermore, the standard deviations become extremely high (see Figure~\ref{fig:weird}, top right for individual runs). It turns out that at lower horizons, parts of the $\mathsf{G}_i$ were not accessible and are only ``unlocked'' at a deeper horizons. Sometimes, these locations lie close to a habitat (hence $g_{\text{LIN}}$ is high), unlocking them causes the attacker's strategy to almost always move to these rich locations. This ``phase transition'' is accentuated as $g_{\text{LIN}}$ has a singularity at $z=0$. Hence, even though the game is \textit{larger}, the NE can be \textit{simpler}. 
To validate our hypothesis, we ran the same experiment (same habitats) using $g_\text{EXP}(z)=\exp(-z)$, avoiding the singularities in $g_\text{LIN}$. This anomaly and large standard deviations then vanish. 

\paragraph{Effect of delay factor in LI and PT.}
Again, we consider (i) the $5\times 5$ grid world with 4-connectivity (denoted GW), and (ii) the city of \textit{Bakhmut}, Ukraine (denoted BK, 721 nodes, 1229 edges, Figure~\ref{fig:physical_graphs}), a frontline in the Russo-Ukrainian 2022 war. The latter simulates logistical interdiction (LI) involving the delivery of Ukrainian supplies to three frontline locations (yellow dots in figure, playing the role of exits) from two entry points (west-most blue dots) while being susceptible to Russian attacks in the contested territory. 
For each domain we study DO+l runtimes on the same physical graph $\mathsf{G}_i$, varying only the delay factor $\gamma$. In GW we consider $\gamma \in [0.8, 1.2]$, i.e., both persistent threats and interdiction, while in Bakhmut we only consider $\gamma \leq 1$, i.e., interdiction.

We report results in the bottom row of Figure~\ref{fig:weird}. For GW, games with high $\gamma$ are more difficult to solve. When $\gamma$ is small, the attacker is incentivized to rush to an exit with less consideration for being caught; when $\gamma$ is closer to $1$, more unpredictable NE (hence larger support) are employed. This occurs to a lesser degree in Bakhmut. Because players start far apart, earlier actions are often inconsequential: players simply move closer to the frontline where meaningful decisions are actually made (illustrations in \href{https://arxiv.org/abs/2405.03070}{appendix}).
For $\gamma > 1$ in grid worlds (PT), running times skyrocket alongside the subgame and support sizes. This is expected as being very unpredictable is necessary to delay departure without being interdicted (see bottom right of Figure~\ref{fig:weird} for individual runs). 
\section{Conclusion}

This paper introduced Layered Graph Security Games (\lgsgs). \lgsgs offer a balance between model expressiveness and computational complexity. We study the complexity of solving \lgsgs and propose a solver for the challenging case of binary utilities. Our experiments demonstrate scalability of our method and highlights the importance of structure over game size when estimating computational costs.

\clearpage
\section*{Acknowledgments}
This research was supported by the Office of Naval Research award N00014-23-1-2374. Christian Kroer was additionally supported by the Office of Naval Research award N00014-22-1-2530, and the National Science Foundation awards IIS-2147361 and IIS-2238960. We would like to thank Daniel Bienstock for the useful discussions on mixed-integer programming, as well as Sven Bedn\'{a}\v{r} for his assistance with formalizing the Bakhmut scenario. 

\bibliographystyle{named}
\bibliography{ijcai24}

\begin{thebibliography}{}

\bibitem[\protect\citeauthoryear{An \bgroup \em et al.\egroup }{2017}]{an2017stackelberg}
Bo~An, Milind Tambe, and Arunesh Sinha.
\newblock Stackelberg security games (ssg) basics and application overview.
\newblock {\em Improving Homeland Security Decisions}, page 485, 2017.

\bibitem[\protect\citeauthoryear{Boeing}{2017}]{boeing2017osmnx}
Geoff Boeing.
\newblock Osmnx: New methods for acquiring, constructing, analyzing, and visualizing complex street networks.
\newblock {\em Computers, Environment and Urban Systems}, 65:126--139, 2017.

\bibitem[\protect\citeauthoryear{Bonato}{2011}]{bonato2011game}
Anthony Bonato.
\newblock {\em The game of cops and robbers on graphs}.
\newblock American Mathematical Soc., 2011.

\bibitem[\protect\citeauthoryear{Bopardikar \bgroup \em et al.\egroup }{2008}]{bopardikar2008discrete}
Shaunak~D Bopardikar, Francesco Bullo, and Joao~P Hespanha.
\newblock On discrete-time pursuit-evasion games with sensing limitations.
\newblock {\em IEEE Transactions on Robotics}, 24(6):1429--1439, 2008.

\bibitem[\protect\citeauthoryear{Brown and Sandholm}{2018}]{brown2018superhuman}
Noam Brown and Tuomas Sandholm.
\newblock Superhuman ai for heads-up no-limit poker: Libratus beats top professionals.
\newblock {\em Science}, 359(6374):418--424, 2018.

\bibitem[\protect\citeauthoryear{Brown and Sandholm}{2019}]{brown2019superhuman}
Noam Brown and Tuomas Sandholm.
\newblock Superhuman ai for multiplayer poker.
\newblock {\em Science}, 365(6456):885--890, 2019.

\bibitem[\protect\citeauthoryear{Conitzer and Sandholm}{2006}]{conitzer2006computing}
Vincent Conitzer and Tuomas Sandholm.
\newblock Computing the optimal strategy to commit to.
\newblock In {\em Proceedings of the 7th ACM conference on Electronic commerce}, pages 82--90, 2006.

\bibitem[\protect\citeauthoryear{Fang \bgroup \em et al.\egroup }{2015}]{fang2015security}
Fei Fang, Peter Stone, and Milind Tambe.
\newblock When security games go green: Designing defender strategies to prevent poaching and illegal fishing.
\newblock In {\em IJCAI}, pages 2589--2595, 2015.

\bibitem[\protect\citeauthoryear{Fang \bgroup \em et al.\egroup }{2017}]{fang2017paws}
Fei Fang, Thanh~H Nguyen, Rob Pickles, Wai~Y Lam, Gopalasamy~R Clements, Bo~An, Amandeep Singh, Brian~C Schwedock, Milind Tambe, and Andrew Lemieux.
\newblock Paws—a deployed game-theoretic application to combat poaching.
\newblock {\em AI Magazine}, 2017.

\bibitem[\protect\citeauthoryear{Farina \bgroup \em et al.\egroup }{2019}]{farina2019efficient}
Gabriele Farina, Chun~Kai Ling, Fei Fang, and Tuomas Sandholm.
\newblock Efficient regret minimization algorithm for extensive-form correlated equilibrium.
\newblock {\em Advances in Neural Information Processing Systems}, 32, 2019.

\bibitem[\protect\citeauthoryear{Farina \bgroup \em et al.\egroup }{2022}]{farina2022kernelized}
Gabriele Farina, Chung-Wei Lee, Haipeng Luo, and Christian Kroer.
\newblock Kernelized multiplicative weights for 0/1-polyhedral games: Bridging the gap between learning in extensive-form and normal-form games.
\newblock In {\em International Conference on Machine Learning}, pages 6337--6357. PMLR, 2022.

\bibitem[\protect\citeauthoryear{Friedman}{2013}]{friedman2013differential}
Avner Friedman.
\newblock {\em Differential games}.
\newblock Courier Corporation, 2013.

\bibitem[\protect\citeauthoryear{{Gurobi Optimization, LLC}}{2023}]{gurobi}
{Gurobi Optimization, LLC}.
\newblock {Gurobi Optimizer Reference Manual}, 2023.

\bibitem[\protect\citeauthoryear{Hagberg \bgroup \em et al.\egroup }{2008}]{hagberg2008exploring}
Aric Hagberg, Pieter Swart, and Daniel S~Chult.
\newblock Exploring network structure, dynamics, and function using networkx.
\newblock Technical report, Los Alamos National Lab.(LANL), Los Alamos, NM (United States), 2008.

\bibitem[\protect\citeauthoryear{Isaacs}{1999}]{isaacs1999differential}
Rufus Isaacs.
\newblock {\em Differential games: a mathematical theory with applications to warfare and pursuit, control and optimization}.
\newblock Courier Corporation, 1999.

\bibitem[\protect\citeauthoryear{Jain \bgroup \em et al.\egroup }{2013}]{jain2013security}
Manish Jain, Bo~An, and Milind Tambe.
\newblock Security games applied to real-world: Research contributions and challenges.
\newblock In {\em Moving Target Defense II: Application of Game Theory and Adversarial Modeling}, pages 15--39. Springer, 2013.

\bibitem[\protect\citeauthoryear{Kiekintveld \bgroup \em et al.\egroup }{2009}]{kiekintveld2009computing}
Christopher Kiekintveld, Manish Jain, Jason Tsai, James Pita, Fernando Ord{\'o}nez, and Milind Tambe.
\newblock Computing optimal randomized resource allocations for massive security games.
\newblock 2009.

\bibitem[\protect\citeauthoryear{Korzhyk \bgroup \em et al.\egroup }{2010}]{korzhyk2010complexity}
Dmytro Korzhyk, Vincent Conitzer, and Ronald Parr.
\newblock Complexity of computing optimal stackelberg strategies in security resource allocation games.
\newblock In {\em Proceedings of the AAAI Conference on Artificial Intelligence}, volume~24, pages 805--810, 2010.

\bibitem[\protect\citeauthoryear{Lanctot \bgroup \em et al.\egroup }{2017}]{lanctot2017unified}
Marc Lanctot, Vinicius Zambaldi, Audrunas Gruslys, Angeliki Lazaridou, Karl Tuyls, Julien P{\'e}rolat, David Silver, and Thore Graepel.
\newblock A unified game-theoretic approach to multiagent reinforcement learning.
\newblock {\em Advances in neural information processing systems}, 30, 2017.

\bibitem[\protect\citeauthoryear{Morav{\v{c}}{\'\i}k \bgroup \em et al.\egroup }{2017}]{moravvcik2017deepstack}
Matej Morav{\v{c}}{\'\i}k, Martin Schmid, Neil Burch, Viliam Lis{\`y}, Dustin Morrill, Nolan Bard, Trevor Davis, Kevin Waugh, Michael Johanson, and Michael Bowling.
\newblock Deepstack: Expert-level artificial intelligence in heads-up no-limit poker.
\newblock {\em Science}, 356(6337):508--513, 2017.

\bibitem[\protect\citeauthoryear{Parsons}{2006}]{parsons2006pursuit}
Torrence~D Parsons.
\newblock Pursuit-evasion in a graph.
\newblock In {\em Theory and Applications of Graphs: Proceedings, Michigan May 11--15, 1976}, pages 426--441. Springer, 2006.

\bibitem[\protect\citeauthoryear{Perolat \bgroup \em et al.\egroup }{2022}]{perolat2022mastering}
Julien Perolat, Bart De~Vylder, Daniel Hennes, Eugene Tarassov, Florian Strub, Vincent de~Boer, Paul Muller, Jerome~T Connor, Neil Burch, Thomas Anthony, et~al.
\newblock Mastering the game of stratego with model-free multiagent reinforcement learning.
\newblock {\em Science}, 378(6623):990--996, 2022.

\bibitem[\protect\citeauthoryear{Pita \bgroup \em et al.\egroup }{2008}]{pita2008armor}
James Pita, Manish Jain, Fernando Ord{\'o}nez, Christopher Portway, Milind Tambe, Craig Western, Praveen Paruchuri, and Sarit Kraus.
\newblock Armor security for los angeles international airport.
\newblock In {\em AAAI}, pages 1884--1885, 2008.

\bibitem[\protect\citeauthoryear{Rass \bgroup \em et al.\egroup }{2017}]{rass2017defending}
Stefan Rass, Sandra K{\"o}nig, and Stefan Schauer.
\newblock Defending against advanced persistent threats using game-theory.
\newblock {\em PloS one}, 12(1):e0168675, 2017.

\bibitem[\protect\citeauthoryear{Shieh \bgroup \em et al.\egroup }{2012}]{shieh2012protect}
Eric Shieh, Bo~An, Rong Yang, Milind Tambe, Craig Baldwin, Joseph DiRenzo, Ben Maule, and Garrett Meyer.
\newblock Protect: A deployed game theoretic system to protect the ports of the united states.
\newblock In {\em Proceedings of the 11th international conference on autonomous agents and multiagent systems-volume 1}, pages 13--20, 2012.

\bibitem[\protect\citeauthoryear{Shoham and Leyton-Brown}{2008}]{shoham2008multiagent}
Yoav Shoham and Kevin Leyton-Brown.
\newblock {\em Multiagent systems: Algorithmic, game-theoretic, and logical foundations}.
\newblock Cambridge University Press, 2008.

\bibitem[\protect\citeauthoryear{Smith and Lim}{2008}]{smith2008algorithms}
J~Cole Smith and Churlzu Lim.
\newblock Algorithms for network interdiction and fortification games.
\newblock {\em Pareto optimality, game theory and equilibria}, pages 609--644, 2008.

\bibitem[\protect\citeauthoryear{Smith and Song}{2020}]{smith2020survey}
J~Cole Smith and Yongjia Song.
\newblock A survey of network interdiction models and algorithms.
\newblock {\em European Journal of Operational Research}, 283(3):797--811, 2020.

\bibitem[\protect\citeauthoryear{Takimoto and Warmuth}{2003}]{takimoto2003path}
Eiji Takimoto and Manfred~K Warmuth.
\newblock Path kernels and multiplicative updates.
\newblock {\em The Journal of Machine Learning Research}, 4:773--818, 2003.

\bibitem[\protect\citeauthoryear{Tsai \bgroup \em et al.\egroup }{2009}]{Tsai2009}
Jason Tsai, Christopher Kiekintveld, Fernando Ordóñez, Milind Tamble, and Shyamsunder Rathi.
\newblock Iris - a tool for strategic security allocation in transportation networks categories and subject descriptors.
\newblock In {\em Proceedings of the 8th International Conference on Autonomous Agents and Multiagent Systems}, 2009.

\bibitem[\protect\citeauthoryear{v. Neumann}{1928}]{v1928theorie}
J~v.~Neumann.
\newblock Zur theorie der gesellschaftsspiele.
\newblock {\em Mathematische annalen}, 100(1):295--320, 1928.

\bibitem[\protect\citeauthoryear{Wang \bgroup \em et al.\egroup }{2019}]{wang2019deep}
Yufei Wang, Zheyuan~Ryan Shi, Lantao Yu, Yi~Wu, Rohit Singh, Lucas Joppa, and Fei Fang.
\newblock Deep reinforcement learning for green security games with real-time information.
\newblock In {\em Proceedings of the AAAI Conference on Artificial Intelligence}, volume~33, pages 1401--1408, 2019.

\bibitem[\protect\citeauthoryear{Washburn and Wood}{1995}]{washburn1995two}
Alan Washburn and Kevin Wood.
\newblock Two-person zero-sum games for network interdiction.
\newblock {\em Operations research}, 43(2):243--251, 1995.

\bibitem[\protect\citeauthoryear{Weintraub \bgroup \em et al.\egroup }{2020}]{weintraub2020introduction}
Isaac~E Weintraub, Meir Pachter, and Eloy Garcia.
\newblock An introduction to pursuit-evasion differential games.
\newblock In {\em 2020 American Control Conference (ACC)}, pages 1049--1066. IEEE, 2020.

\bibitem[\protect\citeauthoryear{Wollmer}{1964}]{wollmer1964removing}
Richard Wollmer.
\newblock Removing arcs from a network.
\newblock {\em Operations Research}, 12(6):934--940, 1964.

\bibitem[\protect\citeauthoryear{Xu}{2016}]{xu2016mysteries}
Haifeng Xu.
\newblock The mysteries of security games: Equilibrium computation becomes combinatorial algorithm design.
\newblock In {\em Proceedings of the 2016 ACM Conference on Economics and Computation}, pages 497--514, 2016.

\bibitem[\protect\citeauthoryear{Xue \bgroup \em et al.\egroup }{2021}]{xue2021solving}
Wanqi Xue, Youzhi Zhang, Shuxin Li, Xinrun Wang, Bo~An, and Chai~Kiat Yeo.
\newblock Solving large-scale extensive-form network security games via neural fictitious self-play.
\newblock {\em arXiv preprint arXiv:2106.00897}, 2021.

\bibitem[\protect\citeauthoryear{Zhang \bgroup \em et al.\egroup }{2017}]{zhang2017optimal}
Youzhi Zhang, Bo~An, Long Tran-Thanh, Zhen Wang, Jiarui Gan, and Nicholas~R Jennings.
\newblock Optimal escape interdiction on transportation networks.
\newblock 2017.

\end{thebibliography}

\clearpage
\appendix
\onecolumn
\def\fscale{0.7}
\section{Proofs}

\begin{cmr}{Proposition~\ref{prop:np-hard-zhang}}
    It is NP-hard to find a NE for a layered graph security game with general utilities given in Equation~\ref{eq:general-min-max}.
\end{cmr}
\begin{proof}
    The proof is very similar the reduction of 3-SAT to the Escape Interdiction Game (EIG) of \citet{zhang2017optimal}, but much simpler because of the generality of $\edgeadj$ admitted.
    Let there be $n$ variables and $m$ clauses. They are denoted by $Y_1,\dots,Y_n$ and $C_1,\dots, C_m$ respectively.
    
    The layered graphs $\mathcal{G}_d$ and $\mathcal{G}_a$ can be broken into two parts, the vertices and edges of which are shown in Figure~\ref{fig:nph-fullprob-sat}. 
    Lines drawn in solid edges belong to the defender, who selects an variable assignment in a ``chain graph'' with vertices given in orange. The chain graph has $n+1$ vertices (including a dummy terminal vertex), the first $n$ of which contains two outgoing edges represents $y_i$ or $\neg y_i$ respectively. Clearly, every path formed this way corresponds to an assignment to $y_1,\dots y_n$.
    The dotted lines are for the attacker, and the reachable vertices given in blue. The attacker (tries to) selects a clause which the variable assignment violates.
    
    We now define $\edgeadj(y_i, c_{j, i})=1$ if and only if the variable $Y_i$ belongs to clause $C_j$, and correspondingly $\edgeadj(\neg y_i, c_{j, i})$ if and only if $\neg Y_i$ belongs to clause $C_j$. All other edges across different layers, or belonging to the same player (e.g., both dotted or solid) have $\edgeadj=0$. All terminal vertices have a value of $1$, i.e., $\targetval(v)=1$ for all $v \in \mathcal{V}^\odot$. We claim that the value of the game is $0$ if and only if the formula is satisfiable.
    
    $(\impliedby)$. If the formula is satisfiable, then the defender can choose that path in solid edges, and whatever the attacker chooses will at least be interdicted once (by the definition of satisfiability and $\edgeadj$). This gives $0$ payoff to the attacker.
    
    $(\implies)$. Now, suppose the value of the game is $0$. We look at the defender's strategy. If it was pure, then it is a single path and therefore corresponds to a variable assignment. Since that strategy constitutes a (defender) NE in a game with value $0$, any best-response from the attacker cannot give any attacker payoff greater than $0$. This also means there is no clause-path the attacker can choose that can ever avoid being interdicted (i.e., all clauses are satisfied). Now, if the defender's strategy is \textit{not} pure, then it must be some mixture of paths $x_d^*$. Take any path $p_d$ that is played with strictly positive probability, i.e., $x_d^*(p_d) > 0$; we claim playing $p_d$ deterministically must also constitute a pure Nash. This is because $0$ is already the best possible outcome for the defender (who is the min-player). That is, if it was not Nash, then there must be a best-response $p_a \in \mathcal{P}_a$ that interdicts $p_d$. This implies that the attacker could have played $p_a$ against $x_d^*$ and obtained some strictly positive payoff, contradicting the claim that $x_d^*$ is a NE (and that $0$ is the value of the game). Either way, there exists some pure strategy NE for the defender, implying the formula is satisfiable.

    \begin{figure}[t]
    \centering
    \begin{tikzpicture}[->,>=,auto,node distance=2.5cm,
                    thick,main node/.style={circle,fill=\gcolor,draw,font=\sffamily\tiny\bfseries,inner sep=1}, 
                    thick,atk node/.style={circle,fill=\pcolor,draw,font=\sffamily\tiny\bfseries,inner sep=1},
                    thick,root node/.style={circle,draw,font=\sffamily\tiny\bfseries,inner sep=1},
                    scale=0.3]

  
  \node[main node] (1) at (0,0) {$\quad$};
  
  \node[main node] (2) at (8,0) {$\quad$};

   \node[main node, draw=none] ({n-1}) at (16,0) {$...$};
  
  \node[main node] (n) at (24,0) {$\quad$};
  
  \node[main node] (t) at (32,0) {$\quad$};

  \foreach \source/\dest in {1/2, 2/{n-1}, {n-1}/n, n/t}
    \path (\source) edge[bend right=60] node[below,font=\sffamily\tiny]{$\neg y_{\source}$} (\dest);
  \foreach \source/\dest in {1/2, 2/{n-1}, {n-1}/n, n/t}
    \path (\source) edge[bend left=60] node[above,font=\sffamily\tiny]{$y_{\source}$} (\dest);


  \node[atk node] (A0) at (0, -16) {$\quad$};
  
  \node[atk node] (A11) at (8, -8) {$\quad$};
  \node[atk node] (A12) at (8, -12) {$\quad$};
  \node[atk node] (A13) at (8, -16) {$\quad$};
  \node[atk node] (A14) at (8, -20) {$\quad$};
  \node[atk node] (A15) at (8, -24) {$\quad$};

  \node[atk node, draw=none] (A21) at (16, -8) {$...$};
  \node[atk node, draw=none] (A22) at (16, -12) {$...$};
  \node[atk node, draw=none] (A23) at (16, -16) {$...$};
  \node[atk node, draw=none] (A24) at (16, -20) {$...$};
  \node[atk node, draw=none] (A25) at (16, -24) {$...$};

  \node[atk node] (A31) at (24, -8) {$\quad$};
  \node[atk node] (A32) at (24, -12) {$\quad$};
  \node[atk node] (A33) at (24, -16) {$\quad$};
  \node[atk node] (A34) at (24, -20) {$\quad$};
  \node[atk node] (A35) at (24, -24) {$\quad$};

  \node[atk node] (A41) at (32, -8) {$\quad$};
  \node[atk node] (A42) at (32, -12) {$\quad$};
  \node[atk node] (A43) at (32, -16) {$\quad$};
  \node[atk node] (A44) at (32, -20) {$\quad$};
  \node[atk node] (A45) at (32, -24) {$\quad$};


  \path (A0) edge[dashed, bend left=50] node[above=-0.0cm,font=\sffamily\tiny]{$c_{1,1}$} (A11);
  \path (A0) edge[dashed, bend left=30] node[above=-0.05cm,font=\sffamily\tiny]{$c_{2,1}$} (A12);
  \path (A0) edge[dashed] node[above=-0.05cm,font=\sffamily\tiny]{$c_{3,1}$} (A13);
  \path (A0) edge[dashed, bend right=30] node[above=-0.05cm,font=\sffamily\tiny]{$c_{4,1}$} (A14);
  \path (A0) edge[dashed, bend right=50] node[above=0.05cm,font=\sffamily\tiny]{$c_{5,1}$} (A15);
  
  \path (A11) edge[dashed] node[above=-0.05cm,font=\sffamily\tiny]{$c_{1,2}$} (A21);
  \path (A21) edge[dashed] node[above=-0.05cm,font=\sffamily\tiny]{$c_{1,n-1}$} (A31);
  \path (A31) edge[dashed] node[above=-0.05cm,font=\sffamily\tiny]{$c_{1,n}$} (A41);
  
  \path (A12) edge[dashed] node[above=-0.05cm,font=\sffamily\tiny]{$c_{2,2}$} (A22);
  \path (A22) edge[dashed] node[above=-0.05cm,font=\sffamily\tiny]{$c_{2,n-1}$} (A32);
  \path (A32) edge[dashed] node[above=-0.05cm,font=\sffamily\tiny]{$c_{2,n}$} (A42);

  \path (A13) edge[dashed] node[above=-0.05cm,font=\sffamily\tiny]{$c_{3,2}$} (A23);
  \path (A23) edge[dashed] node[above=-0.05cm,font=\sffamily\tiny]{$c_{3,n-1}$} (A33);
  \path (A33) edge[dashed] node[above=-0.05cm,font=\sffamily\tiny]{$c_{3,n}$} (A43);
  
  \path (A14) edge[dashed] node[above=-0.05cm,font=\sffamily\tiny]{$c_{4,2}$} (A24);
  \path (A24) edge[dashed] node[above=-0.05cm,font=\sffamily\tiny]{$c_{4,n-1}$} (A34);
  \path (A34) edge[dashed] node[above=-0.05cm,font=\sffamily\tiny]{$c_{4,n}$} (A44);

  \path (A15) edge[dashed] node[above=-0.05cm,font=\sffamily\tiny]{$c_{5,2}$} (A25);
  \path (A25) edge[dashed] node[above=-0.05cm,font=\sffamily\tiny]{$c_{5,n-1}$} (A35);
  \path (A35) edge[dashed] node[above=-0.05cm,font=\sffamily\tiny]{$c_{5,n}$} (A45);

\node[root node] (root) at (-4, -8) {$s$};
\path (root) edge (1);
\path (root) edge[dashed] (A0);

\end{tikzpicture}
    \caption{The graph described in Proposition~\ref{prop:np-hard-zhang}. Full and dashed edges denote edges present in $\mathcal{G}_d$ and $\mathcal{G}_a$ respectively. Edges between orange vertices ``belong'' to the defender and corresspond to variable assignments, while paths in the blue vertices denote a single clauses.}
    \label{fig:nph-fullprob-sat}
    \end{figure}
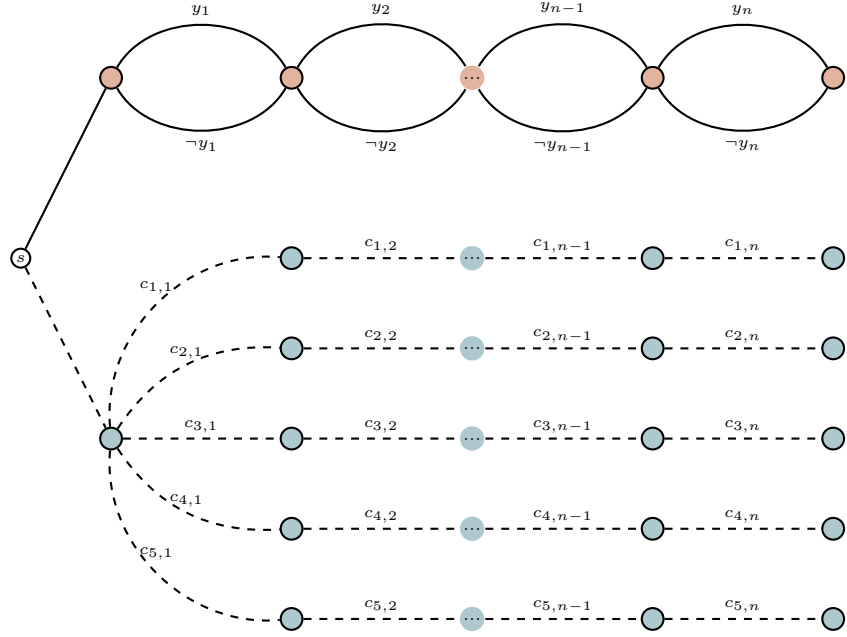
\end{proof}

\begin{cmr}{Proposition~\ref{thm:kuhn-flows} (Kuhn's theorem for $\mathbf{u_{\textsc{lin}}}$)}
    Suppose $x_d \in \Delta_d, x_a \in \Delta_a$ in a layered graph security game. Then $u_{\textsc{Lin}}$ is bilinear in their flows $f_d(x_d)$ and $f_a(x_a)$. Specifically,
    \begin{align*}
    u_\textsc{Lin}(x_d, x_a) = \sum_{e_d\in\mathcal{E}_d}\sum_{e_a\in\mathcal{E}_a} Q(e_d,e_a)  f_d(e_d) f_a(e_a).
    \end{align*}
\end{cmr}
\begin{proof}
We proceed by rearranging the formula for computing the expected utility given distributions over paths $x_d\in\Delta_d$ and $x_a\in\Delta_a$.
    \begin{align*}
    u_\textsc{Lin}(x_d, x_a) &= \sum_{p_d\in\mathcal{P}_d}\sum_{p_a\in\mathcal{P}_a} x_a(p_a)x_d(p_d)u_\textsc{Lin}(p_d,p_a)\\
    &= \sum_{p_d\in\mathcal{P}_d}\sum_{p_a\in\mathcal{P}_a}\sum_{e_d\in p_d}\sum_{e_a\in p_a} x_a(p_a)x_d(p_d)Q(e_d, e_a)\\
    &= \sum_{e_d\in\mathcal{E}_d}\sum_{e_a\in\mathcal{E}_a}\sum_{p_d\in P_d(e_d)}\sum_{p_a\in P_a(e_a)} x_a(p_a)x_d(p_d)Q(e_d, e_a) \\
    &= \sum_{e_d\in\mathcal{E}_d}\sum_{e_a\in\mathcal{E}_a} Q(e_d,e_a)  f_d(e_d) f_a(e_a),
\end{align*}
\end{proof}

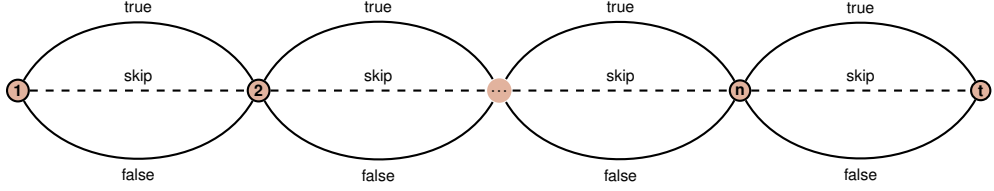
\begin{figure}[t]
\centering
\begin{tikzpicture}[->,>=,auto,node distance=2.5cm,
                    thick,main node/.style={circle,fill=\gcolor,draw,font=\sffamily\tiny\bfseries,inner sep=1}, scale=0.4]

  
  \node[main node] (1) at (0,0) {1};
  
  \node[main node] (2) at (8,0) {2};

   \node[main node, draw=none] ({n-1}) at (16,0) {$...$};
  
  \node[main node] (n) at (24,0) {n};
  
  \node[main node] (t) at (32,0) {t};

  \foreach \source/\dest in {1/2, 2/{n-1}, {n-1}/n, n/t}
    \path (\source) edge[bend right=60] node[below,font=\sffamily\tiny]{false} (\dest);
  \foreach \source/\dest in {1/2, 2/{n-1}, {n-1}/n, n/t}
    \path (\source) edge[bend left=60] node[above,font=\sffamily\tiny]{true} (\dest);

    \foreach \source/\dest in {1/2, 2/{n-1}, {n-1}/n, n/t}
    \path (\source) edge[dashed] node[above=-0.05cm,font=\sffamily\tiny]{skip} (\dest);

\end{tikzpicture}
\caption{The graph described in Proposition~\ref{prop:sat}, representing both $\mathcal{G}_i$ (indicated by full lines) and $\mathcal{G}_{-i}$ (represented by full and dashed lines), serving as a game for the SAT problems we aim to reduce from. Each vertex, denoted by $j$, corresponds to a variable $j$ in the SAT formula, which contains a total of $n$ variables. Paths within these graphs function as encodings for either an assignment (pertaining to player $i$) or a clause (pertaining to player $-i$).}
\label{fig:sat}
\end{figure}

\begin{cmr}{Proposition~\ref{prop:sat}}
    Let $\widetilde{\mathcal{P}}_i \subseteq \mathcal{P}_i$ be of size $k$ (possibly much smaller than $|\mathcal{P}_i|$) and $\widetilde{x}_i$ be a distribution with support $\widetilde{\mathcal{P}}_i$.
    Finding a best response of player $-i$ against $\widetilde{x}_i$ in a \lgsg with $u=u_\textsc{bin}$ is NP-hard in terms of $|\mathcal{G}_a|, |\mathcal{G}_d|$ and $k$.
\end{cmr}
\begin{proof}
    For the defender's best-response, we employ reductions from MAX-SAT. Consider a CNF with $n$ variables and $m$ clauses. The LGSG we construct has $n+1$ vertices organized in a row, one for each variable and a final terminal vertex. For $\mathcal{E}_d$, each non-terminal layer contains two outgoing parallel edges to the next vertex, these signify positive or negative assignments. Clearly, each assignment is a valid defender path in $\mathcal{P}_d$ and vice versa.
    $\mathcal{E}_a$ is the same, except that there is an extra `skip' edge between adjacent variables. See Figure~\ref{fig:sat} for an illustration. 
    $\widetilde{\mathcal{P}}_a$ comprises $m$ paths played uniformly at random. Each of these paths corresponds to a clause: if variable $j$ is True in the clause, the attacker takes the top edge; if False, it takes the bottom path; if $j$ is not in the clause, it takes the skip edge. Since exactly one of the $3$ edges per layer is taken, this describes a path in $\mathcal{P}_a$. 
    We set $\edgeadj(e_d, e_a) = \indicator[e_d=e_a]$ and $\targetval^\odot=m$. 
    Hence, each attacker path in $\widetilde{\mathcal{P}}_a$ interdicted by the defender signifies a satisfied clause, and reduces the attacker payoff by $1$ (starting from $m$).
    Since the defender seeks to interdict as many paths in $\widetilde{\mathcal{P}}_a$ as possible, this corresponds to solving the MAX-SAT problem. A similar reduction from MIN-SAT shows NP-hardness of the attacker best-response.
\end{proof}

\clearpage

\section{Attacker's Best Response MILP}
We overload the function $R$ to denote for a given defender's path $p_d$ the subset of attacker's edges the path $p_d$ can interdict as $R_d(p_d)$, formally $$R_d(p_d) = \{e_a:\exists e_d\in p_d R(e_d,e_a)=1\}.$$
The attacker's best-response MILP has then the following form:
\begin{align*}
    \argmax_{f_a,y,z}&\sum_{v\in \mathcal{V}^L} r(v) \cdot z_{v}\\
& z_v \leq \sum_{e_a\in E^+_a(v)}f_a(e_a) &&\quad\forall v\in \mathcal{V}^L\\
& z_v \leq 1 - \sum_{p_d\in \widetilde{\mathcal{P}}_d} y({p_d})\cdot x_d^*(p_d) &&\quad\forall v\in \mathcal{V}^L \\
    &y({p_d}) \geq f_a(e_a) && \quad\forall p_d \in \widetilde{\mathcal{P}}_d, e_a\in R_d(p_d) \\
    &1 = \sum_{e_a\in \mathcal{E}_a^-(\mathcal{V}^1)} f_a(e_a)\\
    &0 = \smashoperator{\sum_{e_a\in \mathcal{E}_a^-(v)}}f_a(e_a) - \smashoperator{\sum_{e_a\in \mathcal{E}_a^+(v)}}f_a(e_a) &&\quad\forall v \in \mathcal{V}\backslash\{\mathcal{V}^1, \mathcal{V}^{L}\}\\
     &f_a(e_a) \in \{0,1\}  &&\quad\forall e_a\in \mathcal{E}_a \\
    &0 \leq z_v, y({p_d}) \leq 1  &&\quad\forall p_d\in \widetilde{\mathcal{P}}_d, v\in \mathcal{V}^{L}
\end{align*}

The MILP structure for the attacker mirrors that of the defender. Once again, we represent the best-responding path as a binary flow, along with corresponding conservation constraints. The variable $y(p_d)$ serves as an indicator of the interdiction of the attacker's best-response path by the defender's path $p_d$. The first two constraints set the probability $z_v$ of safely reaching the target $v$ in the last layer of the layered graph in relation to the defender's interdicting paths. It is noteworthy that only one $z_v$ will have a non-zero value. The objective then focuses on maximizing the attacker's utility by seeking to maximize $z_v$ weighted by the target's value.
 
This MILP also grows in size with the opponent's subgame $|\widetilde{\mathcal{P}}_d|$, and becomes harder to solve as the subgame $|\widetilde{\mathcal{P}}_d|$ increases.

\section{Implementation Details}

In principle, the subgame LP solved in $\textsc{NashEquilibrium}(\widetilde{\mathcal{P}}_d, \widetilde{\mathcal{P}}_a)$, as well as the best-response MILP oracles that are solved in $\textsc{DefenderBR}(\widetilde{x}_a^)$ and $\textsc{AttackerBR}(\widetilde{x}_d^)$ in Algorithm~\ref{alg:do}, can be reconstructed from scratch in every iteration of the double oracle. However, in practical terms, it proves advantageous to update the existing Gurobi models and re-solve them. This approach allows Gurobi to leverage the previously computed results, resulting in decreased computation time.

The update of the Nash LP is straightforward, involving the addition of an extra best-response constraint or appending a new variable to the existing set of best-response constraints and the probability simplex constraint. However, updating the oracles is a more intricate task. Given that neither oracle relies on the player's own subgame, each oracle needs to be updated on two events: when the opponent's subgame increases and when the opponent's subgame strategy undergoes a change. In most cases, both events occur in each iteration of the algorithm.

Introducing a new attacker's path $p_a$ to the defender's MILP oracle leaves the existing constraints unchanged. However, it necessitates the creation of a new variable $y(p_a)$, the construction of a new constraint
$$y(p_a) \leq \smashoperator{\sum_{e_d\in \{e_d:\exists e_a\in p_a R(e_d,e_a)=1\}}}f_d(e_d),$$
and the inclusion of the variable $y(p_a)$ in the objective. Any change in the attacker's subgame strategy $x^*_a$ then involves a modification of the coefficients in the objective.

In contrast, an increase in the defender's subgame and a change in their subgame strategy affect even the already existing constraints in the attacker's MILP. First, for a new path $p_d$, a new variable $y(p_d)$ is created. Similarly to the defender's MILP, the new set of constraints
$$y({p_d}) \geq f_a(e_a) \qquad\forall e_a\in R_d(p_d)$$
is added to the MILP. The change in the defender's subgame strategy $x^*_d$ then implies modifications of coefficients in all the interdiction-probability constraints
$$z_v \leq 1 - \sum_{p_d\in \widetilde{\mathcal{P}}_d} y({p_d})\cdot x_d^*(p_d) \qquad\forall v\in \mathcal{V}^L.$$

\subsubsection{Speedups for MILP Solvers}
As we mentioned in Section~\ref{sec:do-desc}, we attempted several speedups for the MILP solver, including (a) admitting approximate best-responses (or better responses) rather than solving MILPs to completion, (b) strategy management by periodically removing ``weak'' strategies in $\widetilde{\mathcal{P}}_i$ that absent in $\widetilde{x}_i^*$, (c) tightening of MILPs by adding cuts/implied constraints or lifting, and (d) tuning the MILP solver, using warm-starts, or heuristics. Unfortunately, we find that only (a) yielded consistently better results. We now describe briefly how these were performed (with an focus on (a)).

\begin{enumerate}
    \item We set the appropriate parameter in Gurobi telling it to terminate when the computation time exceeds $1$s. When that happens, it is almost always the case that we have found a better (but not best) response. We add this to the strategy set anyway, as if it was a best response. This works because often times, a relatively good solution to the MILP is found quickly, and Gurobi is spending most of its effort proving optimality. In addition, we set a parameter $\alpha \geq 1$. This is how frequent we force ourselves to solve the MILP exactly. For our experiments, we set $\alpha = 20$. Lastly, we stress that when using early termination it is crucial that we \textit{do not} exit the loop in Algorithm~\ref{alg:do} using strategies from the partially solved MILP. This is because the equilibrium gap will seem tighter than they actually are. One method is to only terminate when we are solving the MILP exactly, which occurs once every $\alpha$ iterations. In practice, when we notice that the equilibrium gap computed using approximate best-responses is very small, e.g., $10\epsilon$, we will double the running time and resolve the MILP. This is repeated until the gap becomes large, or the MILP is solved fully and still gives a gap $< \epsilon$. This resolving does not significantly increase running times, since we do not resolve from scratch: Gurobi is able to continue the MILP solver from when it left off (we have not made any changes the the model).
    \item In our initial implementation, we removed paths (strategies) from $\widetilde{\mathcal{P}}_i$ when it has not been included in any Nash equilibrium for more than $50$ iterations. By keeping the set $\widetilde{\mathcal{P}}_i$ small, we hope to keep the resultant MILPs small. We also made sure we do not cycle around by repeatedly removing the same path---this is done by keep track of paths removed and ensuring that each path is removed from $\widetilde{\mathcal{P}}_i$ at most once.
    Contrary to what we may hope for, apart from the first $50$ or so paths, most paths turn out to be not terribly bad in practice and are frequently included in \textit{some} NE, albeit not all of them. Therefore, almost no paths were removed.
    We noticed that different combinations of paths ``work well'' together. Often times DO will alternate between improving these different combinations, making all of them useful in some sense.  
    \item We attempted to add some additional constraints that we know are true (from our problem formulation) but are not immediately derived from the MILP. One such constraint is to set $y(p_a)$ to be binary (in the defender best response). At first glance, we are increasing the number of binary variables and could make the problem more difficult. However, this should not be the case since we are indicating to Gurobi that $y$ is a variable that could be branched on (instead of just $x$'s). This sometimes yielded faster performance, but we did not see a noticeable difference. Another interesting example is to constrain the number of nonzero $x$'s in each layer to be no more than one (in fact it will be exactly 1). This can be derived from the flow constraints in $\Gamma_i$. Interestingly, when implemented as a Special Ordered Set of type 1 (SOS1), the solver performs significantly faster ($\sim$ 15 \%) most of the time. We suspect the inclusion of this SOS constraint helps Gurobi perform selection for variable branching slightly differently. Unfortunately, this improvement was not consistent over all our experiments (though including it seems to at least cause no harm)
\end{enumerate}

\section{Additional Experimental Data}\label{app:details}

In this section, additional details and experimental results are presented, which were excluded from the main text due to space limitations. The complete code for all experiments will be released upon acceptance of the paper.

\subsection{Detailed Description of Physical Graphs}
Recall that $\mathsf{G}_d$ and $\mathsf{G}_a$ are graphs that we ``unroll'' over time to obtain the layered graphs $\mathcal{G}_d$ and $\mathcal{G}_a$. This unrolling may depend be application specific, and we may add additional vertices to allow for dependencies on a small portion of history. This is particularly important for Anti-Terrorism(AT) and Logistical Interdiction/Persistent Threats(LI/PT).
The physical worlds $\mathsf{G}_d$ and $\mathcal{G}_a$ are designed to be independent of this unrolling routine (both in theory and in implementation). 

We will first discuss the physical worlds and how they are parameterized and generated in the context of our experiments. After that, we will move on to describe the individual unrolling process for PE, AT and LI/PT applications. 

\paragraph{Grid World.}
The grid world is simply a $S \times S$ grid of vertices arranged in a rectangle of length $(S-1) \times (S-1)$. Each vertex is \textit{potentially} adjacent to its 4-neighbors, with no wraparound for vertices at the edges. The vertices are identical, but edges could be different between $\mathsf{G}_d$ and $\mathsf{G}_a$. This models situations where some roads are available to one player but not the other: e.g., security personnel have shortcuts they may take, while the attacker may take illegal and dangerous actions (e.g., driving against traffic) that the defender has to avoid.

The generation process for the Grid World (GW) takes two parameters, an integer $S > 1$, the width and height of the grid, and optional parameters $q_{d, \text{drop}}, q_{a, \text{drop}} \in [0, 1]$, the probability that an edge in the grid is independently dropped. Typically, $q_{i,\text{drop}}$ is quite low, often around $0.1$.
An example of $\mathsf{G}_i$ (for either player) is shown in Figure~\ref{fig:physical_graphs}. There $q_{d,\text{drop}}=0$ and $q_{a,\text{drop}} > 0$, i.e., some edges are off limits to the attacker.

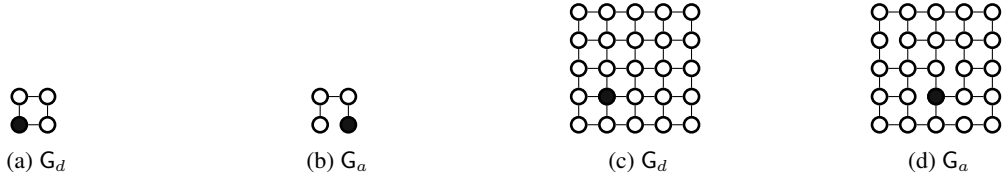
\begin{figure}
    \centering
    \begin{subfigure}[t]{0.22\linewidth}
        \centering
        \begin{tikzpicture}[every node/.style={circle, inner sep=0pt},scale=0.5]
    \definecolor{light}{RGB}{200, 200, 200}
    \definecolor{dark}{RGB}{20, 20, 20}

\node[draw, line width=1 pt, fill=dark, minimum size=2 mm] (1-1) at (0.75*1, 0.75*1) {};
\node[draw, line width=1 pt, minimum size=2 mm] (1-2) at (0.75*1, 0.75*2) {};
\node[draw, line width=1 pt, minimum size=2 mm] (2-1) at (0.75*2, 0.75*1) {};
\node[draw, line width=1 pt, minimum size=2 mm] (2-2) at (0.75*2, 0.75*2) {};

  \foreach \x in {1,...,2} {
    \foreach \y in {1,...,2} {
      \ifnum\x<2
        \draw (\x-\y) -- (\the\numexpr\x+1\relax-\y);
      \fi
      
      \ifnum\y>1
        \draw (\x-\y) -- (\x-\the\numexpr\y-1\relax);
      \fi
    }
  }
\end{tikzpicture}
        \caption{$\mathsf{G}_d$}
        \label{fig:physical-2x2-def-GW}
    \end{subfigure}
    \begin{subfigure}[t]{0.22\linewidth}
        \centering
        \begin{tikzpicture}[every node/.style={circle, inner sep=0pt},scale=0.5]
    \definecolor{light}{RGB}{200, 200, 200}
    \definecolor{dark}{RGB}{20, 20, 20}

\node[draw, line width=1 pt,  minimum size=2 mm] (1-1) at (0.75*1, 0.75*1) {};
\node[draw, line width=1 pt, minimum size=2 mm] (1-2) at (0.75*1, 0.75*2) {};
\node[draw, line width=1 pt, fill=dark, minimum size=2 mm] (2-1) at (0.75*2, 0.75*1) {};
\node[draw, line width=1 pt, minimum size=2 mm] (2-2) at (0.75*2, 0.75*2) {};

  \foreach \x in {1,...,2} {
    \foreach \y in {1,...,2} {
      \ifnum\x<2
        \ifthenelse{\x=1 \AND \y=1}{;}{
                    \draw (\x-\y) -- (\the\numexpr\x+1\relax-\y);
        }
      \fi
      
      \ifnum\y>1
        \draw (\x-\y) -- (\x-\the\numexpr\y-1\relax);
      \fi
    }
  }
\end{tikzpicture}
        \caption{$\mathsf{G}_a$}
        \label{fig:physical-2x2-atk-GW}
    \end{subfigure}
    \begin{subfigure}[t]{0.22\linewidth}
        \centering
        \begin{tikzpicture}[every node/.style={circle, inner sep=0pt}, scale=0.5]
    \definecolor{light}{RGB}{200, 200, 200}
    \definecolor{dark}{RGB}{20, 20, 20}
  \foreach \x in {1,...,5} {
    \foreach \y in {1,...,5} {
      \node[draw, line width=1 pt, minimum size=2 mm] (\x-\y) at (0.75*\x, 0.75*\y) {};
    }
  }

\node[draw, line width=1 pt, fill=dark, minimum size=2 mm] (start) at (0.75*2, 0.75*2) {};

  \foreach \x in {1,...,5} {
    \foreach \y in {1,...,5} {
      \ifnum\x<5
        \draw (\x-\y) -- (\the\numexpr\x+1\relax-\y);
      \fi
      
      \ifnum\y>1
        \draw (\x-\y) -- (\x-\the\numexpr\y-1\relax);
      \fi
    }
  }
\end{tikzpicture}
        \caption{$\mathsf{G}_d$}
        \label{fig:physical-5x5-def-GW}
    \end{subfigure}
    \begin{subfigure}[t]{0.22\linewidth}
        \centering 
        \begin{tikzpicture}[every node/.style={circle, inner sep=0pt}, scale=0.5]
    \definecolor{light}{RGB}{200, 200, 200}
    \definecolor{dark}{RGB}{20, 20, 20}
  \foreach \x in {1,...,5} {
    \foreach \y in {1,...,5} {
      \node[draw, line width=1 pt, minimum size=2 mm] (\x-\y) at (0.75*\x, 0.75*\y) {};
    }
  }

\node[draw, line width=1 pt, fill=dark, minimum size=2 mm] (start) at (0.75*3, 0.75*2) {};

  \foreach \x in {1,...,5} {
    \foreach \y in {1,...,5} {
      \ifnum\x<5
        \ifthenelse{\x=2 \AND \y=2}{;}{
            \ifthenelse{\x=1 \AND \y=4}{;}{
                \ifthenelse{\x=3 \AND \y=3}{;}{
                    \draw (\x-\y) -- (\the\numexpr\x+1\relax-\y);
                }
            }
        }
      \fi
      
      \ifnum\y>1
        \ifthenelse{\x=4 \AND \y=2}{;}{\draw (\x-\y) -- (\x-\the\numexpr\y-1\relax);}
      \fi
    }
  }
\end{tikzpicture}
        \caption{$\mathsf{G}_a$}
        \label{fig:physical-5x5-atk-GW}
    \end{subfigure}
    \caption{Defender and attacker physical graphs for Grid Worlds (GW). Figure~\ref{fig:physical-2x2-def-GW} and \ref{fig:physical-2x2-atk-GW} give graphs for a toy $2 \times 2$ map, while Figure~\ref{fig:physical-5x5-def-GW} and \ref{fig:physical-5x5-atk-GW} give show an instance of size $5 \times 5$. Black dots represent starting positions for each player. We set $q_{a,\text{drop}}=0$ and $q_{a,\text{drop}}>0$.}
    \label{fig:physical-GW}
\end{figure}

\paragraph{Open Street Maps.} Real-world maps, represented as Networkx graphs~\cite{hagberg2008exploring} and obtained through the OSMnx framework~\cite{boeing2017osmnx}, undergo a series of processing steps to improve their applicability to our application domains. The initial step is to download the original graph using the $\textsc{graph\_from\_place}$ method. Subsequently, the $\textsc{consolidate\_intersections}$ method is used to merge nearby intersections, with a specified tolerance parameter set to $30$. This merging process serves a dual purpose: firstly, it models the ability to interdict the attacker at nodes that are in close proximity to each other. Secondly, it contributes to reducing the overall number of nodes in the graphs, streamlining the computation while maintaining the essential features of the real-world layout. Following the intersection consolidation, loops are added to the graph, enabling players to stay at the same vertex for multiple timesteps. Next, the edges of the graph are discretized using the $\textsc{utils\_geo.interpolate\_points}$ method into sub-edges of place-specific unit length. This discretization serves the purpose of equalizing the distance that can be traversed in a single timestep across different locations in the graph.

In Lower Manhattan (place=\enquote{Financial district, NYC, USA}), the unit distance is set to $80$. In the case of Minnewaska State Park (place=\enquote{Minnewaska State Park, NY, USA}), a unit distance of $300$ is set. Lastly, for Bakhmut (place=\enquote{Bakhmut, Ukraine}), the designated unit distance is $200$. These variations in unit distance cater to the unique geographical characteristics of each location, e.g., on Lower Manhattan it is set to a typical width of a Manhattan block of houses.

\subsection{Detailed Description of Application Domains and Unrolling them Over Time}
Recall that we need to go from the physical maps $\mathsf{G}_i$ and the underlying task (e.g., pursuit-evasion) to a \lgsg, which requires the following components. 
\begin{itemize}
\item Layered graphs $\mathcal{G}_d$ and $\mathcal{G}_a$ 
\item Target values $\targetval^\odot(v)$ for terminal states $v \in \mathcal{V}^\odot$
\item Interdiction function $\edgeadj$.
\end{itemize}

In most of our use cases, we find that the the number of layers corresponds to the time horizon $T$. 
Indeed, since trees with depth $T$ are also layered graphs of depth $T$, we can trivially convert $\mathsf{G}_i$ to $\mathcal{G}_i$ by unrolling a search \textit{tree} (including any extra actions needed, e.g., planting an explosive in AT games) from its root (which would correspond to the starting location in the physical graph, taking the union of the vertices produced, and by setting $\targetval$ and $\edgeadj$ accordingly based solely on the leaves and the edges from the second-last layer (both have a 1-1 correspondence to paths for each player).
Of course, this is a bad idea since $|\mathcal{V}^\ell|$ now grows exponentially in $\ell$: we are no longer utilizing the fact that the utility $u$ has some ``nice'' structure. We now describe the unrolling process for PE, AT and LI/PT. 
We remind the reader that we are focusing on binary reward models.

\subsubsection{Pursuit-Evasion (PE)}
Pursuit evasion is the simplest variant of unrolling. 
For each player $i=\{ d, a \}$, we perform the following.
In each layer (except for the first, which is a singleton) we have $|\mathsf{V}|$ vertices, which we call $v_{\ell, \mathsf{v}}$, for layers $\ell \in \{ 2,...T+1\}$ and $\mathsf{v}\in\mathsf{V}$. 
For layer $\ell \geq 2$, we write their outgoing edge sets for player $i$ 
\begin{align}
    \mathcal{E}^\ell_i = \bigg\{ (v_{\ell, \mathsf{v}}, v_{\ell+1, \mathsf{v}'}) \Big\vert 
    \underbrace{(\mathsf{v}, \mathsf{v}') \in \mathsf{E}_i}_{\text{move } \mathsf{v}\rightarrow\mathsf{v}'}
    \vee \underbrace{(\mathsf{v} = \mathsf{v'})}_{\text{stay in } \mathsf{v}}
    \bigg\},
\label{eq:PE-intermediate_edge}
\end{align}
and for the first layer (a singleton containing the source)
\begin{align*}
    \mathcal{E}^1_i = \bigg\{ (v_\text{source}, v_{2, \mathsf{v}}) \Big\vert \mathsf{v} \in \mathsf{V}^{\text{start}}_i \bigg\},
\end{align*}
where $\mathsf{V}^{\text{start}}_i$ is the set of possible starting vertices in the physical graph. Note that in \eqref{eq:PE-intermediate_edge} we allow for either player to ``stay'' in its current vertex if it desires (this is equivalent to having $\mathsf{G}_i$ contains self-loops).
Finally, the set of edges is given by the union over these disjoint sets
\begin{align*}
    \mathcal{E}_i = \bigcup_{\ell=1}^{T} \mathcal{E}_i^\ell.
\end{align*}
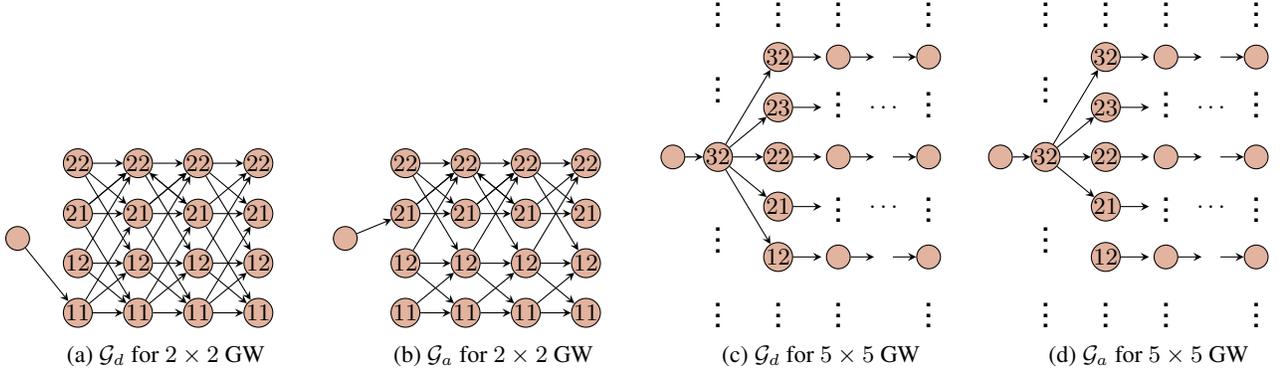
\begin{figure}[t]
    \begin{subfigure}[t]{0.24 \linewidth}
    \begin{tikzpicture}[>=stealth, 
every node/.style={circle,fill=\gcolor,draw,font=\sffamily\small\bfseries,inner sep=0pt}
]
  \definecolor{C4}{RGB}{0, 0, 0}
\definecolor{C3}{RGB}{60, 60, 60}
\definecolor{C2}{RGB}{120, 120, 120}
\definecolor{C1}{RGB}{180, 180, 180}
\definecolor{C0}{RGB}{240, 240, 240}
  
  \node (source) at (0,0.33)     {$\quad$};
  
  \node (m1n1) at (0.8,-.66)        {$11$};
  \node (m1n2) at (0.8,0)         {$12$};
  \node (m1n3) at (0.8,.66)         {$21$};
  \node (m1n4) at (0.8,1.33)         {$22$};
  
  \node (m2n1) at (1.6,-.66)        {$11$};
  \node (m2n2) at (1.6,0)         {$12$};
  \node (m2n3) at (1.6,.66)         {$21$};
  \node (m2n4) at (1.6,1.33)         {$22$};

  \node (m3n1) at (2.4,-.66)        {$11$};
  \node (m3n2) at (2.4,0)         {$12$};
  \node (m3n3) at (2.4,.66)         {$21$};
  \node (m3n4) at (2.4,1.33)         {$22$};
  
  \node (m4n1) at (3.2,-.66)        {$11$};
  \node (m4n2) at (3.2,0)         {$12$};
  \node (m4n3) at (3.2,.66)         {$21$};
  \node (m4n4) at (3.2,1.33)         {$22$};
  

  \draw[->] (source) -- (m1n1);
  

  \draw[->] (m1n1) -- (m2n2);
  \draw[->] (m1n1) -- (m2n3);
  
  \draw[->] (m1n2) -- (m2n1);
  \draw[->] (m1n2) -- (m2n4);
  
  \draw[->] (m1n3) -- (m2n4);
  \draw[->] (m1n3) -- (m2n1);
  \draw[->] (m1n3) -- (m2n4);
  
  \draw[->] (m1n4) -- (m2n2);
  \draw[->] (m1n4) -- (m2n3);

        \draw[->] (m1n1)--(m2n1);
        \draw[->] (m1n2)--(m2n2);
        \draw[->] (m1n3)--(m2n3);
        \draw[->] (m1n4)--(m2n4);
  
  \draw[->] (m2n1) -- (m3n2);
  \draw[->] (m2n1) -- (m3n3);
  
  \draw[->] (m2n2) -- (m3n1);
  \draw[->] (m2n2) -- (m3n4);
  
  \draw[->] (m2n3) -- (m3n4);
  \draw[->] (m2n3) -- (m3n1);
  \draw[->] (m2n3) -- (m3n4);
  
  \draw[->] (m2n4) -- (m3n2);
  \draw[->] (m2n4) -- (m3n3);
  
        \draw[->] (m2n1)--(m3n1);
        \draw[->] (m2n2)--(m3n2);
        \draw[->] (m2n3)--(m3n3);
        \draw[->] (m2n4)--(m3n4);

  \draw[->] (m3n1) -- (m4n2);
  \draw[->] (m3n1) -- (m4n3);
  
  \draw[->] (m3n2) -- (m4n1);
  \draw[->] (m3n2) -- (m4n4);
  
  \draw[->] (m3n3) -- (m4n4);
  \draw[->] (m3n3) -- (m4n1);
  \draw[->] (m3n3) -- (m2n4);
  
  \draw[->] (m3n4) -- (m4n2);
  \draw[->] (m3n4) -- (m4n3);
  
        \draw[->] (m3n1)--(m4n1);
        \draw[->] (m3n2)--(m4n2);
        \draw[->] (m3n3)--(m4n3);
        \draw[->] (m3n4)--(m4n4);


\end{tikzpicture}
    \caption{$\mathcal{G}_d$ for $2\times 2$ GW}
    \label{fig:physical-2x2-def-GW-unroll}
    \end{subfigure}
    \begin{subfigure}[t]{0.24 \linewidth}
    \begin{tikzpicture}[>=stealth, 
every node/.style={circle,fill=\gcolor,draw,font=\sffamily\small\bfseries,inner sep=0pt}
]
  \definecolor{C4}{RGB}{0, 0, 0}
\definecolor{C3}{RGB}{60, 60, 60}
\definecolor{C2}{RGB}{120, 120, 120}
\definecolor{C1}{RGB}{180, 180, 180}
\definecolor{C0}{RGB}{240, 240, 240}
  
  \node (source) at (0,0.33)     {$\quad$};
  
  \node (m1n1) at (0.8,-.66)        {$11$};
  \node (m1n2) at (0.8,0)         {$12$};
  \node (m1n3) at (0.8,.66)         {$21$};
  \node (m1n4) at (0.8,1.33)         {$22$};
  
  \node (m2n1) at (1.6,-.66)        {$11$};
  \node (m2n2) at (1.6,0)         {$12$};
  \node (m2n3) at (1.6,.66)         {$21$};
  \node (m2n4) at (1.6,1.33)         {$22$};

  \node (m3n1) at (2.4,-.66)        {$11$};
  \node (m3n2) at (2.4,0)         {$12$};
  \node (m3n3) at (2.4,.66)         {$21$};
  \node (m3n4) at (2.4,1.33)         {$22$};
  
  \node (m4n1) at (3.2,-.66)        {$11$};
  \node (m4n2) at (3.2,0)         {$12$};
  \node (m4n3) at (3.2,.66)         {$21$};
  \node (m4n4) at (3.2,1.33)         {$22$};
  

  \draw[->] (source) -- (m1n3);
  

  \draw[->] (m1n1) -- (m2n2);
  
  \draw[->] (m1n2) -- (m2n1);
  \draw[->] (m1n2) -- (m2n4);
  
  \draw[->] (m1n3) -- (m2n4);
  \draw[->] (m1n3) -- (m2n4);
  
  \draw[->] (m1n4) -- (m2n2);
  \draw[->] (m1n4) -- (m2n3);
  
        \draw[->] (m1n1)--(m2n1);
        \draw[->] (m1n2)--(m2n2);
        \draw[->] (m1n3)--(m2n3);
        \draw[->] (m1n4)--(m2n4);
  \draw[->] (m2n1) -- (m3n2);
  
  \draw[->] (m2n2) -- (m3n1);
  \draw[->] (m2n2) -- (m3n4);
  
  \draw[->] (m2n3) -- (m3n4);
  \draw[->] (m2n3) -- (m3n4);
  
  \draw[->] (m2n4) -- (m3n2);
  \draw[->] (m2n4) -- (m3n3);

        \draw[->] (m2n1)--(m3n1);
        \draw[->] (m2n2)--(m3n2);
        \draw[->] (m2n3)--(m3n3);
        \draw[->] (m2n4)--(m3n4);
  \draw[->] (m3n1) -- (m4n2);
  
  \draw[->] (m3n2) -- (m4n1);
  \draw[->] (m3n2) -- (m4n4);
  
  \draw[->] (m3n3) -- (m4n4);
  \draw[->] (m3n3) -- (m2n4);
  
  \draw[->] (m3n4) -- (m4n2);
  \draw[->] (m3n4) -- (m4n3);

        \draw[->] (m3n1)--(m4n1);
        \draw[->] (m3n2)--(m4n2);
        \draw[->] (m3n3)--(m4n3);
        \draw[->] (m3n4)--(m4n4);

\end{tikzpicture}
    \caption{$\mathcal{G}_a$ for $2 \times 2$ GW}
    \label{fig:physical-2x2-atk-GW-unroll}
    \end{subfigure}
    \begin{subfigure}[t]{0.24 \linewidth}
    \begin{tikzpicture}[>=stealth, 
every node/.style={circle,fill=\gcolor,draw,font=\sffamily\small\bfseries,inner sep=0pt}
]
  \definecolor{C4}{RGB}{0, 0, 0}
\definecolor{C3}{RGB}{60, 60, 60}
\definecolor{C2}{RGB}{120, 120, 120}
\definecolor{C1}{RGB}{180, 180, 180}
\definecolor{C0}{RGB}{240, 240, 240}
  
  \node (source) at (0.2,0)     {$\quad$};

  \node [draw=none, fill=none] (fake-bottom2) at (0.8, -2) {$\vdots$};
  \node [draw=none, fill=none] (fake-bottom) at (0.8, -1) {$\vdots$};
  \node (mid) at (0.8, -.0) {$32$};
  \node [draw=none, fill=none] (fake-top) at (0.8, 1) {$\vdots$};
  \node [draw=none, fill=none] (fake-top2) at (0.8, 2) {$\vdots$};

  \node [draw=none, fill=none] (fake-bottom22) at (1.6, -2) {$\vdots$};
  \node (m1n1) at (1.6, -1.33) {$12$};
  \node (m1n2) at (1.6, -.66) {$21$};
  \node (m1n3) at (1.6, -.0) {$22$};
  \node (m1n4) at (1.6, .66) {$23$};
  \node (m1n5) at (1.6, 1.33) {$32$};
  \node [draw=none, fill=none] (fake-top22) at (1.6, 2) {$\vdots$};

  \node [draw=none,fill=none](mdn1) at (2.2, -1.33) {};
  \node [draw=none,fill=none](mdn2) at (2.2, -.66) {};
  \node [draw=none,fill=none](mdn3) at (2.2, -.0) {};
  \node [draw=none,fill=none](mdn4) at (2.2, .66) {};
  \node [draw=none,fill=none](mdn5) at (2.2, 1.33) {};

  \node [draw=none, fill=none] (fake-bottom32) at (2.4, -2) {$\vdots$};
  \node (q1) at (2.4, 1.33) {$\quad$};
  \node [draw=none, fill=none] (fake-bottom3) at (2.4, -.6) {$\vdots$};
  \node (q2) at (2.4, -.0) {$\quad$};
  \node [draw=none, fill=none] (fake-top3) at (2.4, .8) {$\vdots$};
  \node (q3) at (2.4, -1.33) {$\quad$};
  \node [draw=none, fill=none] (fake-top32) at (2.4, 2) {$\vdots$};

  \node [draw=none,fill=none](e1) at (2.9, 1.33) {};
  \node [draw=none,fill=none](e2) at (2.9, -.0) {};
  \node [draw=none,fill=none](e3) at (2.9, -1.33) {};

  \node [draw=none, fill=none] (m3) at (3.0, -.66) {$\large{\dots}$};
  \node [draw=none, fill=none] (m3) at (3.0, .66) {$\large{\dots}$};
  
  \node [draw=none,fill=none](g1) at (3.1, 1.33) {};
  \node [draw=none,fill=none](g2) at (3.1, -.0) {};
  \node [draw=none,fill=none](g3) at (3.1, -1.33) {};

  \node [draw=none, fill=none] (fake-bottom32) at (3.6, -2) {$\vdots$};
  \node (r1) at (3.6, 1.33) {$\quad$};
  \node [draw=none, fill=none] (fake-bottom3) at (3.6, -.6) {$\vdots$};
  \node (r2) at (3.6, -.0) {$\quad$};
  \node [draw=none, fill=none] (fake-top3) at (3.6, .8) {$\vdots$};
  \node (r3) at (3.6, -1.33) {$\quad$};
  \node [draw=none, fill=none] (fake-top32) at (3.6, 2) {$\vdots$};

  \draw[->] (source) -- (mid);


  \draw[->] (mid) -- (m1n1);
  \draw[->] (mid) -- (m1n2);
  \draw[->] (mid) -- (m1n3);
  \draw[->] (mid) -- (m1n4);
  \draw[->] (mid) -- (m1n5);
  
  \draw[->] (m1n1) -- (mdn1);
  \draw[->] (m1n2) -- (mdn2);
  \draw[->] (m1n3) -- (mdn3);
  \draw[->] (m1n4) -- (mdn4);
  \draw[->] (m1n5) -- (mdn5);

   \draw[->] (q1) -- (e1);
   \draw[->] (q2) -- (e2);
   \draw[->] (q3) -- (e3);
   
   \draw[->] (g1) -- (r1);
   \draw[->] (g2) -- (r2);
   \draw[->] (g3) -- (r3);

\end{tikzpicture}
    \caption{$\mathcal{G}_d$ for $5\times 5$ GW}
    \label{fig:physical-5x5-def-GW-unroll}
    \end{subfigure}
    \begin{subfigure}[t]{0.24 \linewidth}
    \begin{tikzpicture}[>=stealth, 
every node/.style={circle,fill=\gcolor,draw,font=\sffamily\small\bfseries,inner sep=0pt}
]
  \definecolor{C4}{RGB}{0, 0, 0}
\definecolor{C3}{RGB}{60, 60, 60}
\definecolor{C2}{RGB}{120, 120, 120}
\definecolor{C1}{RGB}{180, 180, 180}
\definecolor{C0}{RGB}{240, 240, 240}
  
  \node (source) at (0.2,0)     {$\quad$};

  \node [draw=none, fill=none] (fake-bottom2) at (0.8, -2) {$\vdots$};
  \node [draw=none, fill=none] (fake-bottom) at (0.8, -1) {$\vdots$};
  \node (mid) at (0.8, -.0) {$32$};
  \node [draw=none, fill=none] (fake-top) at (0.8, 1) {$\vdots$};
  \node [draw=none, fill=none] (fake-top2) at (0.8, 2) {$\vdots$};

  \node [draw=none, fill=none] (fake-bottom22) at (1.6, -2) {$\vdots$};
  \node (m1n1) at (1.6, -1.33) {$12$};
  \node (m1n2) at (1.6, -.66) {$21$};
  \node (m1n3) at (1.6, -.0) {$22$};
  \node (m1n4) at (1.6, .66) {$23$};
  \node (m1n5) at (1.6, 1.33) {$32$};
  \node [draw=none, fill=none] (fake-top22) at (1.6, 2) {$\vdots$};

  \node [draw=none,fill=none](mdn1) at (2.2, -1.33) {};
  \node [draw=none,fill=none](mdn2) at (2.2, -.66) {};
  \node [draw=none,fill=none](mdn3) at (2.2, -.0) {};
  \node [draw=none,fill=none](mdn4) at (2.2, .66) {};
  \node [draw=none,fill=none](mdn5) at (2.2, 1.33) {};

  \node [draw=none, fill=none] (fake-bottom32) at (2.4, -2) {$\vdots$};
  \node (q1) at (2.4, 1.33) {$\quad$};
  \node [draw=none, fill=none] (fake-bottom3) at (2.4, -.6) {$\vdots$};
  \node (q2) at (2.4, -.0) {$\quad$};
  \node [draw=none, fill=none] (fake-top3) at (2.4, .8) {$\vdots$};
  \node (q3) at (2.4, -1.33) {$\quad$};
  \node [draw=none, fill=none] (fake-top32) at (2.4, 2) {$\vdots$};

  \node [draw=none,fill=none](e1) at (2.9, 1.33) {};
  \node [draw=none,fill=none](e2) at (2.9, -.0) {};
  \node [draw=none,fill=none](e3) at (2.9, -1.33) {};

  \node [draw=none, fill=none] (m3) at (3.0, -.66) {$\large{\dots}$};
  \node [draw=none, fill=none] (m3) at (3.0, .66) {$\large{\dots}$};
  
  \node [draw=none,fill=none](g1) at (3.1, 1.33) {};
  \node [draw=none,fill=none](g2) at (3.1, -.0) {};
  \node [draw=none,fill=none](g3) at (3.1, -1.33) {};

  \node [draw=none, fill=none] (fake-bottom32) at (3.6, -2) {$\vdots$};
  \node (r1) at (3.6, 1.33) {$\quad$};
  \node [draw=none, fill=none] (fake-bottom3) at (3.6, -.6) {$\vdots$};
  \node (r2) at (3.6, -.0) {$\quad$};
  \node [draw=none, fill=none] (fake-top3) at (3.6, .8) {$\vdots$};
  \node (r3) at (3.6, -1.33) {$\quad$};
  \node [draw=none, fill=none] (fake-top32) at (3.6, 2) {$\vdots$};

  \draw[->] (source) -- (mid);


  \draw[->] (mid) -- (m1n2);
  \draw[->] (mid) -- (m1n3);
  \draw[->] (mid) -- (m1n4);
  \draw[->] (mid) -- (m1n5);
  
  \draw[->] (m1n1) -- (mdn1);
  \draw[->] (m1n2) -- (mdn2);
  \draw[->] (m1n3) -- (mdn3);
  \draw[->] (m1n4) -- (mdn4);
  \draw[->] (m1n5) -- (mdn5);

   \draw[->] (q1) -- (e1);
   \draw[->] (q2) -- (e2);
   \draw[->] (q3) -- (e3);
   
   \draw[->] (g1) -- (r1);
   \draw[->] (g2) -- (r2);
   \draw[->] (g3) -- (r3);

\end{tikzpicture}
    \caption{$\mathcal{G}_a$ for $5 \times 5$ GW}
    \label{fig:physical-5x5-atk-GW-unroll}
    \end{subfigure}
    \caption{Unrolled layered graphs for the grid world (GW) under pursuit evasion (PE). Vertices are labeled according to their $xy$-coordinate, with the bottom left of the grid being $11$. Figures~\ref{fig:physical-2x2-def-GW-unroll} and \ref{fig:physical-2x2-atk-GW-unroll} show $\mathcal{G}_d$ and $\mathcal{G}_a$ for the $2 \times 2$ grid worlds of Figures~\ref{fig:physical-2x2-def-GW} and \ref{fig:physical-2x2-atk-GW} respectively for a horizon of $T=4$. Figures~\ref{fig:physical-5x5-def-GW-unroll} and \ref{fig:physical-5x5-atk-GW-unroll} show the same for the $5 \times 5$ variant shown in Figures~\ref{fig:physical-5x5-def-GW} and Figures~\ref{fig:physical-5x5-atk-GW}. Note that the $\mathcal{G}_d$ and $\mathcal{G}_a$ differ.} 
    \label{fig:physical-GW-unroll}
\end{figure}
Figure~\ref{fig:physical-GW-unroll} shows examples of how this is done for the grid worlds of Figure~\ref{fig:physical-GW} for grid worlds of size $2 \times 2$ and $5 \times 5$ respectively. 

Next we will specify $\edgeadj$. Recall that in our experimental setup of Section~\ref{sec:experiments}, we set $\edgeadj$ to capture the idea that interdiction occurs when players share a vertex, rather than edge. This is in order to be consistent with other related work, such that that of \citet{zhang2017optimal}. This interdiction function may be written as
\begin{align*}
    \edgeadj \left(e_d:=(u_d,v_d), e_a:=(u_a, v_a) \right) = 
    \begin{cases}
        1 \qquad & v_d = v_a \\
        0 \qquad & \text{otherwise}
    \end{cases} \\
    \intertext{or equivalently}
    \edgeadj \left(e_d, e_a \right) = 
    \begin{cases}
        1 \qquad & e_d^+ = e_a^+ \\
        0 \qquad & \text{otherwise}.
    \end{cases}
\end{align*}

Lastly, we specify $\targetval^\odot(v)$, where $v$ lies in the final layer, i.e., $v^\odot \in \mathcal{V}^\odot=\mathcal{V}^L$. If the goal is simply to evade capture, we can set $v^\odot \equiv 1$, since this means that the evader (attacker) is not concerned with the vertex it ends up in at the end of the game, as long as it was not interdicted. If, however, there is some preference based on some non-negative function $\targetval(\mathsf{v})$ for $\mathsf{v}\in \mathsf{V}$ (which is the case we use in our experiments), we set
\begin{align*}
    \targetval^\odot(v:=(L, \textsf{v})) = \targetval(\mathsf{v}). 
\end{align*}

\subsubsection{Anti-Terrorism (AT)}

For AT, the situation is similar; however, we will introduce $T_{\text{setup}}$ extra vertices in order to keep track of how long the explosive has been set up. These vertices are only reachable by the attacker, and while in them, the only action that may be taken is to advance time. Therefore, the vertices in layers $\ell \in \{2, ..., T+1 \}$ are given by $v_{\ell, \mathsf{v}, w}$, where $w \in \{0,1,\dots, T_\text{setup}\}$, and will be $0$ if the explosive has not yet been placed. This is a significant increase in the size of $\mathcal{G}_d, \mathcal{G}_a$ compared to the previously discussed PE, but still manageable overall (and \textit{much} better than expanding a tree). 

As with before, we write for layers $\ell \geq 2$, their outgoing edge sets for player $i$, \textit{assuming the explosive has not been planted}
\begin{align}
    \mathcal{E}^\ell_i = \bigg\{ (v_{\ell, \mathsf{v}, 0}, v_{\ell+1, \mathsf{v}'}) \Big\vert 
    \underbrace{(\mathsf{v}, \mathsf{v}', 0) \in \mathsf{E}_i}_{\text{move } \mathsf{v}\rightarrow\mathsf{v}'}
    \vee \underbrace{(\mathsf{v} = \mathsf{v'})}_{\text{stay in } \mathsf{v}}
    \bigg\},
\label{eq:PE-intermediate_edge}
\end{align}
and for the first layer (a singleton containing the source)
\begin{align*}
    \mathcal{E}^1_i = \bigg\{ (v_\text{source}, v_{2, \mathsf{v}}, 0) \Big\vert \mathsf{v} \in \mathsf{V}^{\text{start}}_i \bigg\},
\end{align*}
where $\mathsf{V}^{\text{start}}_i$ is the set of possible starting vertices in the physical graph; this essentially begins the game. So far, $\mathcal{E}^\ell_i$'s are formed the same way (though they may differ since $\mathsf{E}_i$ differs). However, recall that the attacker has the additional actions of setting up up the explosive. Hence for $\ell \geq 2$, the attacker has the option of planting the explosive, giving 
\begin{align*}
    \hat{\mathcal{E}}_a = \left \{ (v_{\ell, \mathsf{v}, 0 }, v_{\ell+1, \mathsf{v}, 1 }) \vert \ell \in \{ 2,...,T \} \right \},
\end{align*}
or, if the explosive is already planted, but not gone off the attacker has to wait there, incrementing the waiting count by one
\begin{align*}
    \check{\mathcal{E}} = \left \{ (v_{\ell, \mathsf{v}, w }, v_{\ell+1, \mathsf{v}, w+1 }) \vert \ell \in \{ 2,...,T \}, w \in \{ 1,\dots,T_{\text{setup}} \} \right \}.
\end{align*}
If the explosive has already gone off, i.e., $w = T_\text{setup}$, then we do not increment $w$
\begin{align*}
    \check{\mathcal{E}}' = \left \{ (v_{\ell, \mathsf{v}, w }, v_{\ell+1, \mathsf{v}, w }) \vert \ell \in \{ 2,...,T \}, w = T_\text{setup} \right \}.
\end{align*}

Finally, the set of edges for each player is given by the union over these disjoint sets,
\begin{align*}
    \mathcal{E}_d = \bigcup_{\ell=1}^{T} \mathcal{E}_d^\ell, \quad 
    \mathcal{E}_a = \left( \bigcup_{\ell=1}^{T} \mathcal{E}_a^\ell \right) \cup  \hat{\mathcal{E}}_a \cup \check{\mathcal{E}}_a \cup \check{\mathcal{E}}_a'.
\end{align*}

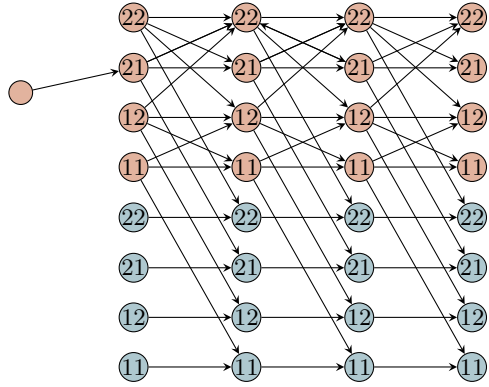
\begin{figure}[ht]
\centering
\begin{tikzpicture}[>=stealth, 
every node/.style={circle,fill=\gcolor,draw,font=\sffamily\small\bfseries,inner sep=0pt},
bomb node/.style={circle,fill=\pcolor,draw,font=\sffamily\small\bfseries,inner sep=0pt}
]
  \definecolor{C4}{RGB}{0, 0, 0}
\definecolor{C3}{RGB}{60, 60, 60}
\definecolor{C2}{RGB}{120, 120, 120}
\definecolor{C1}{RGB}{180, 180, 180}
\definecolor{C0}{RGB}{240, 240, 240}
  
  \node (source) at (0,0.33)     {$\quad$};
  
  \node (m1n1) at (1.5,-.66)        {$11$};
  \node (m1n2) at (1.5,0)         {$12$};
  \node (m1n3) at (1.5,.66)         {$21$};
  \node (m1n4) at (1.5,1.33)         {$22$};
  \node [bomb node](m1d1) at (1.5,-.66-2.66)        {$11$};
  \node [bomb node](m1d2) at (1.5,0 -2.66)         {$12$};
  \node [bomb node](m1d3) at (1.5,.66 -2.66)         {$21$};
  \node [bomb node](m1d4) at (1.5,1.33 -2.66)         {$22$};

  \node (m2n1) at (3,-.66)        {$11$};
  \node (m2n2) at (3,0)         {$12$};
  \node (m2n3) at (3,.66)         {$21$};
  \node (m2n4) at (3,1.33)         {$22$};
  \node [bomb node](m2d1) at (3,-.66-2.66)        {$11$};
  \node [bomb node](m2d2) at (3,0 -2.66)         {$12$};
  \node [bomb node](m2d3) at (3,.66 -2.66)         {$21$};
  \node [bomb node](m2d4) at (3,1.33 -2.66)         {$22$};

  \node (m3n1) at (4.5,-.66)        {$11$};
  \node (m3n2) at (4.5,0)         {$12$};
  \node (m3n3) at (4.5,.66)         {$21$};
  \node (m3n4) at (4.5,1.33)         {$22$};
  \node [bomb node](m3d1) at (4.5,-.66-2.66)        {$11$};
  \node [bomb node](m3d2) at (4.5,0 -2.66)         {$12$};
  \node [bomb node](m3d3) at (4.5,.66 -2.66)         {$21$};
  \node [bomb node](m3d4) at (4.5,1.33 -2.66)         {$22$};
  
  \node (m4n1) at (6,-.66)        {$11$};
  \node (m4n2) at (6,0)         {$12$};
  \node (m4n3) at (6,.66)         {$21$};
  \node (m4n4) at (6,1.33)         {$22$};
  \node [bomb node](m4d1) at (6,-.66-2.66)        {$11$};
  \node [bomb node](m4d2) at (6,0 -2.66)         {$12$};
  \node [bomb node](m4d3) at (6,.66 -2.66)         {$21$};
  \node [bomb node](m4d4) at (6,1.33 -2.66)         {$22$};
  

  \draw[->] (source) -- (m1n3);
  

  \draw[->] (m1n1) -- (m2n2);
  
  \draw[->] (m1n2) -- (m2n1);
  \draw[->] (m1n2) -- (m2n4);
  
  \draw[->] (m1n3) -- (m2n4);
  \draw[->] (m1n3) -- (m2n4);
  
  \draw[->] (m1n4) -- (m2n2);
  \draw[->] (m1n4) -- (m2n3);
  
        \draw[->] (m1n1)--(m2n1);
        \draw[->] (m1n2)--(m2n2);
        \draw[->] (m1n3)--(m2n3);
        \draw[->] (m1n4)--(m2n4);

      \draw[->] (m1n1) -- (m2d1);
      \draw[->] (m1n2) -- (m2d2);
      \draw[->] (m1n3) -- (m2d3);
      \draw[->] (m1n4) -- (m2d4);
      
      \draw[->] (m1d1) -- (m2d1);
      \draw[->] (m1d2) -- (m2d2);
      \draw[->] (m1d3) -- (m2d3);
      \draw[->] (m1d4) -- (m2d4);

  \draw[->] (m2n1) -- (m3n2);
  
  \draw[->] (m2n2) -- (m3n1);
  \draw[->] (m2n2) -- (m3n4);
  
  \draw[->] (m2n3) -- (m3n4);
  \draw[->] (m2n3) -- (m3n4);
  
  \draw[->] (m2n4) -- (m3n2);
  \draw[->] (m2n4) -- (m3n3);

        \draw[->] (m2n1)--(m3n1);
        \draw[->] (m2n2)--(m3n2);
        \draw[->] (m2n3)--(m3n3);
        \draw[->] (m2n4)--(m3n4);

      \draw[->] (m2n1) -- (m3d1);
      \draw[->] (m2n2) -- (m3d2);
      \draw[->] (m2n3) -- (m3d3);
      \draw[->] (m2n4) -- (m3d4);
      
      \draw[->] (m2d1) -- (m3d1);
      \draw[->] (m2d2) -- (m3d2);
      \draw[->] (m2d3) -- (m3d3);
      \draw[->] (m2d4) -- (m3d4);
    
  \draw[->] (m3n1) -- (m4n2);
  
  \draw[->] (m3n2) -- (m4n1);
  \draw[->] (m3n2) -- (m4n4);
  
  \draw[->] (m3n3) -- (m4n4);
  \draw[->] (m3n3) -- (m2n4);
  
  \draw[->] (m3n4) -- (m4n2);
  \draw[->] (m3n4) -- (m4n3);

        \draw[->] (m3n1)--(m4n1);
        \draw[->] (m3n2)--(m4n2);
        \draw[->] (m3n3)--(m4n3);
        \draw[->] (m3n4)--(m4n4);
      \draw[->] (m3n1) -- (m4d1);
      \draw[->] (m3n2) -- (m4d2);
      \draw[->] (m3n3) -- (m4d3);
      \draw[->] (m3n4) -- (m4d4);
      
      \draw[->] (m3d1) -- (m4d1);
      \draw[->] (m3d2) -- (m4d2);
      \draw[->] (m3d3) -- (m4d3);
      \draw[->] (m3d4) -- (m4d4);

\end{tikzpicture}
\caption{$\mathcal{G}_a$ by unrolling the $2\times 2$ grid-world in Figure~\ref{fig:physical-2x2-atk-GW} under the AT domain with $T_\text{setup}=1$ and $T=3$. Vertices are labeled according their physical location. The color represents $w$, the time that has been spent setting up the explosive; orange vertices are when $w=0$, blue vertices are when $w=1$. }
\label{fig:unrolled-BP-atk}.
\end{figure}

We illustrate the unrolled graph in Figure~\ref{fig:unrolled-BP-atk}. For simplicity, we will only show this for the toy grid world of size $2 \time 2$ (Figure~\ref{fig:physical-2x2-atk-GW}), and only for the attacker. The defender graph is essentially the same as Figure~\ref{fig:physical-2x2-def-GW-unroll}, except it has many additional vertices that are not reachable. Also, we assume that $T_\text{setup} = 1$. 

In Figure~\ref{fig:unrolled-BP-atk}, the blue vertices are duplicates of the orange ones. These represent vertices where the explosive has been planted --- once planted, the attacker stays in that physical location for the rest of the game. Also, since $T_\text{setup}=1$, the outgoing edges in the blue vertices go horizontally (if $T_\text{setup} > 1$), then as $w$ increases it will move to the next set of duplicated edges.

Now, we will specify $\edgeadj$. If the explosive has not been planted, the interdiction function is easy. Let $e_d=(u_d, v_d), e_a = (u_a, v_a)$, where $v_d = (\ell_d+1, \mathsf{v}_d, 0)$, $v_a = (\ell_d, \mathsf{v}_d, w)$ and $v_a = (\ell_d+1, \mathsf{v}_d', w')$. Then
\begin{align*}
    \edgeadj \left(e_d, e_a \right) = 
    \begin{cases}
        1 \qquad & \mathsf{v}_d = \mathsf{v}'_d \wedge \ell_d = \ell_a \wedge w = 0 \\
        1 \qquad & \mathsf{v}_d = \mathsf{v}'_d \wedge \ell_d = \ell_a \wedge w \neq w' \\
        1 \qquad & u_a = v_a \wedge \ell_d = \ell_a = 1 \\
        0 \qquad & \text{otherwise}.
    \end{cases}
\end{align*}
The first case is the same as PE. The second case is when the defender interdicts the attacker \textit{before} the explosive goes off (if it has already went off then we will have $w = w'$). The third case is the edge case where both players (choose to) start on the same vertex.

Finally, we need to specify $\targetval$. Assume each physical target is worth $\targetval(\mathsf{v}) \geq 0$. We simply set 
\begin{align*}
    \targetval^\odot(v_{\ell, \mathsf{v}, w}) = 
    \begin{cases}
        \targetval(\mathsf{v}) \qquad & w = T_\text{setup} \\
        0 \qquad & \text{otherwise},
    \end{cases}
\end{align*}
i.e., the explosive must have went off.

\subsubsection{Logistical Interdiction (LI)}
Recall that in LI, the utilities are dependent on the time that the attacker reached an escape point (if at all). Just like AT, we will have to add additional vertices that capture the notion of \textit{time} that has passed since the escape $t_\text{exit}$.  

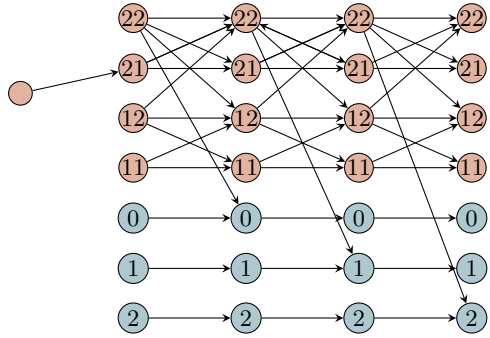
\begin{figure}[ht]
\centering
\begin{tikzpicture}[>=stealth, 
every node/.style={circle,fill=\gcolor,draw,font=\sffamily\small\bfseries,inner sep=0pt},
bomb node/.style={circle,fill=\pcolor,draw,font=\sffamily\small\bfseries,inner sep=0pt,minimum size=4mm}
]
  \definecolor{C4}{RGB}{0, 0, 0}
\definecolor{C3}{RGB}{60, 60, 60}
\definecolor{C2}{RGB}{120, 120, 120}
\definecolor{C1}{RGB}{180, 180, 180}
\definecolor{C0}{RGB}{240, 240, 240}
  
  \node (source) at (0,0.33)     {$\quad$};
  
  \node (m1n1) at (1.5,-.66)        {$11$};
  \node (m1n2) at (1.5,0)         {$12$};
  \node (m1n3) at (1.5,.66)         {$21$};
  \node (m1n4) at (1.5,1.33)         {$22$};
  \node [bomb node](m1d2) at (1.5,0 -2.66)         {$2$};
  \node [bomb node](m1d3) at (1.5,.66 -2.66)         {$1$};
  \node [bomb node](m1d4) at (1.5,1.33 -2.66)         {$0$};

  \node (m2n1) at (3,-.66)        {$11$};
  \node (m2n2) at (3,0)         {$12$};
  \node (m2n3) at (3,.66)         {$21$};
  \node (m2n4) at (3,1.33)         {$22$};
  \node [bomb node](m2d2) at (3,0 -2.66)         {$2$};
  \node [bomb node](m2d3) at (3,.66 -2.66)         {$1$};
  \node [bomb node](m2d4) at (3,1.33 -2.66)         {$0$};

  \node (m3n1) at (4.5,-.66)        {$11$};
  \node (m3n2) at (4.5,0)         {$12$};
  \node (m3n3) at (4.5,.66)         {$21$};
  \node (m3n4) at (4.5,1.33)         {$22$};
  \node [bomb node](m3d2) at (4.5,0 -2.66)         {$2$};
  \node [bomb node](m3d3) at (4.5,.66 -2.66)         {$1$};
  \node [bomb node](m3d4) at (4.5,1.33 -2.66)         {$0$};
  
  \node (m4n1) at (6,-.66)        {$11$};
  \node (m4n2) at (6,0)         {$12$};
  \node (m4n3) at (6,.66)         {$21$};
  \node (m4n4) at (6,1.33)         {$22$};
  \node [bomb node](m4d2) at (6,0 -2.66)         {$2$};
  \node [bomb node](m4d3) at (6,.66 -2.66)         {$1$};
  \node [bomb node](m4d4) at (6,1.33 -2.66)         {$0$};
  

  \draw[->] (source) -- (m1n3);
  

  \draw[->] (m1n1) -- (m2n2);
  
  \draw[->] (m1n2) -- (m2n1);
  \draw[->] (m1n2) -- (m2n4);
  
  \draw[->] (m1n3) -- (m2n4);
  \draw[->] (m1n3) -- (m2n4);
  
  \draw[->] (m1n4) -- (m2n2);
  \draw[->] (m1n4) -- (m2n3);
  
        \draw[->] (m1n1)--(m2n1);
        \draw[->] (m1n2)--(m2n2);
        \draw[->] (m1n3)--(m2n3);
        \draw[->] (m1n4)--(m2n4);

      \draw[->] (m1n4) -- (m2d4);
      
      \draw[->] (m1d2) -- (m2d2);
      \draw[->] (m1d3) -- (m2d3);
      \draw[->] (m1d4) -- (m2d4);

  \draw[->] (m2n1) -- (m3n2);
  
  \draw[->] (m2n2) -- (m3n1);
  \draw[->] (m2n2) -- (m3n4);
  
  \draw[->] (m2n3) -- (m3n4);
  \draw[->] (m2n3) -- (m3n4);
  
  \draw[->] (m2n4) -- (m3n2);
  \draw[->] (m2n4) -- (m3n3);

        \draw[->] (m2n1)--(m3n1);
        \draw[->] (m2n2)--(m3n2);
        \draw[->] (m2n3)--(m3n3);
        \draw[->] (m2n4)--(m3n4);

      \draw[->] (m2n4) -- (m3d3);
      
      \draw[->] (m2d2) -- (m3d2);
      \draw[->] (m2d3) -- (m3d3);
      \draw[->] (m2d4) -- (m3d4);
    
  \draw[->] (m3n1) -- (m4n2);
  
  \draw[->] (m3n2) -- (m4n1);
  \draw[->] (m3n2) -- (m4n4);
  
  \draw[->] (m3n3) -- (m4n4);
  \draw[->] (m3n3) -- (m2n4);
  
  \draw[->] (m3n4) -- (m4n2);
  \draw[->] (m3n4) -- (m4n3);

        \draw[->] (m3n1)--(m4n1);
        \draw[->] (m3n2)--(m4n2);
        \draw[->] (m3n3)--(m4n3);
        \draw[->] (m3n4)--(m4n4);
      \draw[->] (m3n4) -- (m4d2);
      
      \draw[->] (m3d2) -- (m4d2);
      \draw[->] (m3d3) -- (m4d3);
      \draw[->] (m3d4) -- (m4d4);

\end{tikzpicture}
\caption{$\mathcal{G}_a$ by unrolling the $2\times 2$ grid-world in Figure~\ref{fig:physical-2x2-atk-GW} under the LI domain with $T=3$ and some $\gamma \geq 0$. Orange vertices are labeled according their physical location. Blue vertices (sinks) are labeled based on the time has passed when the attacker reached the exit. }
\label{fig:unrolled-LI-atk}.
\end{figure}

Let the exit vertices in the physical graph be $\mathsf{V}_\text{exit} \subseteq \mathsf{V}$. In our layered graphs $\mathcal{G}_a$ and $\mathcal{G}_d$, we will once again have vertices for the physical location of each player. In addition, we have $T$ ``sink'' vertices where the attacker is forced to enter once it exists. These special sink vertices may not be entered by the defender, and there is one of them for each timestep; this will help record when the escape occurred and will be used in $\targetval^\odot$ to compute utilities $u_\textsc{BIN}$. We illustrate the unrolled \textit{attacker} graph in Figure~\ref{fig:unrolled-LI-atk}. We will assume that the escape point is in the top right vertex 22, which makes the problem trivial (the attacker can simply reach the exit by just moving up) but still worth illustrating.

From Figure~\ref{fig:unrolled-LI-atk}, we can see several blue sinks denoting when the escape occurred. The vertex sets are of the form $v_{\ell, \mathsf{v}}$ \textit{or}, for $\ell \geq 2$, the blue sinks $\hat{v}_{\ell, t}$ where $t$ is the time that escape occurred. The edge sets include the usual edges in the physical world

\begin{align}
    \mathcal{E}^\ell_i = \bigg\{ (v_{\ell, \mathsf{v}}, v_{\ell+1, \mathsf{v}'}) \Big\vert 
    \underbrace{(\mathsf{v}, \mathsf{v}') \in \mathsf{E}_i}_{\text{move } \mathsf{v}\rightarrow\mathsf{v}'}
    \vee \underbrace{(\mathsf{v} = \mathsf{v'})}_{\text{stay in } \mathsf{v}}
    \bigg\},
\end{align}

as well as the extra edges from exit vertices leading to blue ``sinks'' for the attacker ($\mathcal{G}_a$ does not have edges leading to the blue sinks. 

\begin{align*}
    \bar{\mathcal{E}}_a = \bigg\{ (v_{\ell, \mathsf{v}}, \hat{v}_{\ell+1, \ell-2}) \Big\vert 
    \mathsf{v} \in \mathsf{V}_{\text{exit}}, 
    \ell \geq 2
    \bigg\}.
\end{align*}

Finally, once the attacker has escaped it will stay there, leaving the $t$ index untouched,
\begin{align*}
    \hat{\mathcal{E}}_a = \bigg\{ (\hat{v}_{\ell, t}, \hat{v}_{\ell+1, t}) \Big\vert 
    \ell \geq 2
    \bigg\}.
\end{align*}
The edge sets are their respective unions for each player,
\begin{align*}
    \mathcal{E}_d = \bigcup_{\ell=1}^{T} \mathcal{E}_d^\ell, \quad 
    \mathcal{E}_a = \left( \bigcup_{\ell=1}^{T} \mathcal{E}_a^\ell \right) \cup  \bar{\mathcal{E}}_a \cup \hat{\mathcal{E}}_a.
\end{align*}

Now, $\edgeadj$ is exactly the same as PE.

\begin{align*}
    \edgeadj \left(e_d, e_a \right) = 
    \begin{cases}
        1 \qquad & e_d^+ = e_a^+ \\
        0 \qquad & \text{otherwise}.
    \end{cases}
\end{align*}

Finally, the target values $\targetval^\odot$ are given by 
\begin{align*}
    \targetval^\odot(\hat{v}_{L, t}) = \gamma^t \\
    \intertext{and}
    \targetval^\odot(v_{L, \mathsf{v}}) = \begin{cases}
        \gamma^{L-2} \qquad & \mathsf{v} \in \mathsf{V}_\text{exit} \\
        0 \qquad & \text{otherwise}.
    \end{cases}
\end{align*}
Where the second instance of $\targetval^\odot$ is a special case which occurs when the attacker reaches the exit just when the game ends.

\subsection{Application Domains Setups}

\begin{figure}
    \centering
\begin{subfigure}[t]{0.45\linewidth}
    \centering
    \begin{tikzpicture}[scale=\fscale]
\pgfplotsset{compat=1.3}
\begin{semilogyaxis}[
    ylabel={\gamesizelabel},
    xlabel={Game size [depth]},
    label style={font=\labfont},
    tick label style={font=\tickfont},
    ylabel style={align=center},
    legend pos=north west,
    ymajorgrids=true,
    grid style=dashed,
    scaled ticks=false, 
    ytick pos=left,
    ymax = 10e24,
    legend columns=3,
]

\addplot[color=\sizescolor,mark=o,error bars/.cd,y dir=both, y explicit]
coordinates {
(6, 1095.5) += (0, 55.255810060613044) -= (0, 55.255810060613044)
(7, 4496.333333333333) += (0, 205.53342046515107) -= (0, 205.53342046515107)
(8, 18579.428571428572) += (0, 981.2916175217865) -= (0, 981.2916175217865)
(9, 74154.05) += (0, 3985.177618358863) -= (0, 3985.177618358863)
(10, 309773.4) += (0, 18256.81102801449) -= (0, 18256.81102801449)
(11, 1301483.7) += (0, 83204.82999790019) -= (0, 83204.82999790019)
(12, 5491216.95) += (0, 377205.5931066078) -= (0, 377205.5931066078)
(13, 23245349.8) += (0, 1702177.7582099685) -= (0, 1702177.7582099685)
}; \addlegendentry{PE (GW)}

\addplot[color=\minnewaskaexpcolor,mark=o,error bars/.cd,y dir=both, y explicit]
coordinates {
(6, 1694.25) += (0, 79.27752303077295) -= (0, 79.27752303077295)
(7, 6918.1) += (0, 376.51226232914445) -= (0, 376.51226232914445)
(8, 28627.2) += (0, 1777.9841106980505) -= (0, 1777.9841106980505)
(9, 119614.85) += (0, 8333.76871695892) -= (0, 8333.76871695892)
(10, 503341.5) += (0, 38774.803783649586) -= (0, 38774.803783649586)
(11, 2129194.45) += (0, 179203.42426639757) -= (0, 179203.42426639757)
(12, 9042422.55) += (0, 823507.1935100777) -= (0, 823507.1935100777)
(13, 38519721.85) += (0, 3766457.6330925957) -= (0, 3766457.6330925957)
(14, 164489212.7) += (0, 17160122.53365262) -= (0, 17160122.53365262)
(15, 703808134.5) += (0, 77936602.74330589) -= (0, 77936602.74330589)
(16, 3016439214.4) += (0, 353061569.46524304) -= (0, 353061569.46524304)
(17, 12946629534.6) += (0, 1596060267.6175475) -= (0, 1596060267.6175475)
(18, 55637189639.0) += (0, 7202745642.270909) -= (0, 7202745642.270909)
(19, 239366196479.3) += (0, 32458095896.376953) -= (0, 32458095896.376953)
}; \addlegendentry{AT (GW))}

\addplot[color=black,mark=o,error bars/.cd,y dir=both, y explicit]
coordinates {
(10, 1244673.2) += (0, 11810.450282560869) -= (0, 11810.450282560869)
(11, 5525040.6) += (0, 54435.79645283732) -= (0, 54435.79645283732)
(12, 24525267.3) += (0, 249252.7185611032) -= (0, 249252.7185611032)
(13, 108888579.5) += (0, 1135237.0415297332) -= (0, 1135237.0415297332)
(14, 483553113.9) += (0, 5147548.203043755) -= (0, 5147548.203043755)
(15, 2147948501.8) += (0, 23257821.84143602) -= (0, 23257821.84143602)
(16, 9543720712.8) += (0, 104777813.97704844) -= (0, 104777813.97704844)
(17, 42415834666.05) += (0, 470911821.8681288) -= (0, 470911821.8681288)
(18, 188559480107.5) += (0, 2112301605.202701) -= (0, 2112301605.202701)
(19, 838444487086.55) += (0, 9459352787.361416) -= (0, 9459352787.361416)
(20, 3729056953790.25) += (0, 42302400918.3021) -= (0, 42302400918.3021)
(21, 16588860310865.95) += (0, 188953173543.98676) -= (0, 188953173543.98676)
(22, 73810918030959.25) += (0, 843135903282.2054) -= (0, 843135903282.2054)
(23, 328477388491609.0) += (0, 3758814416525.232) -= (0, 3758814416525.232)
(24, 1462061551407941.5) += (0, 16743998095691.197) -= (0, 16743998095691.197)
(25, 6508727630079540.0) += (0, 74534918743197.31) -= (0, 74534918743197.31)
(26, 2.8979580041761836e+16) += (0, 331576624825769.25) -= (0, 331576624825769.25)
(27, 1.2904740959802158e+17) += (0, 1474205492654188.8) -= (0, 1474205492654188.8)
(28, 5.7472976746813997e+17) += (0, 6550958837210110.0) -= (0, 6550958837210110.0)
(29, 2.559950599803401e+18) += (0, 2.909673280361316e+16) -= (0, 2.909673280361316e+16)
(30, 1.14037987011396e+19) += (0, 1.2917946337383976e+17) -= (0, 1.2917946337383976e+17)
};\addlegendentry{LI (GW)}

\addplot[color=\sizescolor,mark=square,error bars/.cd,y dir=both, y explicit]
coordinates {
(6, 1342.0) += (0, 0.0) -= (0, 0.0)
(7, 3206.0) += (0, 0.0) -= (0, 0.0)
(8, 7824.0) += (0, 0.0) -= (0, 0.0)
(9, 19568.0) += (0, 0.0) -= (0, 0.0)
(10, 50088.0) += (0, 0.0) -= (0, 0.0)
(11, 130714.0) += (0, 0.0) -= (0, 0.0)
(12, 346464.0) += (0, 0.0) -= (0, 0.0)
(13, 930446.0) += (0, 0.0) -= (0, 0.0)
(14, 2529336.0) += (0, 0.0) -= (0, 0.0)
(15, 6959182.0) += (0, 0.0) -= (0, 0.0)
(16, 19380428.0) += (0, 0.0) -= (0, 0.0)
(17, 54615200.0) += (0, 0.0) -= (0, 0.0)
(18, 155646044.0) += (0, 0.0) -= (0, 0.0)
(19, 448200828.0) += (0, 0.0) -= (0, 0.0)
(20, 1303042068.0) += (0, 0.0) -= (0, 0.0)
};\addlegendentry{PE (MN)}

\addplot[color=\minnewaskaexpcolor,mark=square,error bars/.cd,y dir=both, y explicit]
coordinates {
(6, 1443.0) += (0, 0.0) -= (0, 0.0)
(7, 3432.0) += (0, 0.0) -= (0, 0.0)
(8, 8276.0) += (0, 0.0) -= (0, 0.0)
(9, 20369.0) += (0, 0.0) -= (0, 0.0)
(10, 51285.0) += (0, 0.0) -= (0, 0.0)
(11, 131902.0) += (0, 0.0) -= (0, 0.0)
(12, 345447.0) += (0, 0.0) -= (0, 0.0)
(13, 918545.0) += (0, 0.0) -= (0, 0.0)
(14, 2475204.0) += (0, 0.0) -= (0, 0.0)
(15, 6754240.0) += (0, 0.0) -= (0, 0.0)
(16, 18659377.0) += (0, 0.0) -= (0, 0.0)
(17, 52177600.0) += (0, 0.0) -= (0, 0.0)
(18, 147618008.0) += (0, 0.0) -= (0, 0.0)
(19, 422233844.0) += (0, 0.0) -= (0, 0.0)
(20, 1220032270.0) += (0, 0.0) -= (0, 0.0)
(21, 3558575559.0) += (0, 0.0) -= (0, 0.0)
(22, 10472256655.0) += (0, 0.0) -= (0, 0.0)
(23, 31084654512.0) += (0, 0.0) -= (0, 0.0)
(24, 93063067607.0) += (0, 0.0) -= (0, 0.0)
(25, 281058342216.0) += (0, 0.0) -= (0, 0.0)
};\addlegendentry{AT (MN)}

\addplot[color=black,mark=triangle,error bars/.cd,y dir=both, y explicit]
coordinates {
(10, 8918.0) += (0, 0.0) -= (0, 0.0)
(11, 19899.0) += (0, 0.0) -= (0, 0.0)
(12, 46204.0) += (0, 0.0) -= (0, 0.0)
(13, 112108.0) += (0, 0.0) -= (0, 0.0)
(14, 284352.0) += (0, 0.0) -= (0, 0.0)
(15, 750964.0) += (0, 0.0) -= (0, 0.0)
(16, 2050630.0) += (0, 0.0) -= (0, 0.0)
(17, 5744337.0) += (0, 0.0) -= (0, 0.0)
(18, 16391604.0) += (0, 0.0) -= (0, 0.0)
(19, 47387528.0) += (0, 0.0) -= (0, 0.0)
(20, 138253993.0) += (0, 0.0) -= (0, 0.0)
(21, 405973522.0) += (0, 0.0) -= (0, 0.0)
(22, 1197624155.0) += (0, 0.0) -= (0, 0.0)
(23, 3544697262.0) += (0, 0.0) -= (0, 0.0)
(24, 10516295947.0) += (0, 0.0) -= (0, 0.0)
(25, 31251356401.0) += (0, 0.0) -= (0, 0.0)
(26, 92976564307.0) += (0, 0.0) -= (0, 0.0)
(27, 276829408655.0) += (0, 0.0) -= (0, 0.0)
(28, 824643073120.0) += (0, 0.0) -= (0, 0.0)
(29, 2457255390877.0) += (0, 0.0) -= (0, 0.0)
(30, 7323291372880.0) += (0, 0.0) -= (0, 0.0)
};\addlegendentry{AT (MW)}

\end{semilogyaxis}
\end{tikzpicture}
\end{subfigure}
\hfill
\begin{subfigure}[t]{0.45\linewidth}
    \centering
    \begin{tikzpicture}[scale=\fscale]
\pgfplotsset{compat=1.3}
\begin{semilogyaxis}[
    ylabel={\gamesizelabel},
    xlabel={Game size [depth]},
    label style={font=\labfont},
    tick label style={font=\tickfont},
    ylabel style={align=center},
    legend pos=north west,
    ymajorgrids=true,
    grid style=dashed,
    scaled ticks=false, 
    ytick pos=left,
    legend columns=2,
]

\addplot[color=\sizescolor,mark=square,error bars/.cd,y dir=both, y explicit]
coordinates {
(50, 2.1494522766197663e+32) += (0, 0.0) -= (0, 0.0)
(55, 5.225397220748595e+35) += (0, 0.0) -= (0, 0.0)
(60, 1.3071163659871852e+39) += (0, 0.0) -= (0, 0.0)
(65, 3.365465007189097e+42) += (0, 0.0) -= (0, 0.0)
(70, 8.921980346020252e+45) += (0, 0.0) -= (0, 0.0)
(75, 2.435790625385188e+49) += (0, 0.0) -= (0, 0.0)
(80, 6.846632229401962e+52) += (0, 0.0) -= (0, 0.0)
(85, 1.979668834273525e+56) += (0, 0.0) -= (0, 0.0)
(90, 5.878757434887238e+59) += (0, 0.0) -= (0, 0.0)
(95, 1.788770422497314e+63) += (0, 0.0) -= (0, 0.0)
(100, 5.5613169368594255e+66) += (0, 0.0) -= (0, 0.0)
(105, 1.761258401780068e+70) += (0, 0.0) -= (0, 0.0)
(110, 5.664488833287261e+73) += (0, 0.0) -= (0, 0.0)
(115, 1.8447991944843424e+77) += (0, 0.0) -= (0, 0.0)
(120, 6.068615169315373e+80) += (0, 0.0) -= (0, 0.0)
(125, 2.0121001714187657e+84) += (0, 0.0) -= (0, 0.0)
(130, 6.712161507948501e+87) += (0, 0.0) -= (0, 0.0)
(135, 2.2496399121884463e+91) += (0, 0.0) -= (0, 0.0)
(140, 7.566887186034425e+94) += (0, 0.0) -= (0, 0.0)
(145, 2.552109462040742e+98) += (0, 0.0) -= (0, 0.0)
(150, 8.62522461706051e+101) += (0, 0.0) -= (0, 0.0)
}; \addlegendentry{LI (BK)}

\end{semilogyaxis}
\end{tikzpicture}
\end{subfigure}
    \caption{Game size as a function of the depth of the layered graph (i.e., the horizon) for each scenario we consider: pursuit-evasion (PE), anti-terrorism / anti-poaching (AT), and logistical interdiction / persistent threats (LI) on the grid world (GW), Lower Manhattan (MN), Minnewaska State Park (MW), and the city of Bakhmut (BK).}
    \label{fig:paths}
\end{figure}
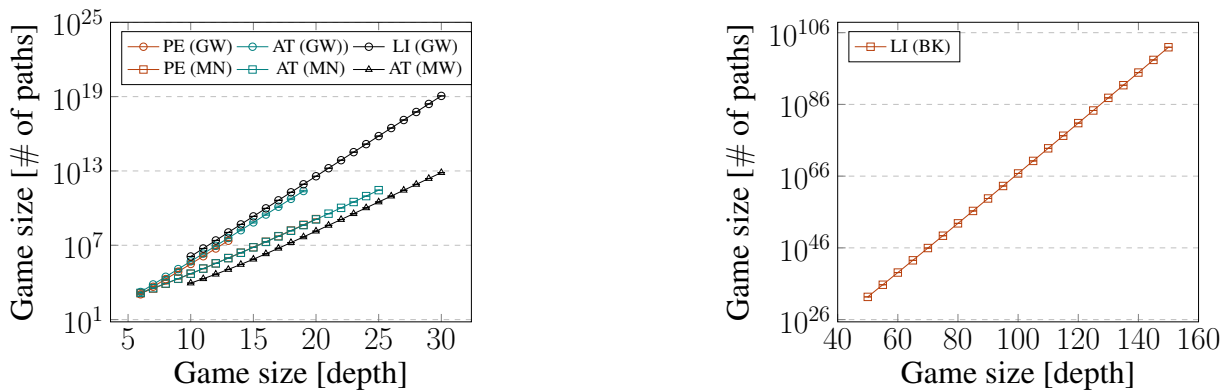

In this subsection we explain how the individual physical graphs are setup to generate \lgsgs for the application domains. For each scenario we depict the corresponding game size as a function of the depth of the layered graphs (i.e., the horizon) in Figure~\ref{fig:paths}.

\paragraph{Pursuit-Evasion on grid world:} In this scenario, players start from opposite corners, and target values are uniformly generated at random from the integral interval $[1,10]$ for each node on the grid. For each player, edges are randomly dropped with a probability of 0.1.

\begin{figure}[t]
    \centering
\begin{subfigure}[t]{0.32\linewidth}
    \centering
    \includegraphics[width=.8\linewidth]{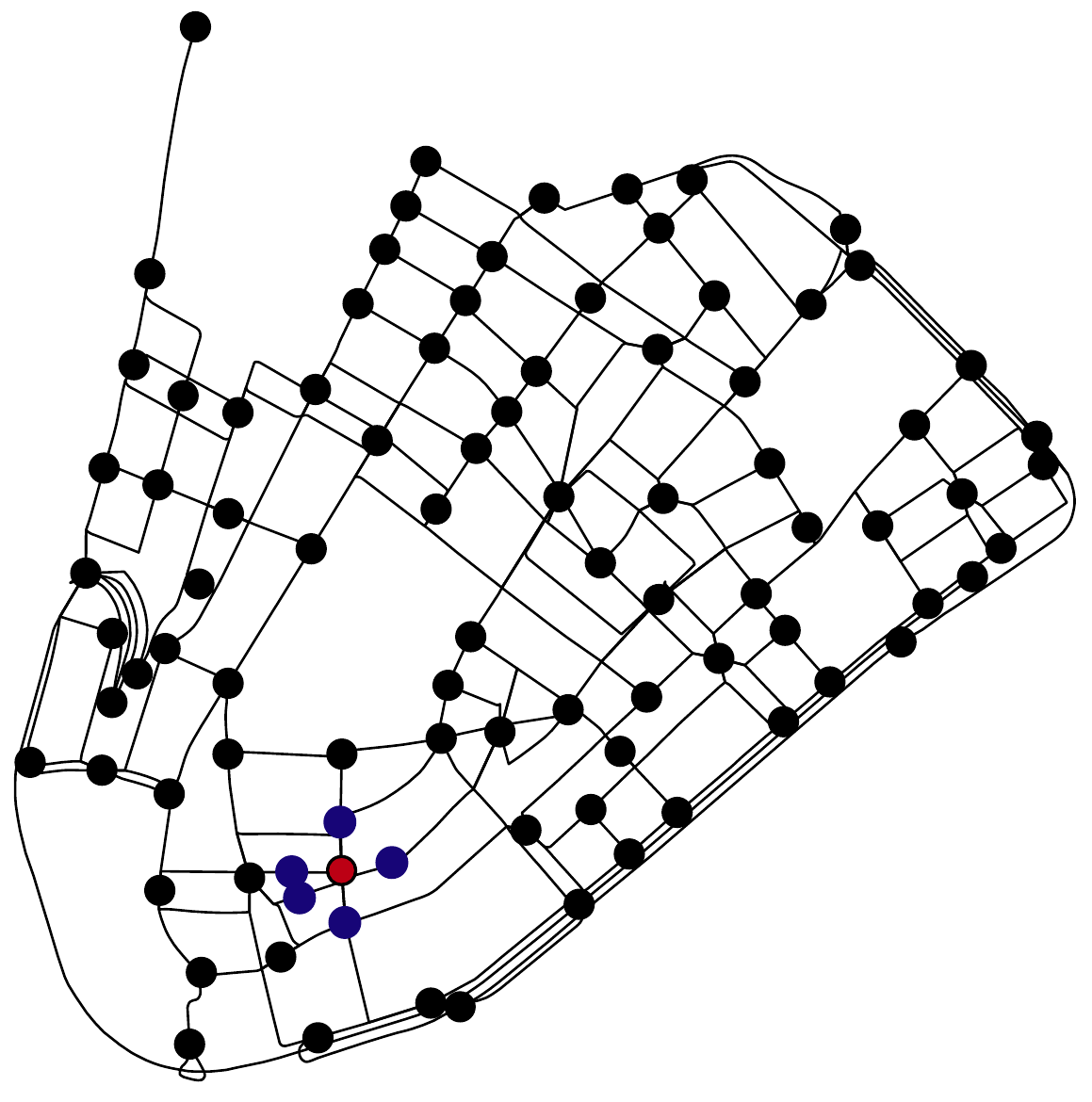}
    \caption{Lower Manhattan, NYC, USA}
\end{subfigure}
\hfill
\begin{subfigure}[t]{0.32\linewidth}
    \centering
    \includegraphics[width=\linewidth]{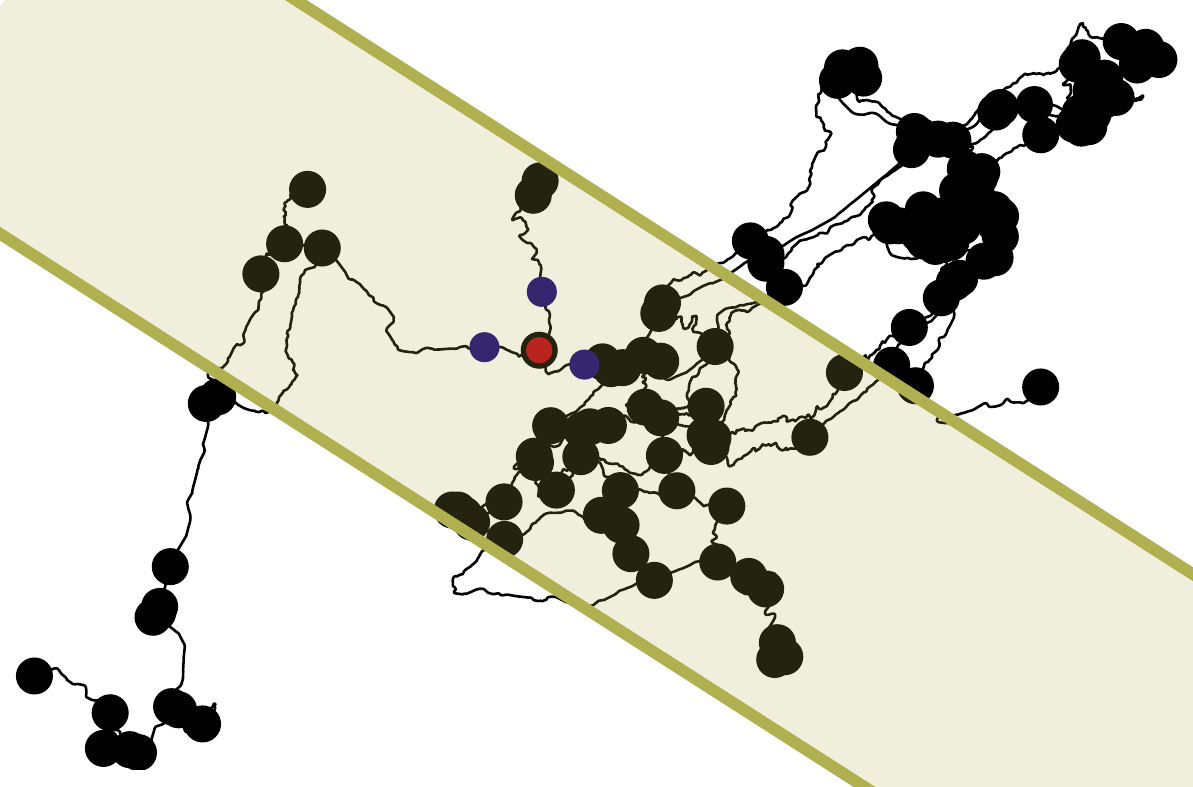}
    \caption{Minnewaska State Park, NY, USA}
\end{subfigure}
\hfill
\begin{subfigure}[t]{0.32\linewidth}
    \centering
    \includegraphics[width=.8\linewidth]{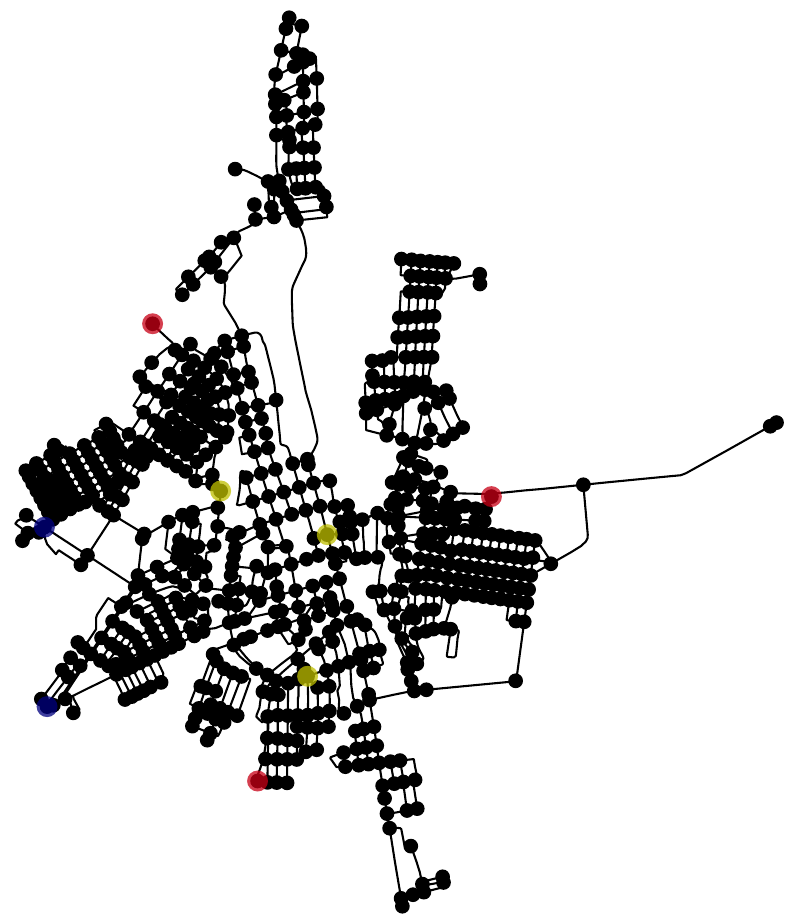}
    \caption{City of Bakhmut, UKR}
\end{subfigure}
    \caption{Starting points in our real-world physical graphs used to generate \lgsgs for our application domains for the defender (red point(s)) and attacker (blue point(s)) in (a) anti-terrorism, (b) anti-terrorism, and (c) logistical interdiction.}
    \label{fig:physical_graphs_setups}
\end{figure}

\paragraph{Pursuit-Evasion on Lower Manhattan:} In this setup, both players can select their starting points from the blue nodes shown in Figure~\ref{fig:physical_graphs_setups}. Similarly, target values are uniformly generated for each node from the integral interval $[1,10]$. No edges are omitted for either player.

\paragraph{Anti-Terrorism on grid world:} In this scenario, the defender starts from a grid corner, while the attacker starts from a node immediately adjacent. Target values are uniformly generated from the integral interval $[1,10]$, and each edge is randomly dropped with a 0.1 probability for each player. The time required to detonate the explosive device is set to $T_{\text{setup}}=2$.

\paragraph{Anti-Terrorism on Lower Manhattan:} Here, the defender starts at the red node, and the attacker may choose a starting node from the adjacent blue nodes in Figure~\ref{fig:physical_graphs_setups}. Similar to previous scenarios, node targets have values generated uniformly from an integral interval $[1,10]$. The time needed to set up the explosive device is $T_{\text{setup}}=2$. No edges are omitted for either player.

\paragraph{Anti-Poaching on Minnewaska State Park:} The defender begins at the red node in Figure~\ref{fig:physical_graphs_setups}, and the attacker selects from the set of blue nodes. Four habitats are sampled from the yellow diagonal area in the figure. Habitat animal scores are uniformly generated from the interval $(0,1]$. The node scores are non-zero only on the intersections. In contrast to the attacker, the defender can traverse selected paved routes (indicated as paved by OSMnx) in one timestep, simulating vehicular movement, during which the attacker cannot be interdicted.  It takes $T_{\text{setup}}=2$ steps for the attacker to poach an animal.

\paragraph{Logistical Interdiction on grid world:} In this setup, the defender starts at the center of the grid (unique in our $5\times 5$ grid), and the attacker starts in a bottom corner. The only non-zero same-valued targets are the top corners, with values rescaled to the interval $(0,10]$ based on the delay factor $\gamma$ and reaching time. The defender can utilize the entire grid (no edges are dropped), while each edge in the attacker's graph is randomly omitted with a probability of 0.1.

\paragraph{Logistical Interdiction on Bakhmut:} This scenario is inspired by the battlefront situation in Bakhmut on March 2, 2023. The Ukrainian supply forces can choose a starting node for the supply run from the blue node in Figure~\ref{fig:physical_graphs_setups}. Russian forces start in one of the red nodes. The objective for Ukrainian forces is to safely reach one of the yellow targets, the only non-zero same-valued targets on the map. Similar to previous scenarios, target values are rescaled to the interval $(0,10]$ based on the delay factor $\gamma$ and reaching time. The graph is the same for both players.

\subsection{Detailed Experimental Setup}
The experiments were conducted on an Intel Xeon Gold 6226 operating at 2.9Ghz, restricted to 8 threads and equipped with 32GB of RAM. The optimization tasks were undertaken with the Gurobi Optimizer version 10.0.3, build v10.0.3rc0 \cite{gurobi}, on a Linux 64-bit platform. The double oracle algorithm was implemented in Python 3.7.9 and configured with a tolerance setting of $\epsilon=10^{-3}$. The real-world graphs were obtained using OSMnx 1.8.1 \cite{boeing2017osmnx}. To ensure statistical robustness, 20 instances of each game were constructed and solved for any game involving randomness. In the reported graphs, we often depict the game size on one axis, defining it as the sum of the number of paths for each player.

\subsection{Additional Experimental Results}

\begin{figure}
    \centering
\begin{subfigure}[t]{0.45\linewidth}
    \centering
    \begin{tikzpicture}[scale=\fscale]
\pgfplotsset{compat=1.3}
\begin{loglogaxis}[
    title={\titfont \bf Anti-poaching on Minnewaska (lin)},
    xlabel={\gamesizelabel},
    ylabel={\computationlabel},
    ylabel style={align=center},
    ymajorgrids=true,
    grid style=dashed,
    scaled ticks=false, 
    ytick pos=left,
    mark size=\fmarksize,
    label style={font=\labfont},
    tick label style={font=\tickfont},
    legend pos=north east,
    legend columns=3,
    ymax = 120000,
]

\legend{\lpname, \doname, \dolname}
\addplot[color=black,mark=\lpmark,error bars/.cd,y dir=both, y explicit]
coordinates {
(8918.0, 139.32399999999998) += (0, 0.7105788596556148) -= (0, 0.7105788596556148)
(19899.0, 671.2575) += (0, 5.134832999641808) -= (0, 5.134832999641808)
(46204.0, 3754.018000000001) += (0, 25.86470676627252) -= (0, 25.86470676627252)
(112108.0, 24050.081) += (0, 106.01395767067662) -= (0, 106.01395767067662)
};\label{bopl:park:lin:lp}

\addplot[color=black,mark=\domark,error bars/.cd,y dir=both, y explicit]
coordinates {
(8918.0, 17.3405) += (0, 0.500401141716902) -= (0, 0.500401141716902)
(19899.0, 15.446999999999997) += (0, 0.2278308467632118) -= (0, 0.2278308467632118)
(46204.0, 17.2615) += (0, 0.555994261140394) -= (0, 0.555994261140394)
(112108.0, 22.3555) += (0, 1.533485447736969) -= (0, 1.533485447736969)
(284352.0, 34.14) += (0, 2.7516724101229015) -= (0, 2.7516724101229015)
(750964.0, 32.866) += (0, 4.578347346046675) -= (0, 4.578347346046675)
(2050630.0, 44.07000000000001) += (0, 7.077102365709912) -= (0, 7.077102365709912)
(5744337.0, 48.414) += (0, 17.151864612461544) -= (0, 17.151864612461544)
(16391604.0, 53.8235) += (0, 15.476626711117092) -= (0, 15.476626711117092)
(47387528.0, 65.70950000000002) += (0, 19.491297632735865) -= (0, 19.491297632735865)
(138253993.0, 119.80350000000001) += (0, 61.72579428951118) -= (0, 61.72579428951118)
(405973522.0, 158.67949999999993) += (0, 92.20044614138831) -= (0, 92.20044614138831)
(1197624155.0, 312.9725) += (0, 203.60746457451043) -= (0, 203.60746457451043)
(3544697262.0, 164.194) += (0, 82.76278111630411) -= (0, 82.76278111630411)
(10516295947.0, 134.84549999999996) += (0, 33.67965736538898) -= (0, 33.67965736538898)
(31251356401.0, 376.90499999999997) += (0, 46.26347864762168) -= (0, 46.26347864762168)
(92976564307.0, 419.68750000000006) += (0, 63.03859208484492) -= (0, 63.03859208484492)
(276829408655.0, 371.26000000000005) += (0, 126.76043815838307) -= (0, 126.76043815838307)
(824643073120.0, 723.53099999999995) += (0, 135.03791163751404) -= (0, 135.03791163751404)
(2457255390877.0, 208.74699999999999) += (0, 40.90180548131804) -= (0, 40.90180548131804)
(7323291372880.0, 235.67450000000002) += (0, 45.21105141721435) -= (0, 45.21105141721435)
}; \label{bopl:park:lin:do}

\addplot[color=black,mark=\dolmark,error bars/.cd,y dir=both, y explicit]
coordinates {
(8917.0, 14.290499999999998) += (0, 0.47686336727896345) -= (0, 0.47686336727896345)
(19885.0, 13.985499999999998) += (0, 0.1586198285209009) -= (0, 0.1586198285209009)
(46100.0, 16.073) += (0, 0.33887282197556684) -= (0, 0.33887282197556684)
(111559.0, 19.865000000000002) += (0, 0.6649877323759636) -= (0, 0.6649877323759636)
(282033.0, 38.4555) += (0, 4.062518824247886) -= (0, 4.062518824247886)
(742608.0, 34.483999999999995) += (0, 4.539047453979049) -= (0, 4.539047453979049)
(2023907.0, 38.518) += (0, 7.301165754071129) -= (0, 7.301165754071129)
(5666415.0, 49.7175) += (0, 9.889790000298287) -= (0, 9.889790000298287)
(16180286.0, 58.68349999999997) += (0, 12.349232924718336) -= (0, 12.349232924718336)
(46845883.0, 62.327999999999996) += (0, 14.529318193153275) -= (0, 14.529318193153275)
(136921875.0, 72.03049999999999) += (0, 17.886670256970103) -= (0, 17.886670256970103)
(402777130.0, 120.55850000000001) += (0, 32.78718044790806) -= (0, 32.78718044790806)
(1189983853.0, 119.93550000000002) += (0, 38.21831532872412) -= (0, 38.21831532872412)
(3526023792.0, 83.4125) += (0, 36.725379736311986) -= (0, 36.725379736311986)
(10468306005.0, 121.95750000000001) += (0, 53.91631721022086) -= (0, 53.91631721022086)
(31119019396.0, 354.18750000000006) += (0, 205.2383250348046) -= (0, 205.2383250348046)
(92583944059.0, 481.12600000000003) += (0, 344.86889236071323) -= (0, 344.86889236071323)
(275594455538.0, 429.7404999999999) += (0, 281.34137302205346) -= (0, 281.34137302205346)
(820613423424.0, 734.8409999999999) += (0, 503.9780435160476) -= (0, 503.9780435160476)
(2443872086675.0, 170.19400000000002) += (0, 117.44476338710308) -= (0, 117.44476338710308)
(7278617927795.0, 61.239) += (0, 12.744199648632726) -= (0, 12.744199648632726)
}; \label{bopl:park:lin:dol}

\end{loglogaxis}







\end{tikzpicture}
\end{subfigure}
\hfill
\begin{subfigure}[t]{0.45\linewidth}
    \centering
    \begin{tikzpicture}[scale=\fscale]
\pgfplotsset{compat=1.3}
\begin{loglogaxis}[
    title={\titfont \bf Anti-poaching on Minnewaska (exp)},
    xlabel={\gamesizelabel},
    ylabel style={align=center},
    legend pos=north east,
    ymajorgrids=true,
    grid style=dashed,
    scaled ticks=false, 
    ytick pos=left,
    ymax = 120000,
    mark size=\fmarksize,
    label style={font=\labfont},
    tick label style={font=\tickfont},
    legend columns=3,
]

\legend{\lpname, \doname, \dolname}
\addplot[color=black,mark=\lpmark,error bars/.cd,y dir=both, y explicit]
coordinates {
(8918.0, 140.6435) += (0, 0.2091905137730471) -= (0, 0.2091905137730471)
(19899.0, 657.832) += (0, 5.6300433110421535) -= (0, 5.6300433110421535)
(46204.0, 3664.024) += (0, 5.63743238493703) -= (0, 5.63743238493703)
(112108.0, 22857.719999999998) += (0, 80.82132257847113) -= (0, 80.82132257847113)
};

\addplot[color=black,mark=\domark,error bars/.cd,y dir=both, y explicit]
coordinates {
(8918.0, 47.9075) += (0, 3.9667313121298724) -= (0, 3.9667313121298724)
(19899.0, 21.783) += (0, 5.032857360439129) -= (0, 5.032857360439129)
(46204.0, 16.1005) += (0, 0.35955489662110074) -= (0, 0.35955489662110074)
(112108.0, 22.061500000000002) += (0, 0.8354237106497453) -= (0, 0.8354237106497453)
(284352.0, 37.654999999999994) += (0, 2.1571052117912193) -= (0, 2.1571052117912193)
(750964.0, 42.593) += (0, 2.1869836352960283) -= (0, 2.1869836352960283)
(2050630.0, 49.22749999999999) += (0, 2.537988208155594) -= (0, 2.537988208155594)
(5744337.0, 80.5235) += (0, 6.179919871521074) -= (0, 6.179919871521074)
(16391604.0, 127.33750000000002) += (0, 10.838200469076714) -= (0, 10.838200469076714)
(47387528.0, 171.17149999999998) += (0, 18.2458945962762) -= (0, 18.2458945962762)
(138253993.0, 258.22549999999995) += (0, 29.32106748860028) -= (0, 29.32106748860028)
(405973522.0, 377.10200000000003) += (0, 54.44220235403683) -= (0, 54.44220235403683)
(1197624155.0, 458.6829999999999) += (0, 46.621739231821884) -= (0, 46.621739231821884)
(3544697262.0, 683.3055000000002) += (0, 59.2363048595023) -= (0, 59.2363048595023)
(10516295947.0, 1024.4015) += (0, 169.71841894315946) -= (0, 169.71841894315946)
(31251356401.0, 1643.4450000000002) += (0, 227.4298417239456) -= (0, 227.4298417239456)
(92976564307.0, 3239.2595) += (0, 430.8779055819305) -= (0, 430.8779055819305)
(276829408655.0, 5715.713499999999) += (0, 887.9136146445763) -= (0, 887.9136146445763)
(824643073120.0, 7585.359499999999) += (0, 1101.500364345563) -= (0, 1101.500364345563)
(2457255390877.0, 14311.551499999996) += (0, 1978.5105990619988) -= (0, 1978.5105990619988)
(7323291372880.0, 22703.656250000004) += (0, 3405.538862660342) -= (0, 3405.538862660342)
}; 

\addplot[color=black,mark=\dolmark,error bars/.cd,y dir=both, y explicit]
coordinates {
(8917.0, 17.698) += (0, 4.584784021774177) -= (0, 4.584784021774177)
(19885.0, 13.497) += (0, 0.14015048303426153) -= (0, 0.14015048303426153)
(46100.0, 15.123500000000002) += (0, 0.17208974861666618) -= (0, 0.17208974861666618)
(111559.0, 20.722) += (0, 0.556026125853434) -= (0, 0.556026125853434)
(282033.0, 31.059500000000003) += (0, 1.62617883961791) -= (0, 1.62617883961791)
(742608.0, 37.393) += (0, 1.0189680280093498) -= (0, 1.0189680280093498)
(2023907.0, 42.1775) += (0, 1.7190281640937033) -= (0, 1.7190281640937033)
(5666415.0, 70.85549999999998) += (0, 3.7754661946363783) -= (0, 3.7754661946363783)
(16180286.0, 117.102) += (0, 8.73802264757157) -= (0, 8.73802264757157)
(46845883.0, 126.00499999999997) += (0, 11.811154261967793) -= (0, 11.811154261967793)
(136921875.0, 155.2765) += (0, 16.853181056515243) -= (0, 16.853181056515243)
(402777130.0, 226.353) += (0, 29.89080725697592) -= (0, 29.89080725697592)
(1189983853.0, 338.651) += (0, 28.75277093191468) -= (0, 28.75277093191468)
(3526023792.0, 495.7205000000001) += (0, 43.316425968344305) -= (0, 43.316425968344305)
(10468306005.0, 666.2090000000001) += (0, 82.51859012527514) -= (0, 82.51859012527514)
(31119019396.0, 1018.854736842105) += (0, 94.04956389702878) -= (0, 94.04956389702878)
(92583944059.0, 2178.3650000000002) += (0, 238.1927686841501) -= (0, 238.1927686841501)
(275594455538.0, 3929.9010000000003) += (0, 591.1302373526319) -= (0, 591.1302373526319)
(820613423424.0, 5862.337) += (0, 926.059796279547) -= (0, 926.059796279547)
(2443872086675.0, 9397.674499999997) += (0, 1287.2883178390873) -= (0, 1287.2883178390873)
(7278617927795.0, 11837.223999999998) += (0, 1669.0014278317951) -= (0, 1669.0014278317951)
}; 

\end{loglogaxis}







\end{tikzpicture}
\end{subfigure}
\\
\begin{subfigure}[t]{0.45\linewidth}
    \centering
    \input{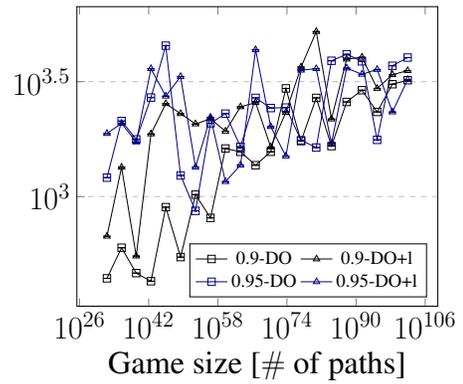}
\end{subfigure}
\hfill
\begin{subfigure}[t]{0.45\linewidth}
    \centering
    \begin{tikzpicture}[scale=\fscale]
\pgfplotsset{compat=1.3}
\begin{loglogaxis}[
    title={\titfont \bf Logistical interdiction on Bakhmut},
    xlabel={\gamesizelabel},
    ylabel style={align=center},
    legend pos=south east,
    ymajorgrids=true,
    grid style=dashed,
    scaled ticks=false, 
    ytick pos=left,
    mark size=\fmarksize,
    label style={font=\labfont},
    tick label style={font=\tickfont},
    legend columns=2,
]

\legend{0.9-DO, 0.9-DO+l, 0.95-DO, 0.95-DO+l}


\addplot[color=black,mark=\domark,error bars/.cd,y dir=both, y explicit]
coordinates {
(2.1494522766197663e+32, 440.59) += (0, 0.0) -= (0, 0.0)
(5.225397220748595e+35, 598.61) += (0, 0.0) -= (0, 0.0)
(1.3071163659871852e+39, 464.17) += (0, 0.0) -= (0, 0.0)
(3.365465007189097e+42, 427.94) += (0, 0.0) -= (0, 0.0)
(8.921980346020252e+45, 899.19) += (0, 0.0) -= (0, 0.0)
(2.435790625385188e+49, 544.99) += (0, 0.0) -= (0, 0.0)
(6.846632229401962e+52, 1019.15) += (0, 0.0) -= (0, 0.0)
(1.979668834273525e+56, 807.36) += (0, 0.0) -= (0, 0.0)
(5.878757434887238e+59, 1618.51) += (0, 0.0) -= (0, 0.0)
(1.788770422497314e+63, 1564.96) += (0, 0.0) -= (0, 0.0)
(5.5613169368594255e+66, 1368.98) += (0, 0.0) -= (0, 0.0)
(1.761258401780068e+70, 1566.53) += (0, 0.0) -= (0, 0.0)
(5.664488833287261e+73, 2955.3) += (0, 0.0) -= (0, 0.0)
(1.8447991944843424e+77, 1742.65) += (0, 0.0) -= (0, 0.0)
(6.068615169315373e+80, 2687.67) += (0, 0.0) -= (0, 0.0)
(2.0121001714187657e+84, 1657.82) += (0, 0.0) -= (0, 0.0)
(6.712161507948501e+87, 2580.71) += (0, 0.0) -= (0, 0.0)
(2.2496399121884463e+91, 2902.74) += (0, 0.0) -= (0, 0.0)
(7.566887186034425e+94, 2334.64) += (0, 0.0) -= (0, 0.0)
(2.552109462040742e+98, 3082.39) += (0, 0.0) -= (0, 0.0)
(8.62522461706051e+101, 3208.9) += (0, 0.0) -= (0, 0.0)
}; 

\addplot[color=black,mark=\dolmark,error bars/.cd,y dir=both, y explicit]
coordinates {
(2.1494522766197663e+32, 671.72) += (0, 0.0) -= (0, 0.0)
(5.225397220748595e+35, 1340.95) += (0, 0.0) -= (0, 0.0)
(1.3071163659871852e+39, 550.72) += (0, 0.0) -= (0, 0.0)
(3.365465007189097e+42, 1873.22) += (0, 0.0) -= (0, 0.0)
(8.921980346020252e+45, 2535.17) += (0, 0.0) -= (0, 0.0)
(2.435790625385188e+49, 2294.65) += (0, 0.0) -= (0, 0.0)
(6.846632229401962e+52, 2069.86) += (0, 0.0) -= (0, 0.0)
(1.979668834273525e+56, 2178.51) += (0, 0.0) -= (0, 0.0)
(5.878757434887238e+59, 1925.11) += (0, 0.0) -= (0, 0.0)
(1.788770422497314e+63, 2457.58) += (0, 0.0) -= (0, 0.0)
(5.5613169368594255e+66, 2564.37) += (0, 0.0) -= (0, 0.0)
(1.761258401780068e+70, 1642.93) += (0, 0.0) -= (0, 0.0)
(5.664488833287261e+73, 2323.28) += (0, 0.0) -= (0, 0.0)
(1.8447991944843424e+77, 3659.16) += (0, 0.0) -= (0, 0.0)
(6.068615169315373e+80, 5222.63) += (0, 0.0) -= (0, 0.0)
(2.0121001714187657e+84, 2175.1) += (0, 0.0) -= (0, 0.0)
(6.712161507948501e+87, 3962.75) += (0, 0.0) -= (0, 0.0)
(2.2496399121884463e+91, 4038.6) += (0, 0.0) -= (0, 0.0)
(7.566887186034425e+94, 2954.05) += (0, 0.0) -= (0, 0.0)
(2.552109462040742e+98, 3391.35) += (0, 0.0) -= (0, 0.0)
(8.62522461706051e+101, 3530.92) += (0, 0.0) -= (0, 0.0)
}; 

\addplot[color=navyblue,mark=\domark,error bars/.cd,y dir=both, y explicit]
coordinates {
(2.1494522766197663e+32, 1208.31) += (0, 0.0) -= (0, 0.0)
(5.225397220748595e+35, 2131.99) += (0, 0.0) -= (0, 0.0)
(1.3071163659871852e+39, 1772.88) += (0, 0.0) -= (0, 0.0)
(3.365465007189097e+42, 2696.59) += (0, 0.0) -= (0, 0.0)
(8.921980346020252e+45, 4535.1) += (0, 0.0) -= (0, 0.0)
(2.435790625385188e+49, 1235.17) += (0, 0.0) -= (0, 0.0)
(6.846632229401962e+52, 866.03) += (0, 0.0) -= (0, 0.0)
(1.979668834273525e+56, 2084.57) += (0, 0.0) -= (0, 0.0)
(5.878757434887238e+59, 2295.33) += (0, 0.0) -= (0, 0.0)
(1.788770422497314e+63, 1646.77) += (0, 0.0) -= (0, 0.0)
(5.5613169368594255e+66, 2689.08) += (0, 0.0) -= (0, 0.0)
(1.761258401780068e+70, 2427.57) += (0, 0.0) -= (0, 0.0)
(5.664488833287261e+73, 2439.28) += (0, 0.0) -= (0, 0.0)
(1.8447991944843424e+77, 1758.53) += (0, 0.0) -= (0, 0.0)
(6.068615169315373e+80, 1634.35) += (0, 0.0) -= (0, 0.0)
(2.0121001714187657e+84, 3899.09) += (0, 0.0) -= (0, 0.0)
(6.712161507948501e+87, 4160.7) += (0, 0.0) -= (0, 0.0)
(2.2496399121884463e+91, 3877.88) += (0, 0.0) -= (0, 0.0)
(7.566887186034425e+94, 1764.97) += (0, 0.0) -= (0, 0.0)
(2.552109462040742e+98, 3708.34) += (0, 0.0) -= (0, 0.0)
(8.62522461706051e+101, 4025.73) += (0, 0.0) -= (0, 0.0)
}; 

\addplot[color=navyblue,mark=\dolmark,error bars/.cd,y dir=both, y explicit]
coordinates {
(2.1494522766197663e+32, 1880.17) += (0, 0.0) -= (0, 0.0)
(5.225397220748595e+35, 2082.07) += (0, 0.0) -= (0, 0.0)
(1.3071163659871852e+39, 1726.98) += (0, 0.0) -= (0, 0.0)
(3.365465007189097e+42, 3602.87) += (0, 0.0) -= (0, 0.0)
(8.921980346020252e+45, 2734.92) += (0, 0.0) -= (0, 0.0)
(2.435790625385188e+49, 3329.89) += (0, 0.0) -= (0, 0.0)
(6.846632229401962e+52, 1343.11) += (0, 0.0) -= (0, 0.0)
(1.979668834273525e+56, 2213.86) += (0, 0.0) -= (0, 0.0)
(5.878757434887238e+59, 1160.93) += (0, 0.0) -= (0, 0.0)
(1.788770422497314e+63, 1370.3) += (0, 0.0) -= (0, 0.0)
(5.5613169368594255e+66, 4354.08) += (0, 0.0) -= (0, 0.0)
(1.761258401780068e+70, 2015.13) += (0, 0.0) -= (0, 0.0)
(5.664488833287261e+73, 1499.72) += (0, 0.0) -= (0, 0.0)
(1.8447991944843424e+77, 3548.18) += (0, 0.0) -= (0, 0.0)
(6.068615169315373e+80, 3601.17) += (0, 0.0) -= (0, 0.0)
(2.0121001714187657e+84, 1702.34) += (0, 0.0) -= (0, 0.0)
(6.712161507948501e+87, 3622.86) += (0, 0.0) -= (0, 0.0)
(2.2496399121884463e+91, 3404.79) += (0, 0.0) -= (0, 0.0)
(7.566887186034425e+94, 3573.84) += (0, 0.0) -= (0, 0.0)
(2.552109462040742e+98, 2333.75) += (0, 0.0) -= (0, 0.0)
(8.62522461706051e+101, 3228.87) += (0, 0.0) -= (0, 0.0)
}; 

\end{loglogaxis}











\end{tikzpicture}
\end{subfigure}
    \caption{Computation times for anti-poaching on Minnewaska State Park and logistical interdiction on grid world and the city of Bakhmut. The complete LP is referred to as \lpname, while the double oracle is designated as \doname for the version incorporating exact best-responses and \dolname for the version involving approximate best-responses. For logistic interdiction, we further compare how the algorithms scale for different values of fixed delay factor $\gamma\in\{0.9, 0.95\}$. The standard Nash LP is unable to compute equilibria even for the smallest logistical interdiction scenarios.}
    \label{fig:runtime2}
\end{figure}

In this section, we present additional empirical results that could not be accommodated within the main text's experimental section. Figure~\ref{fig:runtime2} illustrates how all the algorithms, including the full LP solver and the double oracle with exact or approximate best-responses, scale with respect the number of paths in the corresponding layered graph. We conducted these experiments in scenarios where we controlled computational difficulty through a different parameter or when not all the algorithms were compared in the main text.

The observed results are generally as expected, with a few noteworthy observations. Particularly in the case of logistic interdiction on the map of Bakhmut, the difficulty does not immediately increase with size, although a slight upward trend is noticeable. Additionally, a higher delay factor tends to correlate with longer solving times, particularly for smaller layered graphs, although this is not as pronounced as in grid worlds.

\begin{figure}
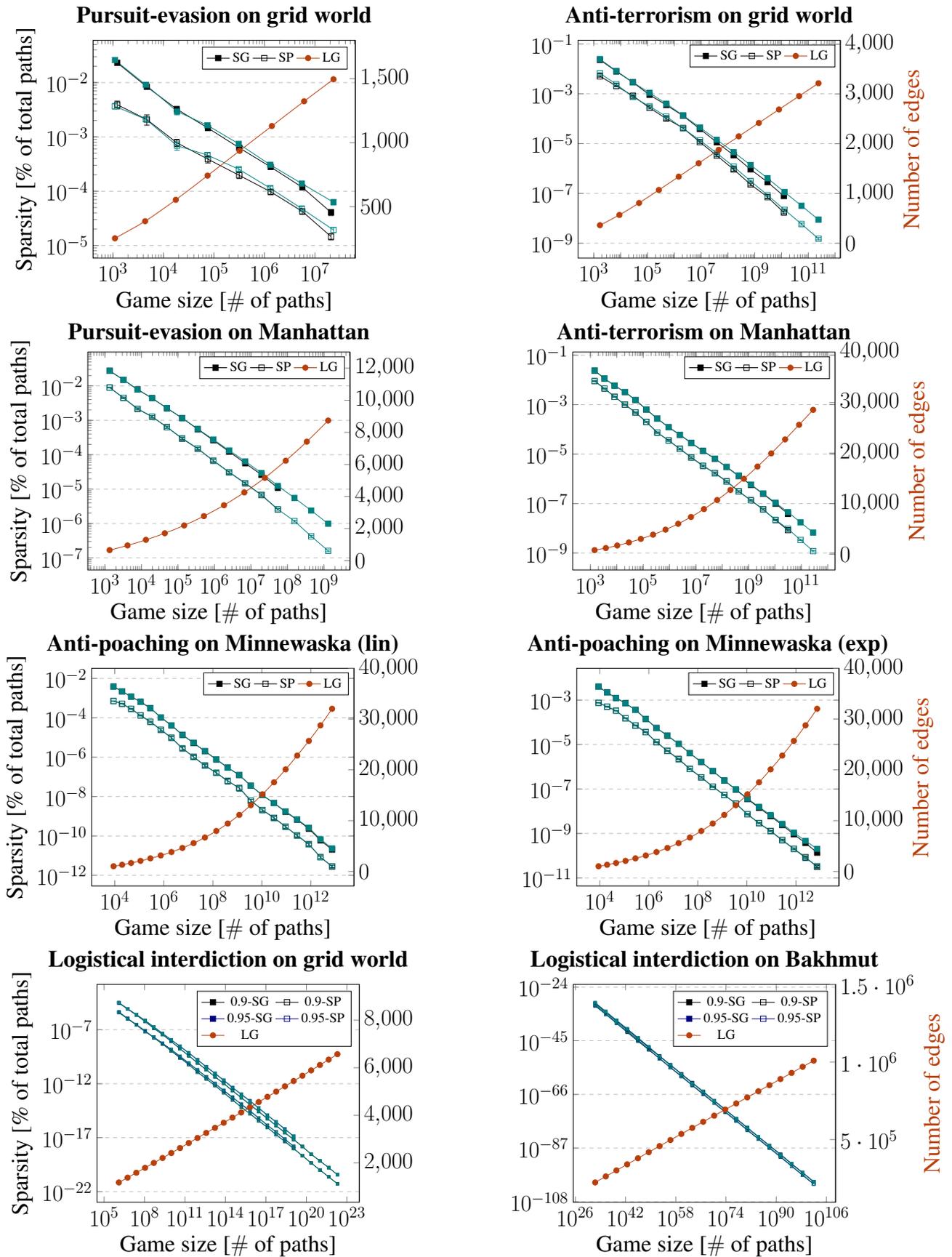

    \centering
\begin{subfigure}[t]{0.47\linewidth}
    \centering
    \begin{tikzpicture}[scale=\fscale]
\pgfplotsset{compat=1.3}
\begin{loglogaxis}[
    title={\titfont \bf Pursuit-evasion on grid world},
    xlabel={\gamesizelabel},
    ylabel={\sparsitylabel},
    label style={font=\labfont},
    tick label style={font=\tickfont},
    ylabel style={align=center},
    ymajorgrids=true,
    grid style=dashed,
    scaled ticks=false, 
    ytick pos=left,
]

\addplot[color=black,mark=square*,error bars/.cd,y dir=both, y explicit]
coordinates {
(1218.25, 0.023136783116676617) += (0, 0.0014232123707259773) -= (0, 0.0014232123707259773)
(4690.7, 0.008360009773112333) += (0, 0.0008981649189390862) -= (0, 0.0008981649189390862)
(18341.333333333332, 0.003241546084559402) += (0, 0.00036826221921117455) -= (0, 0.00036826221921117455)
(75720.26315789473, 0.0014574368466705903) += (0, 0.0001570201588751909) -= (0, 0.0001570201588751909)
(316726.05263157893, 0.0006287133286441903) += (0, 5.257340796489231e-05) -= (0, 5.257340796489231e-05)
(1332211.6315789474, 0.00027884792106495465) += (0, 2.7726609816014805e-05) -= (0, 2.7726609816014805e-05)
(5626509.2105263155, 0.0001175251534375624) += (0, 1.1680336480270165e-05) -= (0, 1.1680336480270165e-05)
(20802633.545454547, 4.0705678912562396e-05) += (0, 5.06259993827696e-06) -= (0, 5.06259993827696e-06)
}; \label{plot_one}
\addplot[color=black,mark=square,error bars/.cd,y dir=both, y explicit]
coordinates {
(1218.25, 0.003903423293885401) += (0, 0.0006376284522985141) -= (0, 0.0006376284522985141)
(4690.7, 0.0020861617197999774) += (0, 0.00044439423347570397) -= (0, 0.00044439423347570397)
(18341.333333333332, 0.000781361022546474) += (0, 0.00011951219912234135) -= (0, 0.00011951219912234135)
(75720.26315789473, 0.00038921278593182575) += (0, 5.7639044056397265e-05) -= (0, 5.7639044056397265e-05)
(316726.05263157893, 0.0001960667960398447) += (0, 2.4977721702925037e-05) -= (0, 2.4977721702925037e-05)
(1332211.6315789474, 9.732805372611934e-05) += (0, 1.0644499264695235e-05) -= (0, 1.0644499264695235e-05)
(5626509.2105263155, 4.2371483871208446e-05) += (0, 4.750220898037465e-06) -= (0, 4.750220898037465e-06)
(20802633.545454547, 1.4569235083147282e-05) += (0, 1.926230683596864e-06) -= (0, 1.926230683596864e-06)
}; \label{plot_two}

\addplot[color=\minnewaskaexpcolor,mark=square*,error bars/.cd,y dir=both, y explicit]
coordinates {
(1095.5, 0.0260063408797304) += (0, 0.0017717253119821932) -= (0, 0.0017717253119821932)
(4496.333333333333, 0.009090399471429292) += (0, 0.0004986577766935102) -= (0, 0.0004986577766935102)
(18579.428571428572, 0.0029462706018502836) += (0, 0.00041017087018517415) -= (0, 0.00041017087018517415)
(74154.05, 0.0016342301669724155) += (0, 0.00010049903029231217) -= (0, 0.00010049903029231217)
(309773.4, 0.0007442625833940139) += (0, 4.577354207626687e-05) -= (0, 4.577354207626687e-05)
(1301483.7, 0.00030643688923782085) += (0, 1.996225811799811e-05) -= (0, 1.996225811799811e-05)
(5491216.95, 0.00013843324943026343) += (0, 1.1906850218274654e-05) -= (0, 1.1906850218274654e-05)
(23245349.8, 6.275886380062369e-05) += (0, 6.056078256004004e-06) -= (0, 6.056078256004004e-06)
};

\addplot[color=\minnewaskaexpcolor,mark=square,error bars/.cd,y dir=both, y explicit]
coordinates {
(1095.5, 0.0036646600842490627) += (0, 0.00042230640732529654) -= (0, 0.00042230640732529654)
(4496.333333333333, 0.002109064559457439) += (0, 0.00020348794541698263) -= (0, 0.00020348794541698263)
(18579.428571428572, 0.0006891941146467264) += (0, 0.00012180503606582812) -= (0, 0.00012180503606582812)
(74154.05, 0.00046287146245581323) += (0, 4.087830316240969e-05) -= (0, 4.087830316240969e-05)
(309773.4, 0.0002544296975506981) += (0, 1.9666803089496127e-05) -= (0, 1.9666803089496127e-05)
(1301483.7, 0.00011390074738469453) += (0, 1.0081531913826484e-05) -= (0, 1.0081531913826484e-05)
(5491216.95, 4.774635665242332e-05) += (0, 4.762864280844832e-06) -= (0, 4.762864280844832e-06)
(23245349.8, 1.9332146084757142e-05) += (0, 2.0447882919878004e-06) -= (0, 2.0447882919878004e-06)
};

\end{loglogaxis}

\begin{semilogxaxis}[
  axis y line*=right,
  axis x line=none,
  ymax=1800,
  legend pos=north east,
  label style={font=\labfont},
  tick label style={font=\tickfont},
  legend columns=3
]
\addlegendimage{/pgfplots/refstyle=plot_one}\addlegendentry{\subgamename}
\addlegendimage{/pgfplots/refstyle=plot_two}\addlegendentry{\supportname}
\addplot[smooth,mark=*,\sizescolor]
  coordinates{
(1218.25, 253.25)
(4690.7, 387.1)
(18341.333333333332, 553.9333333333333)
(75720.26315789473, 743.421052631579)
(316726.05263157893, 936.8421052631579)
(1332211.6315789474, 1130.4736842105262)
(5626509.2105263155, 1324.1052631578948)
(20802633.545454547, 1495.7272727272727)
}; \addlegendentry{\graphsizename}
\end{semilogxaxis}
\end{tikzpicture}
\end{subfigure}
\hfill
\begin{subfigure}[t]{0.47\linewidth}
    \centering
    \begin{tikzpicture}[scale=\fscale]
\pgfplotsset{compat=1.3}
\begin{loglogaxis}[
    title={\titfont \bf Anti-terrorism on grid world},
    xlabel={\gamesizelabel},
    label style={font=\labfont},
    tick label style={font=\tickfont},
    ylabel style={align=center},
    legend pos=north east,
    ymajorgrids=true,
    grid style=dashed,
    ytick pos=left,
    legend columns = 3,
    xtick={1000,10000,100000,1000000,10000000,100000000,1000000000,10000000000,100000000000,1000000000000},
    xticklabels={$10^3$,,$10^5$,,$10^7$,,$10^9$,,$10^{11}$}
]

\addplot[color=black,mark=square*,error bars/.cd,y dir=both, y explicit]
coordinates {
(1694.25, 0.023118818610948326) += (0, 0.0010386562040405145) -= (0, 0.0010386562040405145)
(6918.1, 0.007941658554121716) += (0, 0.0002712276869363806) -= (0, 0.0002712276869363806)
(28627.2, 0.002883053733911604) += (0, 0.00017688754084429494) -= (0, 0.00017688754084429494)
(119614.85, 0.0009036988536371222) += (0, 5.500663977809869e-05) -= (0, 5.500663977809869e-05)
(503341.5, 0.00035331981940501524) += (0, 3.246914017110816e-05) -= (0, 3.246914017110816e-05)
(2129194.45, 0.00013110976617837334) += (0, 1.2367080630783937e-05) -= (0, 1.2367080630783937e-05)
(9042422.55, 3.850534877357736e-05) += (0, 3.230680814496195e-06) -= (0, 3.230680814496195e-06)
(38519721.85, 1.1312707749334538e-05) += (0, 9.275813093323522e-07) -= (0, 9.275813093323522e-07)
(164489212.7, 3.417421746637419e-06) += (0, 4.273453337878623e-07) -= (0, 4.273453337878623e-07)
(703808134.5, 9.208002561534517e-07) += (0, 8.424400958319839e-08) -= (0, 8.424400958319839e-08)
(3016439214.4, 2.8413015945839076e-07) += (0, 3.315637453107939e-08) -= (0, 3.315637453107939e-08)
(12233356088.705883, 7.795989576963934e-08) += (0, 7.788393548319634e-09) -= (0, 7.788393548319634e-09)
}; \label{plot_5}
\addplot[color=black,mark=square,error bars/.cd,y dir=both, y explicit]
coordinates {
(1694.25, 0.005067420692117928) += (0, 0.00038708952246671537) -= (0, 0.00038708952246671537)
(6918.1, 0.0020577594489970576) += (0, 0.00010281529532346044) -= (0, 0.00010281529532346044)
(28627.2, 0.0008316113887059157) += (0, 5.132311597970281e-05) -= (0, 5.132311597970281e-05)
(119614.85, 0.00027348875479717394) += (0, 2.8644040623449414e-05) -= (0, 2.8644040623449414e-05)
(503341.5, 0.0001006818411377886) += (0, 1.4428811576220048e-05) -= (0, 1.4428811576220048e-05)
(2129194.45, 4.23013381267675e-05) += (0, 4.972061875648779e-06) -= (0, 4.972061875648779e-06)
(9042422.55, 1.1459476124939913e-05) += (0, 1.1851203296476838e-06) -= (0, 1.1851203296476838e-06)
(38519721.85, 3.295188961067028e-06) += (0, 3.7147610807971533e-07) -= (0, 3.7147610807971533e-07)
(164489212.7, 9.192185084731103e-07) += (0, 1.2748037805401095e-07) -= (0, 1.2748037805401095e-07)
(703808134.5, 2.2838122111096713e-07) += (0, 2.6992280832965928e-08) -= (0, 2.6992280832965928e-08)
(3016439214.4, 7.144058975711671e-08) += (0, 9.911439167565622e-09) -= (0, 9.911439167565622e-09)
(12233356088.705883, 1.7169437185211586e-08) += (0, 2.103865405518254e-09) -= (0, 2.103865405518254e-09)
}; \label{plot_6}

\addplot[color=\minnewaskaexpcolor,mark=square*,error bars/.cd,y dir=both, y explicit]
coordinates {
(1694.25, 0.02473903769399635) += (0, 0.0014305457169460141) -= (0, 0.0014305457169460141)
(6918.1, 0.008295852687310576) += (0, 0.00047097173671065015) -= (0, 0.00047097173671065015)
(28627.2, 0.0029085225542840585) += (0, 0.0001691421159026744) -= (0, 0.0001691421159026744)
(119614.85, 0.0010907343458169377) += (0, 7.112402383254502e-05) -= (0, 7.112402383254502e-05)
(503341.5, 0.0003987815852808034) += (0, 2.7022825633608e-05) -= (0, 2.7022825633608e-05)
(2129194.45, 0.00013579150490142185) += (0, 8.66346104017897e-06) -= (0, 8.66346104017897e-06)
(9042422.55, 4.439973889884938e-05) += (0, 3.157155473727073e-06) -= (0, 3.157155473727073e-06)
(38519721.85, 1.425110725189768e-05) += (0, 9.993050630545283e-07) -= (0, 9.993050630545283e-07)
(164489212.7, 4.52284796546007e-06) += (0, 3.1584766835490455e-07) -= (0, 3.1584766835490455e-07)
(703808134.5, 1.37734357716832e-06) += (0, 1.0415092313969572e-07) -= (0, 1.0415092313969572e-07)
(3016439214.4, 4.056776414216335e-07) += (0, 3.0372081450566046e-08) -= (0, 3.0372081450566046e-08)
(12946629534.6, 1.128837926397556e-07) += (0, 8.732174782970081e-09) -= (0, 8.732174782970081e-09)
(55637189639.0, 3.1781898720144275e-08) += (0, 2.7492414377774357e-09) -= (0, 2.7492414377774357e-09)
(239366196479.3, 8.82873115756093e-09) += (0, 7.954367210335098e-10) -= (0, 7.954367210335098e-10)
};

\addplot[color=\minnewaskaexpcolor,mark=square,error bars/.cd,y dir=both, y explicit]
coordinates {
(1694.25, 0.0066768451338583765) += (0, 0.0004921110303579202) -= (0, 0.0004921110303579202)
(6918.1, 0.0024038156790489947) += (0, 0.00022837125958688495) -= (0, 0.00022837125958688495)
(28627.2, 0.0007476639873896878) += (0, 6.67871384825748e-05) -= (0, 6.67871384825748e-05)
(119614.85, 0.00031311385650649654) += (0, 2.895453831085811e-05) -= (0, 2.895453831085811e-05)
(503341.5, 0.00012207396589633407) += (0, 9.923491007566928e-06) -= (0, 9.923491007566928e-06)
(2129194.45, 4.097842315009115e-05) += (0, 2.8522192892930433e-06) -= (0, 2.8522192892930433e-06)
(9042422.55, 1.3517895185822338e-05) += (0, 1.0030982892316512e-06) -= (0, 1.0030982892316512e-06)
(38519721.85, 4.1317280437599655e-06) += (0, 3.2019212648021775e-07) -= (0, 3.2019212648021775e-07)
(164489212.7, 1.195331070912313e-06) += (0, 8.844041897569672e-08) -= (0, 8.844041897569672e-08)
(703808134.5, 3.1413836965881687e-07) += (0, 2.4824961856223172e-08) -= (0, 2.4824961856223172e-08)
(3016439214.4, 8.476766204611325e-08) += (0, 6.982298272363902e-09) -= (0, 6.982298272363902e-09)
(12946629534.6, 2.19878548479237e-08) += (0, 2.1085623800936963e-09) -= (0, 2.1085623800936963e-09)
(55637189639.0, 5.862594742408781e-09) += (0, 5.783253239144083e-10) -= (0, 5.783253239144083e-10)
(239366196479.3, 1.5106353805805077e-09) += (0, 1.6400516774793314e-10) -= (0, 1.6400516774793314e-10)
};

\end{loglogaxis}

\begin{semilogxaxis}[
  axis y line*=right,
  axis x line=none,
  ymin=-300,
  ymax=4080,
  ylabel = \textcolor{\sizescolor}{Number of edges},
  label style={font=\labfont},
  tick label style={font=\tickfont},
  legend columns = 3,
]
\addlegendimage{/pgfplots/refstyle=plot_5}\addlegendentry{\subgamename}
\addlegendimage{/pgfplots/refstyle=plot_6}\addlegendentry{\supportname}
\addplot[smooth,mark=*,\sizescolor]
  coordinates{
(1694.25, 361.6)
(6918.1, 567.7)
(28627.2, 809.15)
(119614.85, 1070.6)
(503341.5, 1339.15)
(2129194.45, 1609.05)
(9042422.55, 1879.0)
(38519721.85, 2148.95)
(164489212.7, 2418.9)
(703808134.5, 2688.85)
(3016439214.4, 2958.8)
(12233356088.705883, 3217.529411764706)
};\addlegendentry{\graphsizename}
\end{semilogxaxis}
\end{tikzpicture}
\end{subfigure}
\\
\begin{subfigure}[t]{0.47\linewidth}
    \centering
    \begin{tikzpicture}[scale=\fscale]
\pgfplotsset{compat=1.3}
\begin{loglogaxis}[
    title={\titfont \bf Pursuit-evasion on Manhattan},
    xlabel={\gamesizelabel},
    ylabel={\sparsitylabel},
    label style={font=\labfont},
    tick label style={font=\tickfont},
    ylabel style={align=center},
    legend pos=north east,
    ymajorgrids=true,
    grid style=dashed,
    scaled ticks=false,
    ytick pos=left,
]

\addplot[color=black,mark=square*,error bars/.cd,y dir=both, y explicit]
coordinates {
(1342.0, 0.02738450074515648) += (0, 0.0010715140763854296) -= (0, 0.0010715140763854296)
(3206.0, 0.015018714909544604) += (0, 0.0006300682149836946) -= (0, 0.0006300682149836946)
(7824.0, 0.007866820040899798) += (0, 0.00030295157461421855) -= (0, 0.00030295157461421855)
(19568.0, 0.004400040883074408) += (0, 0.00027985570929567875) -= (0, 0.00027985570929567875)
(50088.0, 0.0022310733109726882) += (0, 0.00014152075175685725) -= (0, 0.00014152075175685725)
(130714.0, 0.0011601664703092247) += (0, 5.7823117806696214e-05) -= (0, 5.7823117806696214e-05)
(346464.0, 0.0005514281426064469) += (0, 2.6647834929202172e-05) -= (0, 2.6647834929202172e-05)
(930446.0, 0.00026170245237230313) += (0, 1.3799401006276844e-05) -= (0, 1.3799401006276844e-05)
(2529336.0, 0.00012305601153820607) += (0, 6.8222632847823985e-06) -= (0, 6.8222632847823985e-06)
(6959182.0, 5.663021889641628e-05) += (0, 3.484839229517093e-06) -= (0, 3.484839229517093e-06)
(19380428.0, 2.6150093279673704e-05) += (0, 1.6097888753921267e-06) -= (0, 1.6097888753921267e-06)
(54615200.0, 1.0810562858537163e-05) += (0, 4.97475813662821e-07) -= (0, 4.97475813662821e-07)
}; \label{plot_15}
\addplot[color=black,mark=square,error bars/.cd,y dir=both, y explicit]
coordinates {
(1342.0, 0.008941877794336809) += (0, 0.00024773098591817017) -= (0, 0.00024773098591817017)
(3206.0, 0.0044915782907049276) += (0, 0.00020541080631207342) -= (0, 0.00020541080631207342)
(7824.0, 0.0021344580777096114) += (0, 0.0001288503179749068) -= (0, 0.0001288503179749068)
(19568.0, 0.0012699304987735078) += (0, 9.761791548716775e-05) -= (0, 9.761791548716775e-05)
(50088.0, 0.0006338843635202044) += (0, 4.5293009706250616e-05) -= (0, 4.5293009706250616e-05)
(130714.0, 0.0002930061049313768) += (0, 2.1928880283399275e-05) -= (0, 2.1928880283399275e-05)
(346464.0, 0.00015008774360395307) += (0, 1.271112034542713e-05) -= (0, 1.271112034542713e-05)
(930446.0, 6.711834969466257e-05) += (0, 5.836090125134776e-06) -= (0, 5.836090125134776e-06)
(2529336.0, 3.085790104596622e-05) += (0, 2.8216146606428676e-06) -= (0, 2.8216146606428676e-06)
(6959182.0, 1.477185105950671e-05) += (0, 1.2855265960509324e-06) -= (0, 1.2855265960509324e-06)
(19380428.0, 6.7155379643834505e-06) += (0, 4.4995370814583804e-07) -= (0, 4.4995370814583804e-07)
(54615200.0, 2.568207346942552e-06) += (0, 1.3326030083885746e-07) -= (0, 1.3326030083885746e-07)
};\label{plot_16}

\addplot[color=\minnewaskaexpcolor,mark=square*,error bars/.cd,y dir=both, y explicit]
coordinates {
(1342.0, 0.02738450074515648) += (0, 0.0010715140763854296) -= (0, 0.0010715140763854296)
(3206.0, 0.015018714909544605) += (0, 0.0006300682149836946) -= (0, 0.0006300682149836946)
(7824.0, 0.007834867075664621) += (0, 0.00028999197996865383) -= (0, 0.00028999197996865383)
(19568.0, 0.0043693785772690105) += (0, 0.00027260565675060755) -= (0, 0.00027260565675060755)
(50088.0, 0.0022729995208433155) += (0, 0.00014161262935374284) -= (0, 0.00014161262935374284)
(130714.0, 0.0011632265862876203) += (0, 6.191904940102514e-05) -= (0, 6.191904940102514e-05)
(346464.0, 0.0005664369169668421) += (0, 3.260834753229244e-05) -= (0, 3.260834753229244e-05)
(930446.0, 0.0002763190985828302) += (0, 1.6188831282467176e-05) -= (0, 1.6188831282467176e-05)
(2529336.0, 0.0001332167810049752) += (0, 7.841260547531893e-06) -= (0, 7.841260547531893e-06)
(6959182.0, 6.393711214910028e-05) += (0, 3.920053325333883e-06) -= (0, 3.920053325333883e-06)
(19380428.0, 2.9313078121907315e-05) += (0, 1.7623001959000841e-06) -= (0, 1.7623001959000841e-06)
(54615200.0, 1.2406802501867614e-05) += (0, 5.40157207577821e-07) -= (0, 5.40157207577821e-07)
(155646044.0, 5.524714781700457e-06) += (0, 2.640635581997197e-07) -= (0, 2.640635581997197e-07)
(448200828.0, 2.3788443335941358e-06) += (0, 9.900575022088871e-08) -= (0, 9.900575022088871e-08)
(1303042068.0, 9.837612955264745e-07) += (0, 3.8150294071271094e-08) -= (0, 3.8150294071271094e-08)
};

\addplot[color=\minnewaskaexpcolor,mark=square,error bars/.cd,y dir=both, y explicit]
coordinates {
(1342.0, 0.008941877794336809) += (0, 0.00024773098591817017) -= (0, 0.00024773098591817017)
(3206.0, 0.004491578290704927) += (0, 0.0002054108063120734) -= (0, 0.0002054108063120734)
(7824.0, 0.0021728016359918202) += (0, 0.00013437052224893835) -= (0, 0.00013437052224893835)
(19568.0, 0.0012827064595257562) += (0, 0.00010071067690733703) -= (0, 0.00010071067690733703)
(50088.0, 0.0006249001756907843) += (0, 4.457459716977617e-05) -= (0, 4.457459716977617e-05)
(130714.0, 0.0003002738803800664) += (0, 2.4216998245863317e-05) -= (0, 2.4216998245863317e-05)
(346464.0, 0.00015138657984667956) += (0, 1.2323132355519516e-05) -= (0, 1.2323132355519516e-05)
(930446.0, 6.711834969466255e-05) += (0, 6.030261871124526e-06) -= (0, 6.030261871124526e-06)
(2529336.0, 3.077882891003806e-05) += (0, 2.823188673545786e-06) -= (0, 2.823188673545786e-06)
(6959182.0, 1.4965839375949646e-05) += (0, 1.3157598850757915e-06) -= (0, 1.3157598850757915e-06)
(19380428.0, 6.774876179205122e-06) += (0, 4.6745519770286405e-07) -= (0, 4.6745519770286405e-07)
(54615200.0, 2.57986787560972e-06) += (0, 1.39240963661162e-07) -= (0, 1.39240963661162e-07)
(155646044.0, 1.1712472435213321e-06) += (0, 7.195889592019994e-08) -= (0, 7.195889592019994e-08)
(448200828.0, 4.250326820012033e-07) += (0, 2.9350787054478847e-08) -= (0, 2.9350787054478847e-08)
(1303042068.0, 1.608453329644286e-07) += (0, 9.933601095721384e-09) -= (0, 9.933601095721384e-09)
};

\end{loglogaxis}

\begin{semilogxaxis}[
  axis y line*=right,
  axis x line=none,
  scaled y ticks=false,
  ymax=13000,
  label style={font=\labfont},
  tick label style={font=\tickfont},
  legend columns=3
]

\addlegendimage{/pgfplots/refstyle=plot_15}\addlegendentry{\subgamename}
\addlegendimage{/pgfplots/refstyle=plot_16}\addlegendentry{\supportname}

\addplot[smooth,mark=*,\sizescolor]
  coordinates{
(1342.0, 662.0)
(3206.0, 952.0)
(7824.0, 1302.0)
(19568.0, 1712.0)
(50088.0, 2198.0)
(130714.0, 2778.0)
(346464.0, 3456.0)
(930446.0, 4254.0)
(2529336.0, 5168.0)
(6959182.0, 6232.0)
(19380428.0, 7420.0)
(54615200.0, 8736.0)
}; \addlegendentry{\graphsizename}
\end{semilogxaxis}
\end{tikzpicture}
\end{subfigure}
\hfill
\begin{subfigure}[t]{0.47\linewidth}
    \centering
    \begin{tikzpicture}[scale=\fscale]
\pgfplotsset{compat=1.3}
\begin{loglogaxis}[
    title={\titfont \bf Anti-terrorism on Manhattan},
    xlabel={\gamesizelabel},
    label style={font=\labfont},
    tick label style={font=\tickfont},
    ylabel style={align=center},
    legend pos=north east,
    ymajorgrids=true,
    grid style=dashed,
    xtick={1000,10000,100000,1000000,10000000,100000000,1000000000,10000000000,100000000000,1000000000000},
    xticklabels={$10^3$,,$10^5$,,$10^7$,,$10^9$,,$10^{11}$},
]

\addplot[color=black,mark=square*,error bars/.cd,y dir=both, y explicit]
coordinates {
(1443.0, 0.023804573804573806) += (0, 0.0007612638718587774) -= (0, 0.0007612638718587774)
(3432.0, 0.011305361305361304) += (0, 0.0003167968074603098) -= (0, 0.0003167968074603098)
(8276.0, 0.005860318994683421) += (0, 0.00021265517270380036) -= (0, 0.00021265517270380036)
(20369.0, 0.003232853846531494) += (0, 0.000138276933326781) -= (0, 0.000138276933326781)
(51285.0, 0.001544311202105879) += (0, 6.856931428892277e-05) -= (0, 6.856931428892277e-05)
(131902.0, 0.0006364573698655063) += (0, 3.911458178826368e-05) -= (0, 3.911458178826368e-05)
(345447.0, 0.00027254542665010843) += (0, 1.7117206475055636e-05) -= (0, 1.7117206475055636e-05)
(918545.0, 0.00012193196849365029) += (0, 7.679863987715358e-06) -= (0, 7.679863987715358e-06)
(2475204.0, 6.0621265964340715e-05) += (0, 4.356367989612997e-06) -= (0, 4.356367989612997e-06)
(6754240.0, 2.8174894584734928e-05) += (0, 2.2965785175805696e-06) -= (0, 2.2965785175805696e-06)
(18659377.0, 1.3457040929072816e-05) += (0, 8.714486444161051e-07) -= (0, 8.714486444161051e-07)
(52177600.0, 6.533454969182177e-06) += (0, 4.7421924831925317e-07) -= (0, 4.7421924831925317e-07)
(147618008.0, 2.98540812175165e-06) += (0, 1.8455111306020542e-07) -= (0, 1.8455111306020542e-07)
(422233844.0, 1.2992800264490405e-06) += (0, 7.520757128781134e-08) -= (0, 7.520757128781134e-08)
(1220032270.0, 5.695750982062139e-07) += (0, 3.246869590146081e-08) -= (0, 3.246869590146081e-08)
(3558575559.0, 2.458286990106313e-07) += (0, 1.274616748814802e-08) -= (0, 1.274616748814802e-08)
(10472256655.0, 9.95462313953884e-08) += (0, 4.939867380059127e-09) -= (0, 4.939867380059127e-09)
(31084654512.0, 3.849058918138465e-08) += (0, 1.7975935975312533e-09) -= (0, 1.7975935975312533e-09)
};\label{plot_3}
\addplot[color=black,mark=square,error bars/.cd,y dir=both, y explicit]
coordinates {
(1443.0, 0.008974358974358974) += (0, 0.00041905511511455997) -= (0, 0.00041905511511455997)
(3432.0, 0.004487179487179488) += (0, 0.00018233178686911851) -= (0, 0.00018233178686911851)
(8276.0, 0.002066215563073949) += (0, 9.63864240005251e-05) -= (0, 9.63864240005251e-05)
(20369.0, 0.0010015219205655652) += (0, 4.5944142716135445e-05) -= (0, 4.5944142716135445e-05)
(51285.0, 0.0004835721945988106) += (0, 2.309739025138221e-05) -= (0, 2.309739025138221e-05)
(131902.0, 0.00019939045655107578) += (0, 1.3212875900356606e-05) -= (0, 1.3212875900356606e-05)
(345447.0, 7.454110181880288e-05) += (0, 4.4195938110861e-06) -= (0, 4.4195938110861e-06)
(918545.0, 3.587194965951587e-05) += (0, 2.694022943130219e-06) -= (0, 2.694022943130219e-06)
(2475204.0, 1.6301686648858038e-05) += (0, 1.3219560973010934e-06) -= (0, 1.3219560973010934e-06)
(6754240.0, 7.2324939593499775e-06) += (0, 5.815338417170923e-07) -= (0, 5.815338417170923e-07)
(18659377.0, 3.341483480397014e-06) += (0, 2.4501407928658584e-07) -= (0, 2.4501407928658584e-07)
(52177600.0, 1.7152954524546936e-06) += (0, 1.419602375063506e-07) -= (0, 1.419602375063506e-07)
(147618008.0, 7.912313787624068e-07) += (0, 6.540342567487868e-08) -= (0, 6.540342567487868e-08)
(422233844.0, 3.108467070204822e-07) += (0, 2.273522620232468e-08) -= (0, 2.273522620232468e-08)
(1220032270.0, 1.3634885247748403e-07) += (0, 9.560400038942898e-09) -= (0, 9.560400038942898e-09)
(3558575559.0, 5.84925053715854e-08) += (0, 4.066177726244252e-09) -= (0, 4.066177726244252e-09)
(10472256655.0, 2.1440108200773814e-08) += (0, 1.2037862758493731e-09) -= (0, 1.2037862758493731e-09)
(31084654512.0, 8.432885533025264e-09) += (0, 5.092064164175095e-10) -= (0, 5.092064164175095e-10)
};\label{plot_4}

\addplot[color=\minnewaskaexpcolor,mark=square*,error bars/.cd,y dir=both, y explicit]
coordinates {
(1443.0, 0.023804573804573802) += (0, 0.0007612638718587774) -= (0, 0.0007612638718587774)
(3432.0, 0.011305361305361306) += (0, 0.0003167968074603098) -= (0, 0.0003167968074603098)
(8276.0, 0.005860318994683423) += (0, 0.00021265517270380033) -= (0, 0.00021265517270380033)
(20369.0, 0.003232853846531493) += (0, 0.000138276933326781) -= (0, 0.000138276933326781)
(51285.0, 0.001544311202105879) += (0, 6.856931428892277e-05) -= (0, 6.856931428892277e-05)
(131902.0, 0.0006410062015738958) += (0, 4.059843569903964e-05) -= (0, 4.059843569903964e-05)
(345447.0, 0.00027240068664657674) += (0, 1.693930024591562e-05) -= (0, 1.693930024591562e-05)
(918545.0, 0.00012231300589519298) += (0, 7.9340655230621e-06) -= (0, 7.9340655230621e-06)
(2475204.0, 6.0479863477919396e-05) += (0, 4.216644136900928e-06) -= (0, 4.216644136900928e-06)
(6754240.0, 2.796021462074193e-05) += (0, 2.139903731412359e-06) -= (0, 2.139903731412359e-06)
(18659377.0, 1.3266788060501702e-05) += (0, 8.408192859110735e-07) -= (0, 8.408192859110735e-07)
(52177600.0, 6.5717855938180365e-06) += (0, 4.272525534006154e-07) -= (0, 4.272525534006154e-07)
(147618008.0, 3.006746981709711e-06) += (0, 1.8665803942623827e-07) -= (0, 1.8665803942623827e-07)
(422233844.0, 1.3590810119901236e-06) += (0, 8.181493271646615e-08) -= (0, 8.181493271646615e-08)
(1220032270.0, 5.855172994727425e-07) += (0, 3.0622201518238655e-08) -= (0, 3.0622201518238655e-08)
(3558575559.0, 2.5369701585139223e-07) += (0, 1.2438923411331074e-08) -= (0, 1.2438923411331074e-08)
(10472256655.0, 1.0750786932465607e-07) += (0, 5.415573964693651e-09) -= (0, 5.415573964693651e-09)
(31084654512.0, 4.512408738933153e-08) += (0, 2.2260581095947015e-09) -= (0, 2.2260581095947015e-09)
(93063067607.0, 1.7373051579514008e-08) += (0, 8.334525768102719e-10) -= (0, 8.334525768102719e-10)
(281058342216.0, 6.702815999296574e-09) += (0, 2.312503716137742e-10) -= (0, 2.312503716137742e-10)
};

\addplot[color=\minnewaskaexpcolor,mark=square,error bars/.cd,y dir=both, y explicit]
coordinates {
(1443.0, 0.008974358974358974) += (0, 0.00041905511511455997) -= (0, 0.00041905511511455997)
(3432.0, 0.004487179487179489) += (0, 0.00018233178686911854) -= (0, 0.00018233178686911854)
(8276.0, 0.002066215563073949) += (0, 9.63864240005251e-05) -= (0, 9.63864240005251e-05)
(20369.0, 0.0010015219205655652) += (0, 4.5944142716135445e-05) -= (0, 4.5944142716135445e-05)
(51285.0, 0.0004835721945988106) += (0, 2.309739025138221e-05) -= (0, 2.309739025138221e-05)
(131902.0, 0.00019863231793301083) += (0, 1.2788212359481581e-05) -= (0, 1.2788212359481581e-05)
(345447.0, 7.381740180114462e-05) += (0, 4.139398954534487e-06) -= (0, 4.139398954534487e-06)
(918545.0, 3.576308183050369e-05) += (0, 2.5375842308221452e-06) -= (0, 2.5375842308221452e-06)
(2475204.0, 1.6321887004061077e-05) += (0, 1.254306885300185e-06) -= (0, 1.254306885300185e-06)
(6754240.0, 7.410160136447623e-06) += (0, 6.217272489016251e-07) -= (0, 6.217272489016251e-07)
(18659377.0, 3.4165127806785835e-06) += (0, 2.5857015325202427e-07) -= (0, 2.5857015325202427e-07)
(52177600.0, 1.6779230934347304e-06) += (0, 1.451849390208764e-07) -= (0, 1.451849390208764e-07)
(147618008.0, 7.7226350324413e-07) += (0, 6.658264651279475e-08) -= (0, 6.658264651279475e-08)
(422233844.0, 3.081230977780171e-07) += (0, 2.4184283324156386e-08) -= (0, 2.4184283324156386e-08)
(1220032270.0, 1.353242894140825e-07) += (0, 1.0056732893416808e-08) -= (0, 1.0056732893416808e-08)
(3558575559.0, 5.703124652972978e-08) += (0, 3.8476615301055655e-09) -= (0, 3.8476615301055655e-09)
(10472256655.0, 2.20439593494665e-08) += (0, 1.304051364185595e-09) -= (0, 1.304051364185595e-09)
(31084654512.0, 9.505226012809761e-09) += (0, 4.957103579958828e-10) -= (0, 4.957103579958828e-10)
(93063067607.0, 3.378014814428763e-09) += (0, 2.0329295409757558e-10) -= (0, 2.0329295409757558e-10)
(281058342216.0, 1.197992842689489e-09) += (0, 7.314107376594974e-11) -= (0, 7.314107376594974e-11)
};

\end{loglogaxis}

\begin{semilogxaxis}[
  axis y line*=right,
  axis x line=none,
  scaled y ticks=false,
  ymax=40000,
  ylabel = \textcolor{\sizescolor}{Number of edges},
  label style={font=\labfont},
  tick label style={font=\tickfont},
  legend columns=3
]
\addlegendimage{/pgfplots/refstyle=plot_3}\addlegendentry{\subgamename}
\addlegendimage{/pgfplots/refstyle=plot_4}\addlegendentry{\supportname}
\addplot[smooth,mark=*,\sizescolor]
  coordinates{
(1443.0, 805.0)
(3432.0, 1212.0)
(8276.0, 1714.0)
(20369.0, 2313.0)
(51285.0, 3021.0)
(131902.0, 3861.0)
(345447.0, 4850.0)
(918545.0, 6010.0)
(2475204.0, 7358.0)
(6754240.0, 8920.0)
(18659377.0, 10705.0)
(52177600.0, 12702.0)
(147618008.0, 14917.0)
(422233844.0, 17342.0)
(1220032270.0, 19954.0)
(3558575559.0, 22723.0)
(10472256655.0, 25608.0)
(31084654512.0, 28574.0)
}; \addlegendentry{\graphsizename}
\end{semilogxaxis}
\end{tikzpicture}
\end{subfigure}
\\
\begin{subfigure}[t]{0.47\linewidth}
    \centering
    \input{figs/e_bopl_park_lin_sizes}
\end{subfigure}
\hfill
\begin{subfigure}[t]{0.47\linewidth}
    \centering
    \input{figs/e_bopl_park_exp_sizes}
\end{subfigure}
\begin{subfigure}[t]{0.47\linewidth}
    \centering
    \input{figs/e_li_grid_sizes}
\end{subfigure}
\hfill
\begin{subfigure}[t]{0.47\linewidth}
    \centering
    \input{figs/e_li_bakhmut_sizes}
\end{subfigure}
    \caption{Comparison of sparsity metrics between vanilla double oracle with exact best response and double oracle with \textcolor{\minnewaskaexpcolor}{approximate best responses} for each scenario we consider. Similarly as in the previous figure, for logistic interdiction, we further compare how the algorithms scale for different values of fixed delay factor $\gamma\in\{0.9, 0.95\}$. We depict also the size of the layered graph for each scenario by the number of its edges, corresponding to the number of binary variables in the best-response MILPs.}
    \label{fig:sparsity:complete}
\end{figure}

Subsequently, in Figure~\ref{fig:sparsity:complete}, we contrast the memory requirements of the double oracle variants by assessing the relative sizes of the subgames and supports in all scenarios considered in this paper. Overall, the distinction between both variants is minimal across all games, as indicated by the \textcolor{\minnewaskaexpcolor}{teal lines (DO with approximate BRs)} overlapping the black lines (DO with exact BRs). For logistic interdiction, we additionally examine the sparsity of the double oracle across two different delay factors. As anticipated, we note that a higher delay factor results in larger subgames and supports, as the attacker is motivated to prioritize safety over speed more than with lower values.

\begin{figure}[t]
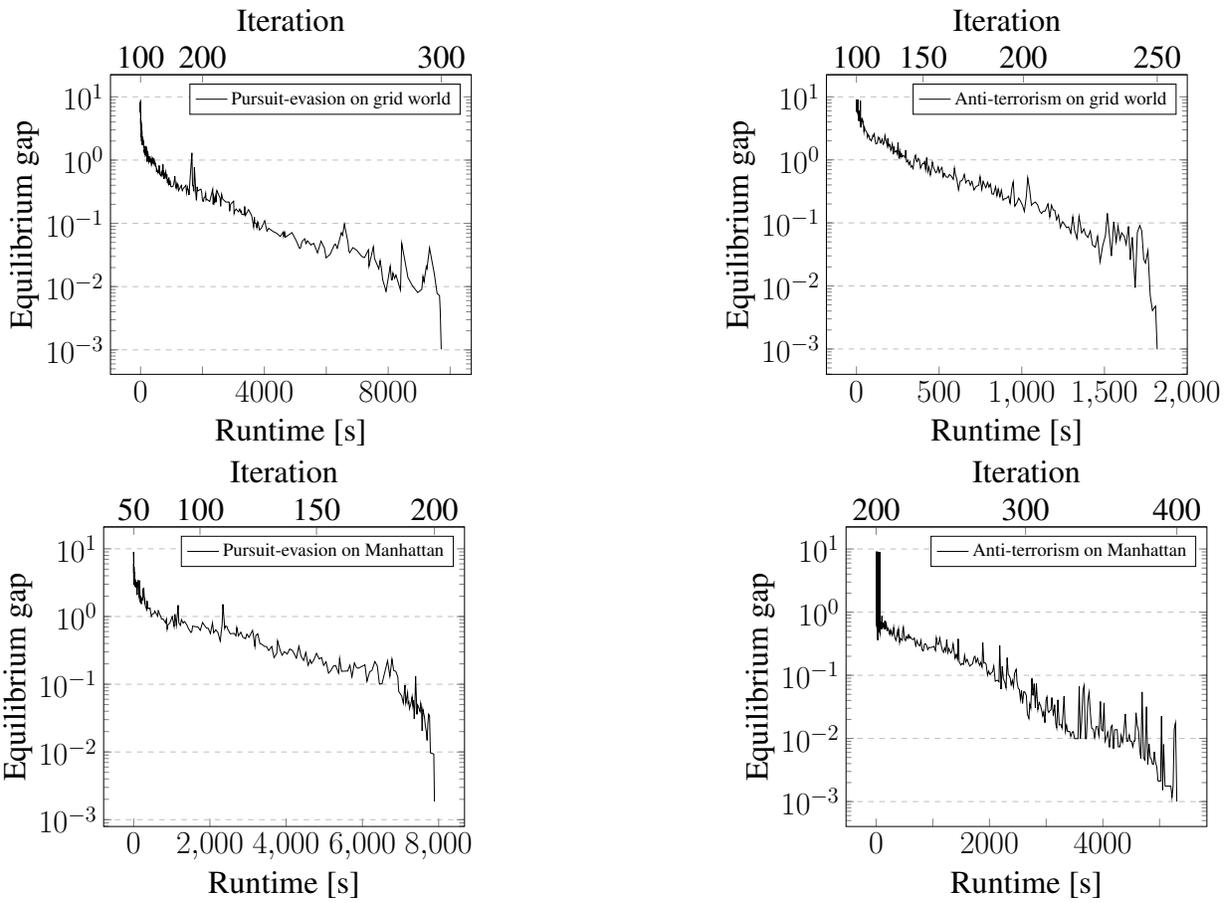

    \centering
\begin{subfigure}[t]{0.45\linewidth}
    \centering
    \input{figs/e_bipu_grid_gap}
\end{subfigure}
\hfill
\begin{subfigure}[t]{0.45\linewidth}
    \centering
    \input{figs/e_bopl_grid_gap}
\end{subfigure}
\\
\begin{subfigure}[t]{0.45\linewidth}
    \centering
    \input{figs/e_bipu_osm_gap}
\end{subfigure}
\hfill
\begin{subfigure}[t]{0.45\linewidth}
    \centering
    \input{figs/e_bopl_osm_gap}
\end{subfigure}
    \caption{Vanilla double oracle with exact best-responses showcasing the evolution of equilibrium gaps over runtime and iterations in pursuit-evasion and anti-terrorism application domains across grid world and Lower Manhattan.}
    \label{fig:eqm_gaps}
\end{figure}

For completeness, Figure~\ref{fig:eqm_gaps} illustrates the typical evolution of equilibrium gaps over computation time and iterations in the double oracle with exact best responses for pursuit-evasion and anti-terrorism scenarios on both grid world and Lower Manhattan. An immediate trend observed is the slowdown of iterations (depicted on the top axis), particularly noticeable on the grid world, attributed to the increasing size of the subgame despite the the number of binary variables in the best-response MILPs being constant. Additionally, equilibrium gaps often experience a sudden drop as they approach the value of $\epsilon=10^{-3}$.

\begin{figure}
    \centering
\hspace{.5cm}
\begin{subfigure}[t]{0.4\linewidth}
    \centering
    \includegraphics[width=\linewidth]{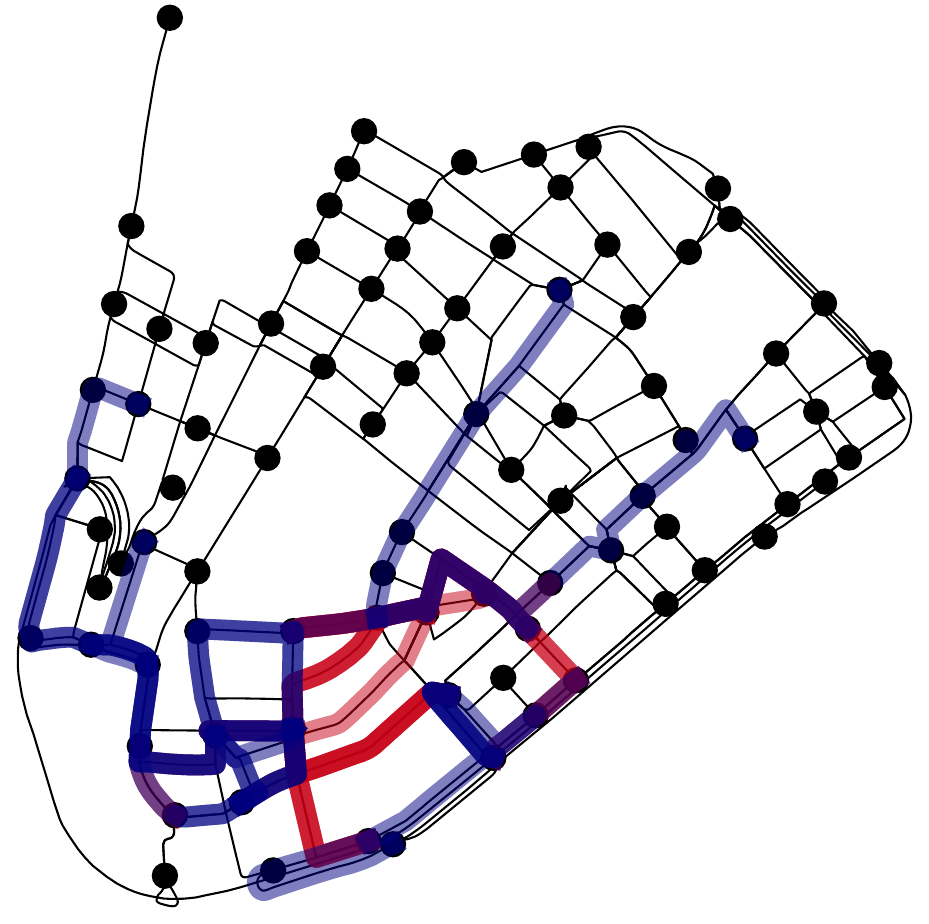}
    \caption{Pursuit-evasion}
\end{subfigure}
\hfill
\begin{subfigure}[t]{0.4\linewidth}
    \centering
    \includegraphics[width=\linewidth]{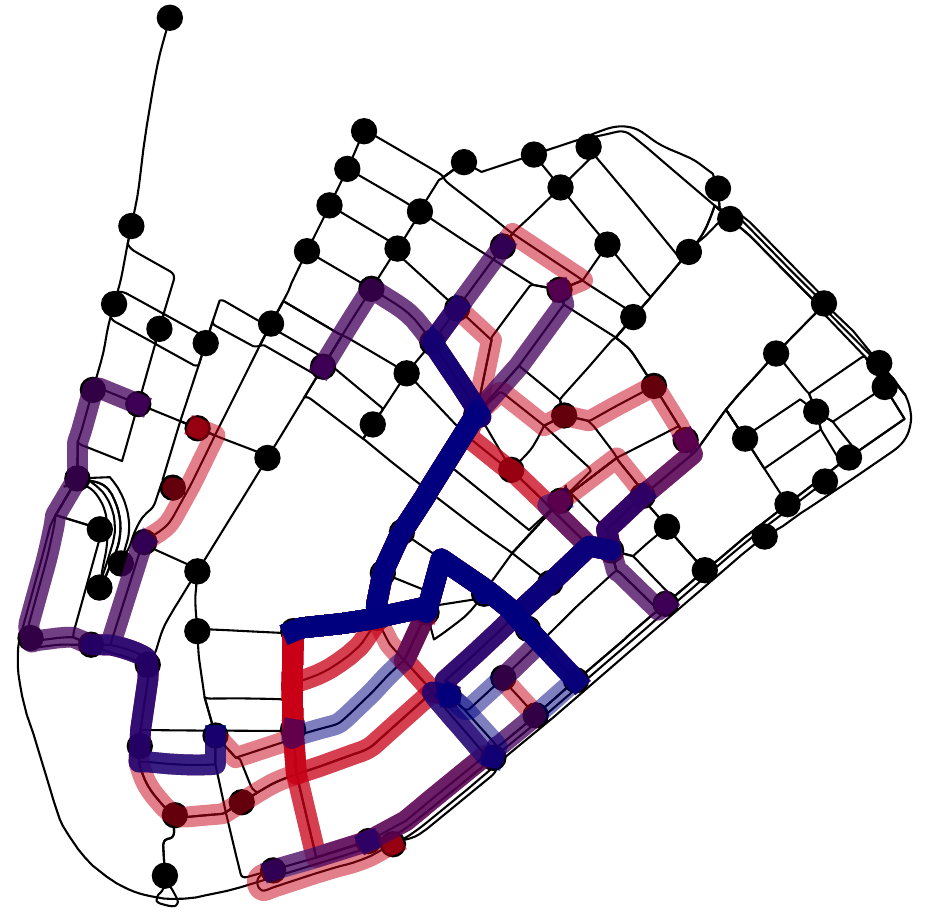}
    \caption{Anti-terrrorism}
\end{subfigure}
\hspace{.5cm}
    \caption{Illustration of defender's (red) and attacker's (blue) equilibrium paths in Lower Manhattan at a depth of 17.}
    \label{fig:manhattan:eqms}
\end{figure}

\begin{figure}
    \centering
\begin{subfigure}[t]{0.45\linewidth}
    \centering
    \includegraphics[width=\linewidth]{figs/minnewaska/eqm_4_20.pdf}
    \caption{Depth 20}
\end{subfigure}
\begin{subfigure}[t]{0.45\linewidth}
    \centering
    \includegraphics[width=\linewidth]{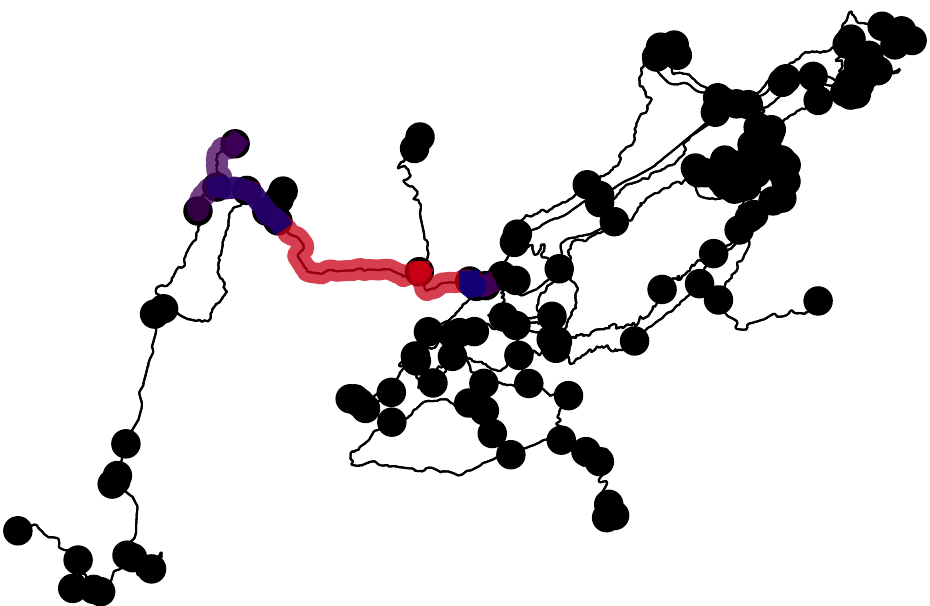}
    \caption{Depth 21}
\end{subfigure}
    \caption{Illustration of defender's (red) and attacker's (blue) equilibrium paths for anti-poaching in Minnewaska State Park using $g_{\text{LIN}}$. At depth 21, a high-value target becomes accessible.}
    \label{fig:minnewaska:eqms}
\end{figure}

\begin{figure*}[t]
    \centering
\begin{subfigure}[t]{0.24\linewidth}
    \centering
    \includegraphics[width=\linewidth]{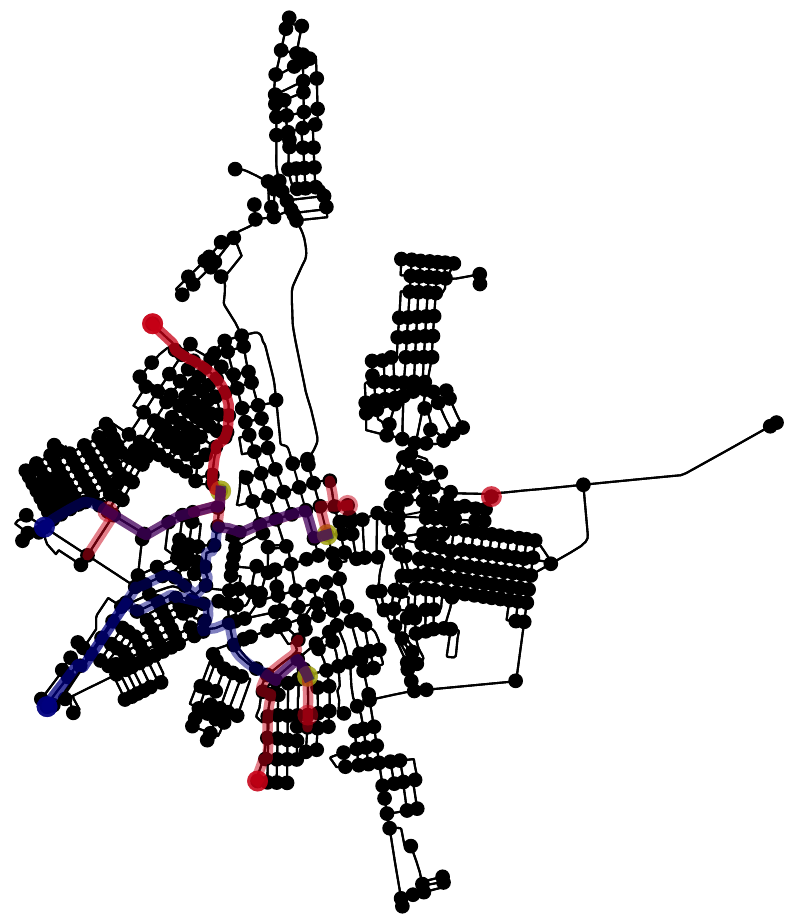}
    \caption{$\gamma = 0.9$, depth 85}
\end{subfigure}
\begin{subfigure}[t]{0.24\linewidth}
    \centering
    \includegraphics[width=\linewidth]{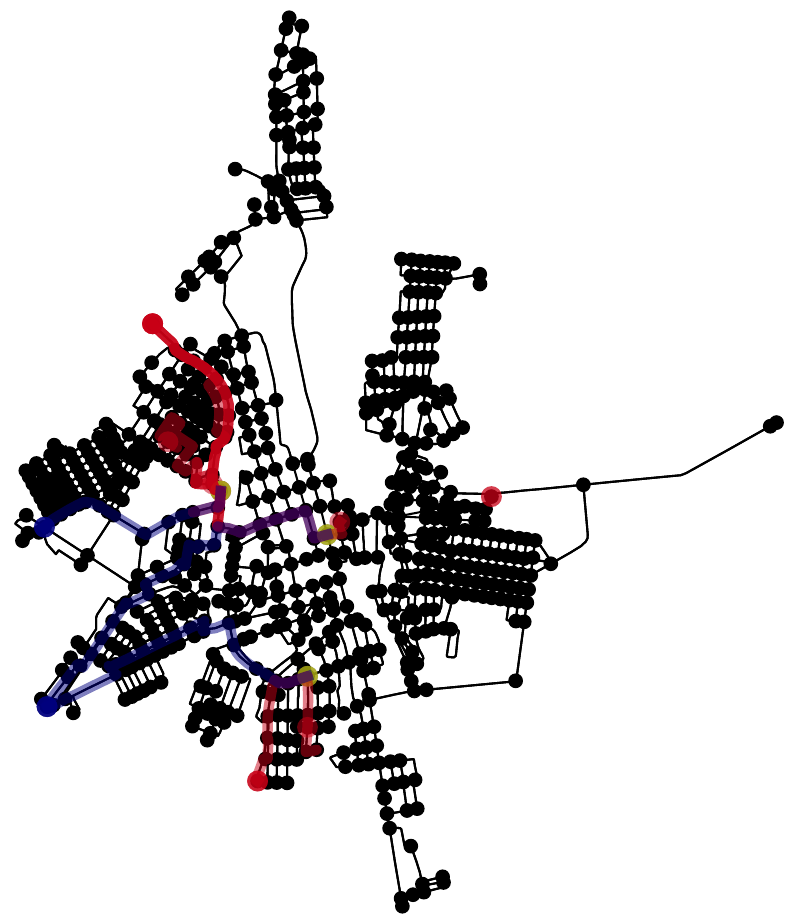}
    \caption{$\gamma = 0.9$, depth 95}
\end{subfigure}
\begin{subfigure}[t]{0.24\linewidth}
    \centering
    \includegraphics[width=\linewidth]{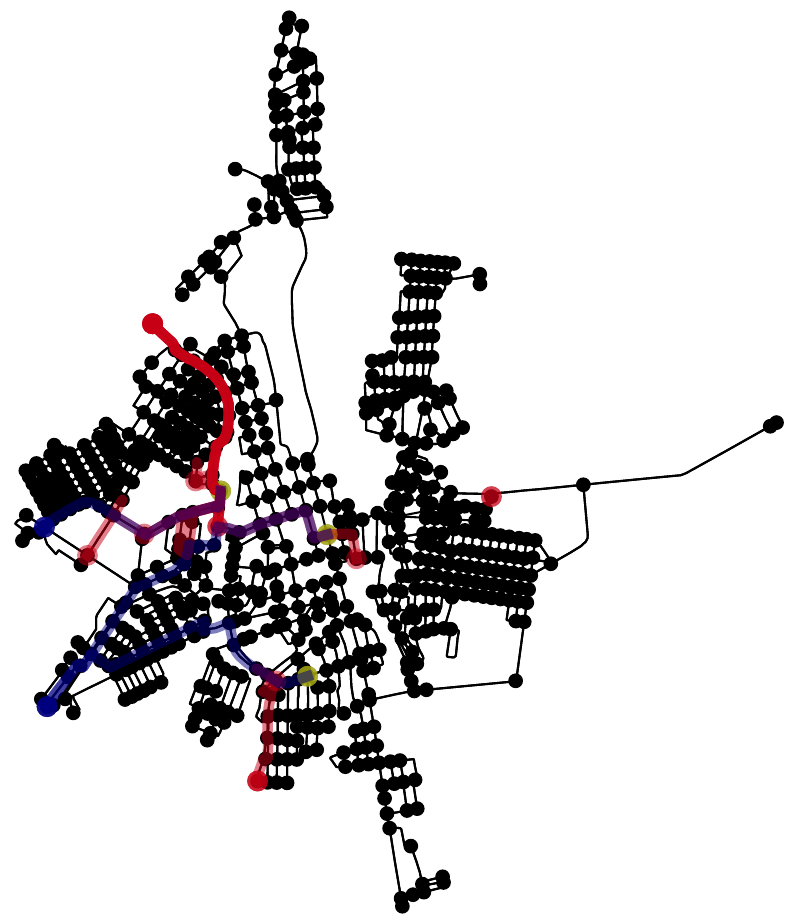}
    \caption{$\gamma = 0.9$, depth 105}
\end{subfigure}
\begin{subfigure}[t]{0.24\linewidth}
    \centering
    \includegraphics[width=\linewidth]{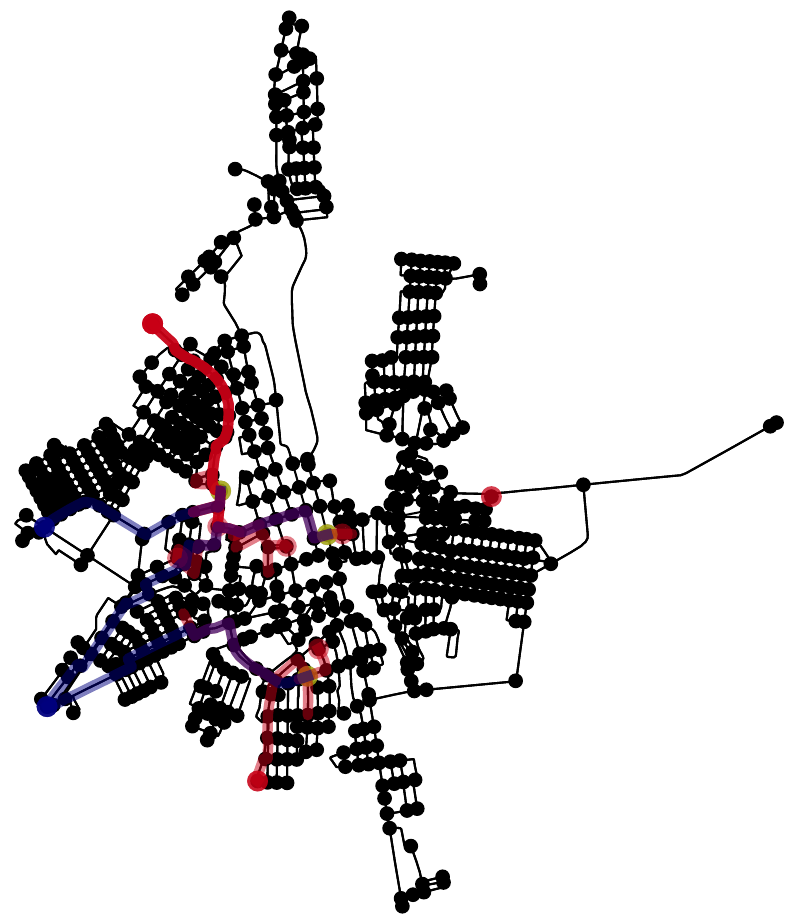}
    \caption{$\gamma = 0.9$, depth 110}
\end{subfigure}
\\
\begin{subfigure}[t]{0.24\linewidth}
    \centering
    \includegraphics[width=\linewidth]{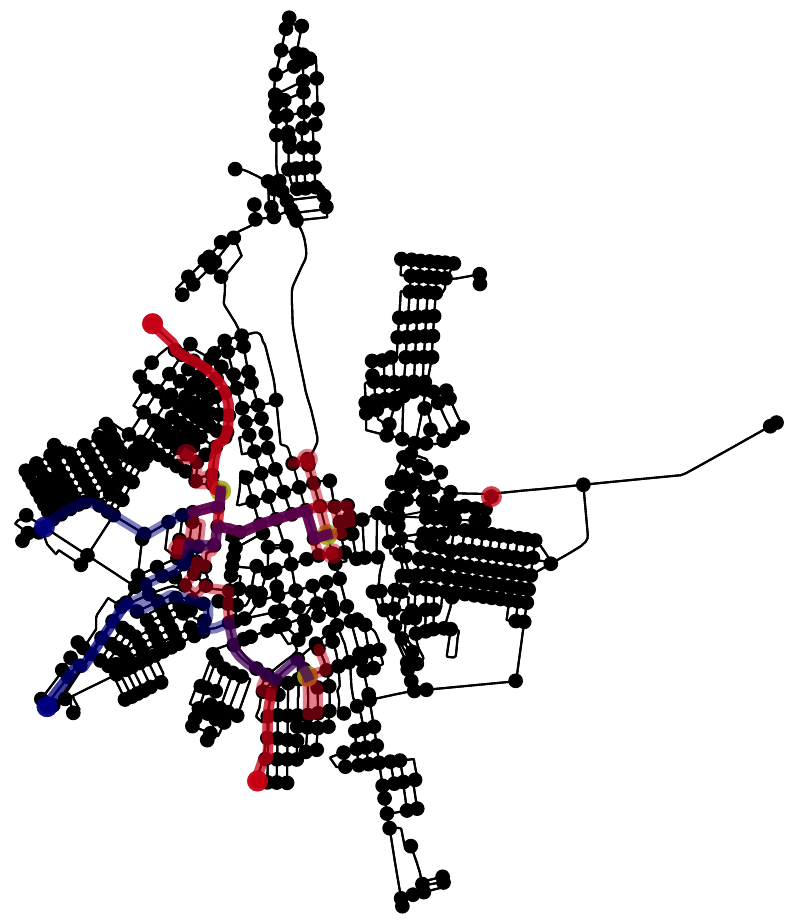}
    \caption{$\gamma = 0.99$, depth 85}
\end{subfigure}
\begin{subfigure}[t]{0.24\linewidth}
    \centering
    \includegraphics[width=\linewidth]{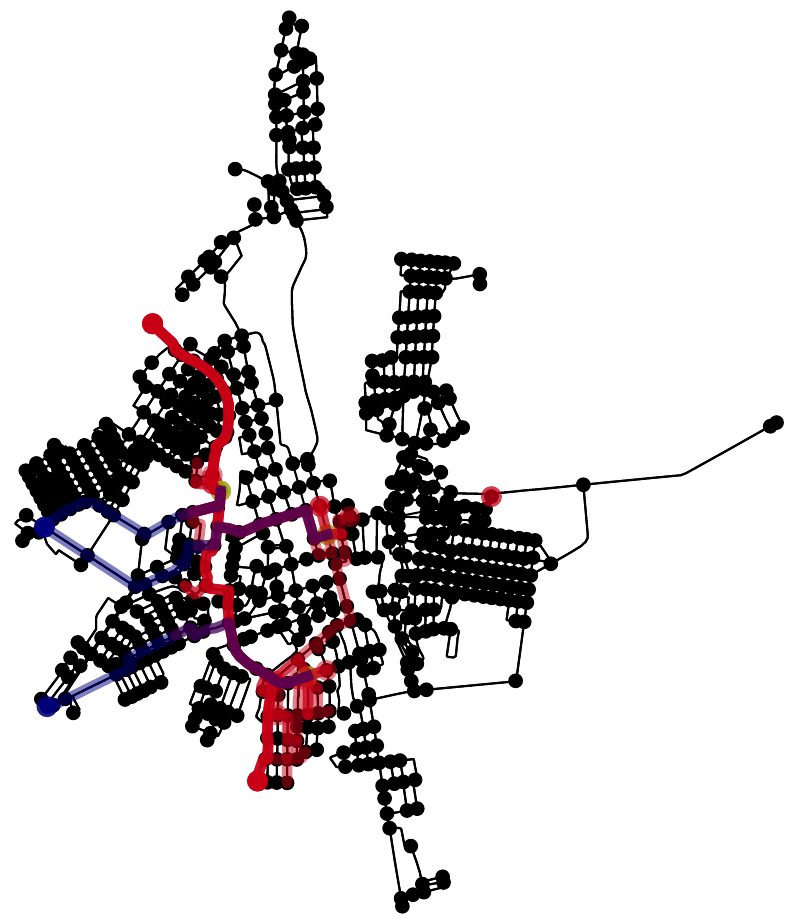}
    \caption{$\gamma = 0.99$, depth 95}
\end{subfigure}
\begin{subfigure}[t]{0.24\linewidth}
    \centering
    \includegraphics[width=\linewidth]{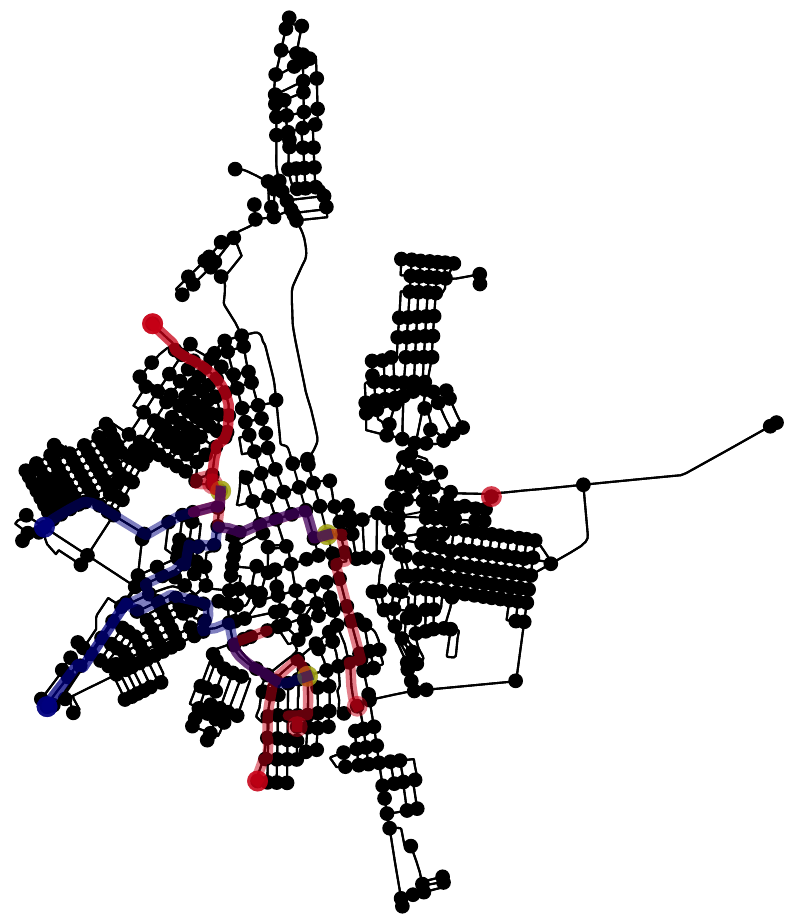}
    \caption{$\gamma = 0.99$, depth 105}
\end{subfigure}
\begin{subfigure}[t]{0.24\linewidth}
    \centering
    \includegraphics[width=\linewidth]{figs/bakhmut/eqm_0.99_110.pdf}
    \caption{$\gamma = 0.99$, depth 110}
\end{subfigure}
    \caption{Illustration of defender's (red) and attacker's (blue) equilibrium paths for logistical interdiction in the city of Bakhmut at depths $85, 95, 105,$ and $110$. The top row represents scenarios with a delay factor of 0.9, while the bottom row has a delay factor set to 0.99.}
    \label{fig:bakhmut:eqms}
\end{figure*}

Finally, in Figures~\ref{fig:manhattan:eqms},~\ref{fig:minnewaska:eqms} and~\ref{fig:bakhmut:eqms} we showcase qualitative results as typical solutions in the scenarios under consideration. First, in Figure~\ref{fig:manhattan:eqms}, we highlight the distinctions between solutions on the exact same grid world with identical target values for both pursuit-evasion and anti-terrorism. We find the defender moves around the attacker's starting nodes, patrolling the area. In contrast, for anti-terrorism, it is more advantageous for the defender to prioritize high-value targets, considering the attacker's need to set up the explosive.

In Figure~\ref{fig:minnewaska:eqms}, we illustrate the typical impact of increasing the horizon for the anti-poaching scenario on the map of Minnewaska State Park, where the values of individual nodes represent the sums of the animal scores, attenuated by $g_{\text{LIN}}(z) = 1/z$, with $z$ being the node's Euclidean distance to the habitat. At depth 20, the attacker still prioritizes nodes in the center of the graph and deploys a complex strategy targeting multiple nodes in the area. Upon reaching depth 21, the node closest to the upper-left high-value animal habitat becomes accessible, resulting in a shift in equilibrium towards this node and noticeably simplifying the strategy.

Figure~\ref{fig:bakhmut:eqms} shows the equilibrium strategies for the logistical interdiction scenario on the map of the city of Bakhmut. The depiction illustrates how the strategies evolve with an increase in the horizon from 85 to 110 and a rise in the delay factor from 0.9 to 0.99. The results indicate that an increase in either parameter generally leads to more complex strategies by the players, with the distinction being more noticeable on the side of the delay factor.

\end{document}